\documentclass{article}

\usepackage[english]{babel}

\usepackage[letterpaper,top=2cm,bottom=2cm,left=3cm,right=3cm,marginparwidth=1.75cm]{geometry}

\usepackage{amsmath}
\usepackage{amssymb}
\usepackage{graphicx}
\usepackage{authblk} 
\usepackage{natbib}

\usepackage[colorlinks=true, allcolors=blue]{hyperref}

\title{Random gas motions inside sub-parsec scale supercritical filaments}
\author[1,2]{\thanks{Corresponding author: zhangchao@tynu.edu.cn} Chao Zhang}
\author[3,4]{\thanks{Corresponding author: liutie@shao.ac.cn} Tie Liu}
\author[5]{Mika Juvela}
\author[6,7]{Paolo Padoan}
\author[8]{Hong-Li Liu}
\author[9,10]{Di Li}
\author[11,12]{Guido Garay}
\author[13]{Neal J. Evans}
\author[14]{Fengwei Xu}
\author[15]{Paul F. Goldsmith}
\author[16]{Qizhou Zhang}
\author[17,18]{Kee-Tae Kim}
\author[3]{Yankun Zhang}
\author[10]{Zhiyuan Ren}
\author[19]{Mengke Zhao}

\affil[1]{Department of Physics, Taiyuan Normal University, Jinzhong 030619, China}
\affil[2]{Institute of Computational and Applied Physics, Taiyuan Normal University, Jinzhong 030619, China}
\affil[3]{State Key Laboratory of Radio Astronomy and Technology, Shanghai Astronomical Observatory, Chinese Academy of Sciences, 80 Nandan Road, Shanghai 200030, China}
\affil[4]{University of Chinese Academy of Sciences, Beijing 100080, China}
\affil[5]{Department of Physics, University of Helsinki, PO Box 64, Helsinki 00014, Finland}
\affil[6]{Department of Physics and Astronomy, Dartmouth College, 6127 Wilder Laboratory, Hanover, NH 03755, USA}
\affil[7]{Institut de Ciències del Cosmos, Universitat de Barcelona, Martí i Franqués 1, Barcelona 08028, Spain}
\affil[8]{Department of Astronomy, Yunnan University, Kunming 650091, China}
\affil[9]{New Cornerstone Science Laboratory, Department of Astronomy, Tsinghua University, Beijing 100084, China}
\affil[10]{National Astronomical Observatories, Chinese Academy of Sciences, Beijing 100012, China}
\affil[11]{Departamento de Astronomia, Universidad de Chile, Las Condes, Santiago 3659, Chile}
\affil[12]{Chinese Academy of Sciences South America Center for Astronomy, National Astronomical Observatories, CAS, Beijing 100101, China}
\affil[13]{Department of Astronomy, The University of Texas at Austin, 2515 Speedway, Stop C1400, Austin, TX 78712-1205, USA}
\affil[14]{Max Planck Institute for Astronomy, Königstuhl 17, Heidelberg 69117, Germany}
\affil[15]{Jet Propulsion Laboratory, California Institute of Technology, 4800 Oak Grove Drive, Pasadena, California 90290, USA}
\affil[16]{Radio and Geoastronomy Division, Center for Astrophysics, Harvard \& Smithsonian, 60 Garden Street, Cambridge, MA 02138, USA}
\affil[17]{Korea Astronomy and Space Science Institute, 776 Daedeokdaero, Yuseong-gu, Daejeon 34055, Republic of Korea}
\affil[18]{University of Science and Technology, Korea (UST), 217 Gajeong-ro, Yuseong-gu, Daejeon 34113, Republic of Korea}
\affil[19]{School of Astronomy and Space Science, Nanjing University, Nanjing 210023, China}

\begin{document}
\maketitle

\begin{abstract}
Supercritical gas filaments in molecular clouds host the dense cores in which new stars form. The mechanisms governing their formation and subsequent gas accretion remain poorly understood. In this study, we conduct a statistical analysis of a large sample of sub-parsec supercritical filaments using H$^{13}$CO$^+$ $J=1\text{–}0$ data from the ALMA Three-millimeter Observations of Massive Star-forming regions (ATOMS) Survey. We identified velocity-coherent filaments in position–position–velocity (PPV) space and systematically examined velocity gradients both along and perpendicular to their skeletons. Our analysis uncovers a remarkable result: at scales of $\sim$0.1–1 pc, the local velocity gradients within these supercritical filaments show no preferred alignment with the filament skeletons and exhibit no correlation with the local gravitational field. This random orientation suggests the presence of chaotic gas motions deep inside these dense structures. These findings may indicate that turbulence—rather than gravity—dominates gas dynamics and structural evolution at small scales, even in regions on the verge of star formation, challenging the paradigm of gravity-dominated structure formation within molecular clouds. This scenario should be further tested by more state-of-the-art simulations. This study offers key observational insights into the roles of turbulence and gravity in establishing the initial conditions for star formation.
\end{abstract}

\section{Introduction}

The formation of stars is a fundamental physical process in the cosmic ecosystem, with filamentary structures within molecular clouds serving as the primary sites where stars form \citep{Men2010,2014prpl.conf...27A,Knyves2015}. The internal dynamical evolution mechanisms of these filaments directly determine the initial conditions for star formation. In particular, understanding how sub-parsec-scale, supercritical filamentary structures give rise to stars through processes such as fragmentation and collapse remains a central challenge in contemporary astrophysics.

Theoretical models suggest that supersonic turbulence — which is ubiquitous in molecular clouds — generates intricate filamentary networks through shock compression \citep{Vazquez1994,Padoan1995,Padoan1999,Jappsen2005}. The subsequent evolution of these filaments into prestellar cores and stars is currently explained by two primary physical paradigms: one posits that supersonic turbulence plays a dominant role in the fragmentation process \citep{Padoan2002,Hennebelle2008,Hopkins2012,Padoan2020}, while the other highlights the role of gravitational fragmentation, typically assuming initially static (rather than turbulent) conditions \citep{Inutsuka1992,Inutsuka1997,Inutsuka2001}. In the turbulent star formation scenario, structures such as filaments, clumps, and cores all emerge from compressive processes driven by random gas motions \citep{Padoan2002,Padoan2020}. In contrast, when strong gravity or magnetic fields dominate, the gas motions within molecular clouds are expected to be more ordered  \citep{Tang2019,Wang2020,Wang2024}. Within the framework of the global hierarchical collapse (GHC) model, gravity-driven filamentary accretion flows are predicted to be anisotropic and often exhibit conveyor-belt-like behavior \citep{Enrique2024}. Nevertheless, observational constraints on the gas motions within dense structures in molecular clouds — which could help distinguish between these theoretical scenarios — remain scarce.

In recent years, advances in observational techniques have provided deeper insights into the kinematic properties of filamentary structures. High-resolution observations have revealed complex internal motions, including longitudinal collapse along the filament axis \citep{Kirk2013,Fern2014,Tackenberg2014,Gong2018,Dutta2018,Lu2018,Chen2020}, radial contraction perpendicular to the filament axis \citep{Kirk2013,Fern2014,Dhabal2018}, and mass accretion from secondary filaments onto the main filament \citep{Palmeirim2013,Dhabal2018,Arzoumanian2018,Shimajiri2019}. Although significant efforts have been made to investigate the kinematic properties of filamentary structures, most studies have focused on large-scale (1–10 pc) global kinematic analyses \citep{Friesen2013,Kirk2013}. However, statistical investigations at small sub-parsec scales remain exceedingly limited \citep{Chen2020ApJ...891...84C,Chen24}. Furthermore, hub structures formed at the intersections of filaments are regarded as favorable sites for mass accumulation and the formation of massive stars (with masses exceeding 8 solar masses), often characterized by enhanced dynamical activity and higher star formation efficiency \citep{Myers2009,Schneider2010,Sugitani2011,Peretto2014,Rayner2017,Baug2018,Trevi2019}. However, systematic studies of the structure and kinematics of these hub regions are still lacking.

\section{Materials and methods}\label{sec:2}
In this work, we conducted a statistical study of sub-parsec scale filaments in a large sample of massive clumps from the ATOMS survey. The ATOMS survey \citep{Liu20_1} — short for ALMA Three-millimeter Observations of Massive Star-forming regions — observed 146 active Galactic star-forming regions (see Supplementary Materials A1; The Supplementary Materials contains nine sections, including observational data, filament extraction, filament properties, simulation data, and error analysis.), most of which exhibit filament-hub systems \citep{Zhou2022}. 

The ATOMS sources were selected from a sample of UC H{\sc ii} region candidates with bright CS emission $T_{\rm b}$ $>$ 2 K \citep{Bronfman96}, indicative of reasonably dense gas. Most ATOMS sources are gravitationally bound, exhibiting a virial parameter below 2. The observations utilized both the 12-meter array and the 7-meter ACA (Atacama Compact Array). The data from both arrays were combined during processing to achieve both good sensitivity and spatial scale information. The spectral resolution and beam size for H$^{13}$CO$^+$ J=1-0 (86.754288 GHz) line data are $\sim$0.211 km~s$^{-1}$ and $\sim$2.5 arcsec, respectively. The typical rms level is 8 mJy~beam$^{-1}$ per channel. More details on the observations and data reduction can be found in Supplementary Materials and in \citep{Liu20_1}.

\begin{figure*}
\centering
    \includegraphics[width=\textwidth]{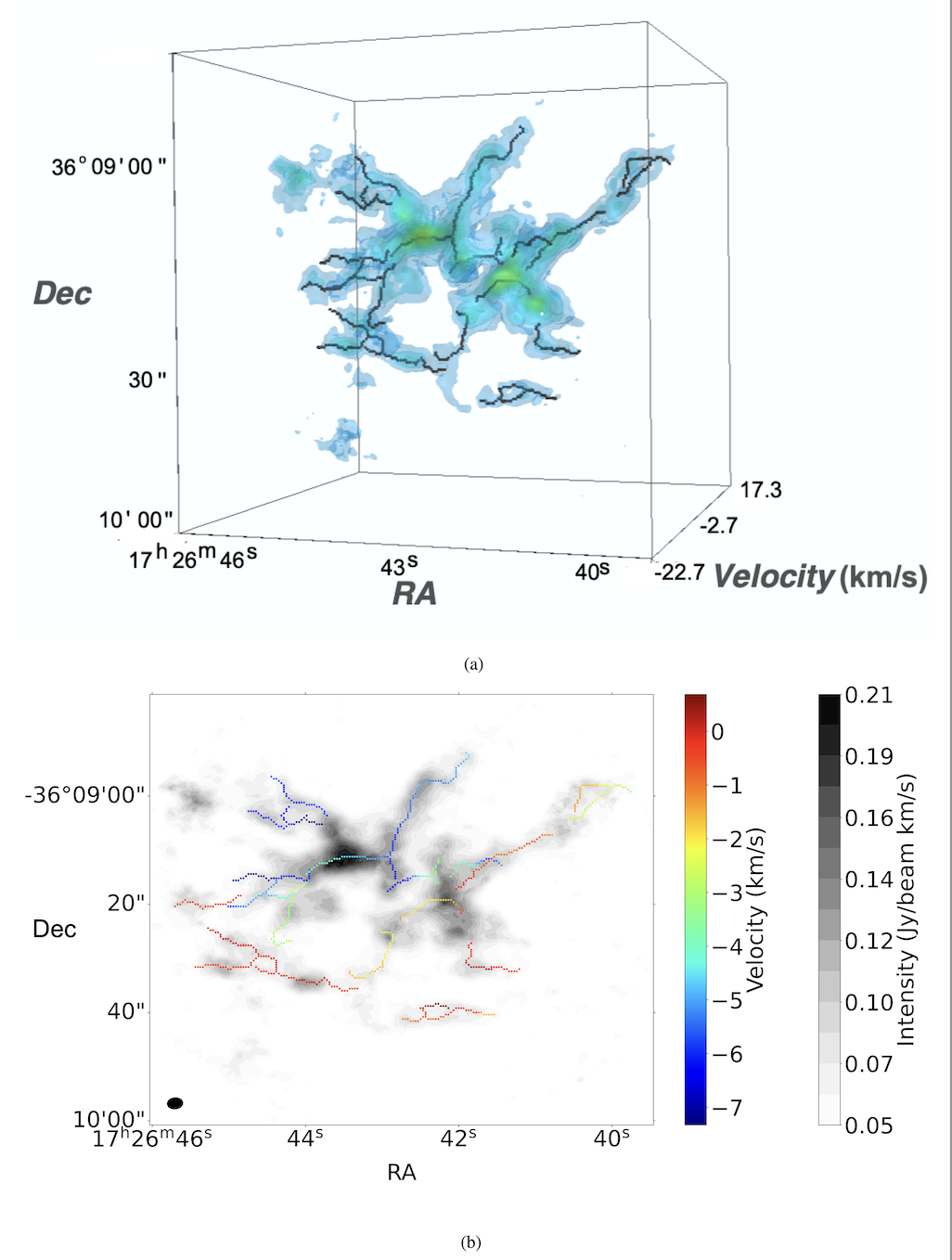}
    \caption{Source I17233-3606 as an example of a three-dimensional filamentary network. (a) The color image represents the three-dimensional PPV map of I17233-3606, and the black lines indicate the skeletons of the extracted filamentary structures. (b) The projection of the three-dimensional data from panel (a) onto the position-position plane. The gray contour map represents the intensity of I17233-3606 and the lines indicate the skeletons of the filamentary structures color coded by their velocities. The black ellipse in the lower left corner indicates the size of the beam. } 
    \label{fig:3dskeleton}
\end{figure*}

Using the H$^{13}$CO$^+$ J=1-0 molecular line data, which is generally optically thin \citep{Zhou2022, Zhang2025}, we extracted and analyzed filaments within these clumps, focusing on their local velocity gradient fields. Gas filaments were extracted from the H$^{13}$CO$^+$ J=1-0 line data in position-position-velocity (PPV) space, using the CRISPy algorithm \citep{Chen2020}. The algorithm identifies emission ridges in PPV space and maps them back to the original grid as ``skeletons", by applying key parameters such as an intensity threshold (e.g., 5-$\sigma$) and a smoothing bandwidth. After initial extraction, the data were cleaned by removing excessively short filaments and pruning minor branches to ensure the significance and continuity of the analyzed structures. Figure~\ref{fig:3dskeleton} shows the identification of the filamentary network within an exemplar source I17233-3606. Panel (a) presents the PPV image of H$^{13}$CO$^+$ J=1-0 line emission, while panel (b) shows the integrated intensity map, overlaid with filament skeletons.  In total, within the 147 sources of the ATOMS survey, we identified 837 filamentary structures that are coherent in velocity (hereafter velocity-coherent), characterized by continuous and smooth velocity along their lengths. Among them, 214 filaments have an aspect ratio (length-to-width ratio) greater than 5.

Following the identification of the filament skeletons, their intensity and velocity gradients were measured. To interpret the gas kinematic patterns, we adopted the vector field decomposition technique, which decomposes the velocity gradient vector at each pixel into two orthogonal components: one parallel to the local filament skeleton direction and one perpendicular to it \citep{Chen2020}. This decomposition allows for distinguishing between different kinematic modes, such as gas flow along the filament and contraction motions perpendicular to it (i.e., towards the filament's interior), thereby providing a foundation for subsequent analysis of gas accumulation and motion mechanisms.

\section{Results}\label{sec:3}
\subsection{Sub-parsec
scale supercritical filamentary structures}

The lengths of these filamentary structures range from 0.02 pc to 1.6 pc, with a median value of 0.23 pc. Their gas masses are derived from H$^{13}$CO$^+$ J=1-0 line emission \citep{Sanhueza2012,Xu2023} (see Supplementary Materials). Figure~\ref{fig:L_to_M} illustrates the mass–length relationship of the filaments, which follows a scaling trend similar to that observed in Hub-Filament Systems (HFSs) reported in the literature \citep{Hacar2025}. Among the filaments in our sample, 823 (98\%) have line masses exceeding the critical line masses ($m_\mathrm{crit}$)\footnote{$m_\mathrm{crit}(T)=\frac{2c_\mathrm{s}^2}{G}\sim16.6\left(\frac{T}{10\mathrm{~K}}\right)M_\odot\mathrm{pc}^{-1}$. For T = 10 K, $m_\mathrm{crit}$ = 16.6 $M_\odot$. Further details are provided in the Supplementary Materials.}, confirming their supercritical nature. These structures are gravitationally bound and, in the absence of additional supporting forces, are expected to be either contracting or undergoing fragmentation (see Supplementary Materials). Moreover, the mass–length relation of the filaments in this sample aligns well with that of massive star-forming (MSF) ATLASGAL clumps from ref. \citep{Urquhart18} (shown as grey dots in Figure~\ref{fig:L_to_M}), suggesting that filaments contain a significant fraction of the mass within MSF clumps. Additionally, the data broadly follow the $L \propto M^{0.5}$ scaling relation, which is interpreted as evidence for turbulence-driven fragmentation of the filaments \citep{Ge2022,Hacar23, Feng2024, Hacar2025}.

\begin{figure*}
    \centering
    \includegraphics[width=\linewidth]{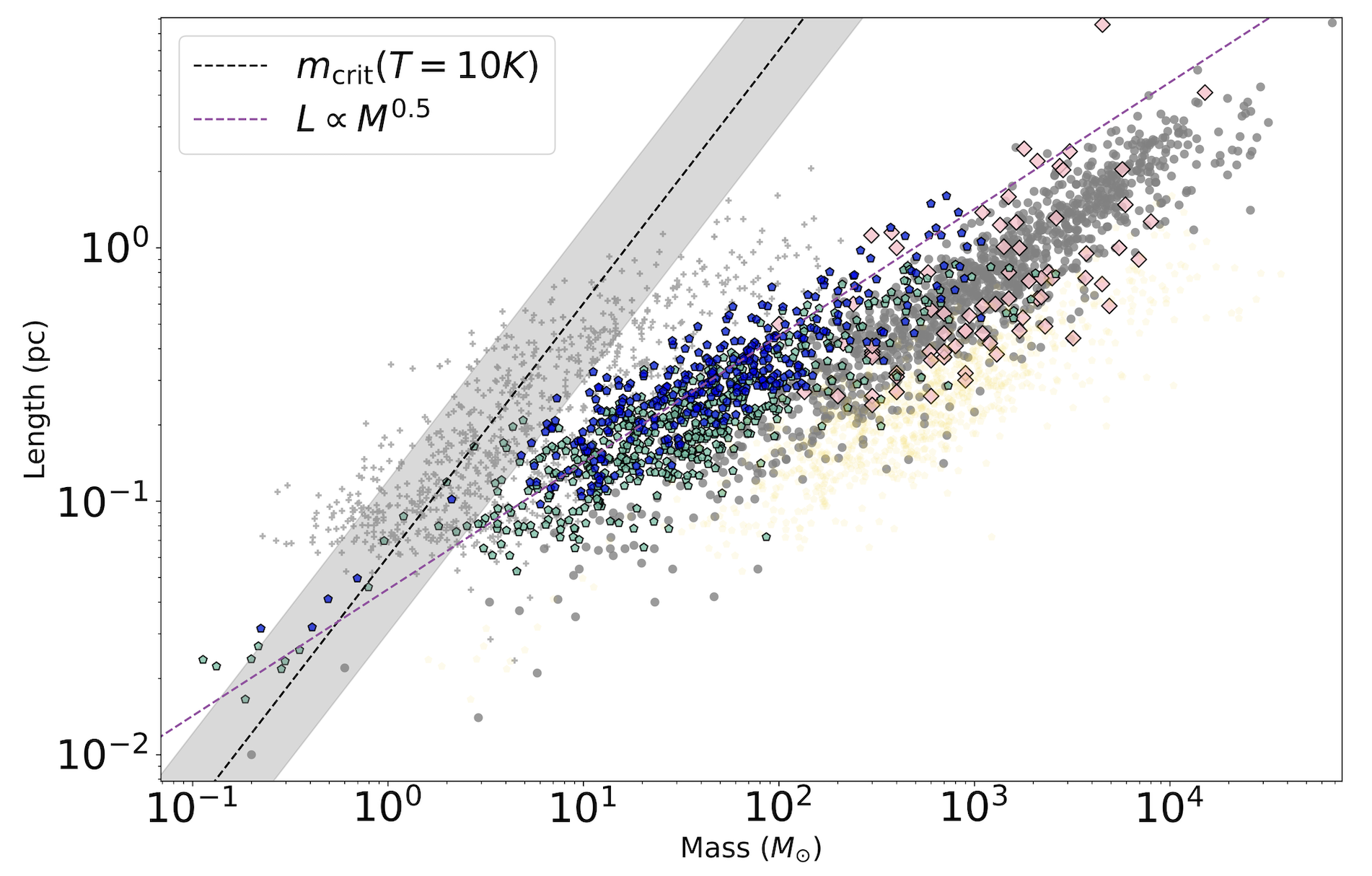}
    \caption{The length of each filament plotted against the filament's mass. The blue and green pentagons represent filaments with aspect ratios greater than 5 and below 5, respectively. The value of $m_\mathrm{crit}$ at 10 K is marked by the black dashed line, with the shaded region representing the transcritical 0.5$m_\mathrm{crit}$$\leq$ $m$ $\leq$2$m_\mathrm{crit}$. The purple dashed lines represents the correlations $L \propto M^{0.5}$. The pink diamonds represent HFS\citep{Hacar2025}. The gray circles represent MSF\citep{Urquhart18}. The gold pentagons denote filament masses derived assuming a lower H$^{13}$CO$^+$ abundance of 9$\times$10$^{-12}$. The gray plus signs represents the simulation data.}
    \label{fig:L_to_M}
\end{figure*}

\subsection{Random Velocity gradients not dominated by local gravity}

In order to investigate the gas motion within these filaments, we calculate the local velocity gradients. Local velocity gradients are often interpreted as evidence for gas flows along or across filaments \citep{2013ApJ...766..115K,2019ApJ...875...24C,2019ApJ...878...10T,Wang2024,Chen24,2025A&A...695A..51B}. Such gradients could, in principle, also trace other dynamical processes like rotation \citep{2015A&A...584A..67B,2021ApJ...908...92H}. Following these previous works, we aim to investigate whether the gas motion traced by local velocity gradients follows an ordered pattern and is governed by gravity.  In this study, we measured the velocity gradients of filaments identified in the ATOMS survey using the same methodology as \citep{Chen2020} (see Supplementary Materials). We also derived the intensity gradients, which trace spatial variations in column density and are expected to align closely with the local gravitational field within filaments \citep{Wang2024}. The gravitational acceleration ($\text{g}$) computed from the 2D surface density provides a reasonable approximation to the projected 3D gravitational field, as shown in \citep{He2023}. Although the magnitude may be systematically overestimated, the direction of g, indicating the orientation of self-gravity, remains highly reliable \citep{He2023}. Our calculation of gravitational acceleration follows the approach outlined in \citep{He2023} (see Supplementary Materials).

\begin{figure*}
    \centering
    \includegraphics[width=0.9\linewidth]{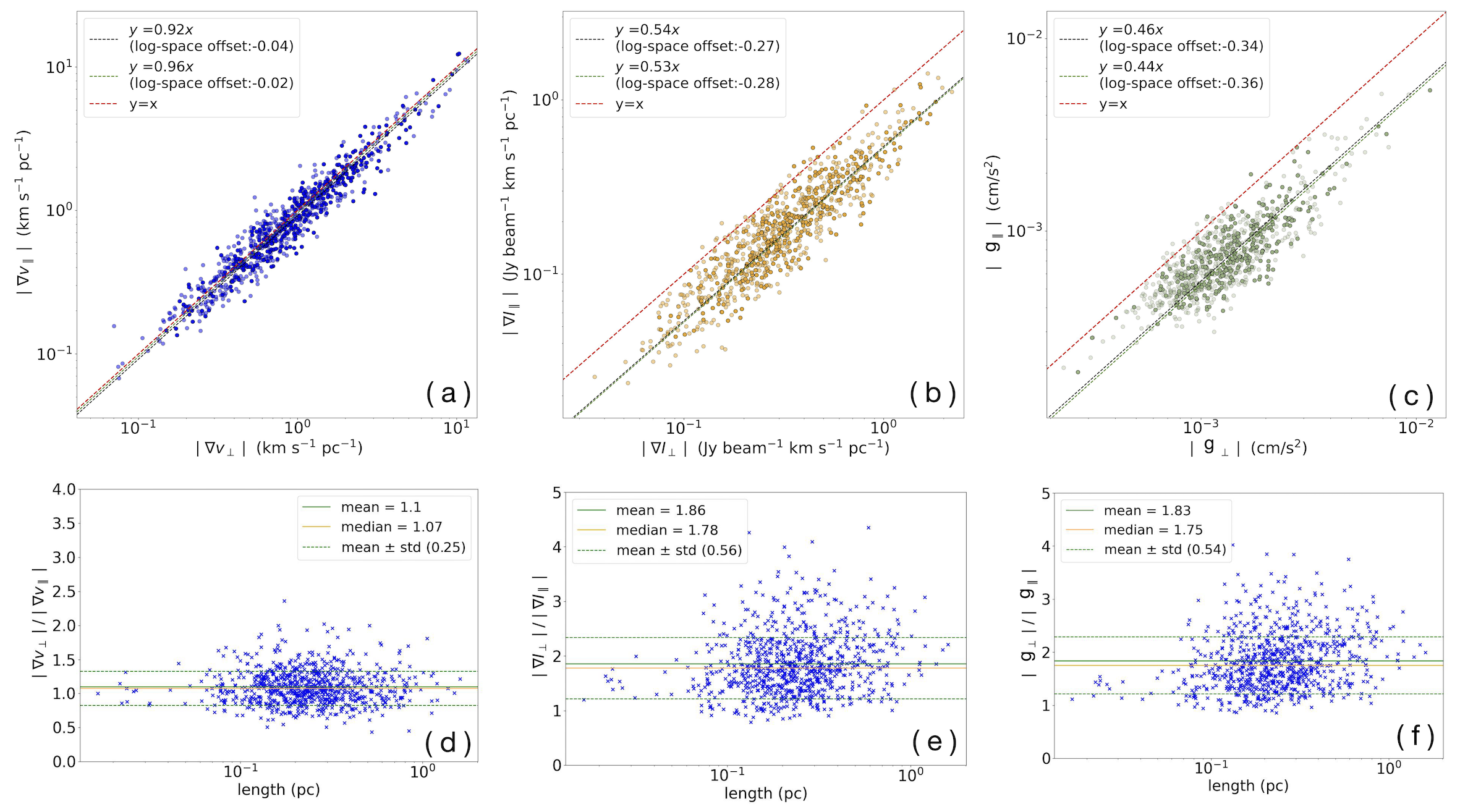}
    \caption{Panel (a) (b) and (c): show the correlation of gradients. Filaments with aspect ratios exceeding 5 are depicted in darker shades, whereas those with aspect ratios below 5 are shown in lighter tones. (a) The median $\mid \nabla v_{\parallel} \mid$ of each filament plotted against their median $\mid\nabla v_{\perp} \mid$ counterparts; (b) The median $\mid\nabla I_{\parallel}\mid$ plotted against their median $\mid\nabla I_{\perp}\mid$ counterparts; (c) The median $\mid \text{g}_{\parallel}\mid$ plotted against their median $\mid \text{g}_{\perp}\mid$ counterparts. The best-fitting linear regression models for all filaments are shown as black dashed lines, while those for filaments with aspect ratios $>$ 5 are represented by green dashed lines. The red dashed line represents a slope of unity. Panel (d) (e) and (f): show the ratio of gradients versus the logarithm of the filament length. (d) The vertical coordinate is the ratio of $\mid\nabla v_{\perp} \mid$ to $\mid \nabla v_{\parallel} \mid$; (e) The vertical coordinate is the ratio of $\mid\nabla I_{\perp} \mid$ to $\mid \nabla I_{\parallel} \mid$; (f) The vertical coordinate is the ratio of $\mid \text{g}_{\perp} \mid$ to $\mid  \text{g}_{\parallel} \mid$. The solid green and yellow lines denote the mean and median values of the ratios, respectively. The green dashed lines represent the mean ± 1 standard deviation (the value in parentheses indicates the standard deviation value).}
    \label{fig:velo_par_div}
\end{figure*}

We applied a signal-to-noise ratio threshold greater than 5 and a step of one beam, to construct pixel-by-pixel maps of the velocity gradients ($\nabla v$), intensity gradients ($\nabla I$), and gravitational acceleration ($\text{g}$) within the filaments. These gradients were then decomposed into components parallel ($\nabla v_{\parallel}$, $\nabla I_{\parallel}$, $\text{g}{_\parallel}$) and perpendicular ($\nabla v{\perp}$, $\nabla I_{\perp}$, $\text{g}_{\perp}$) to the filament spine (see Supplementary Materials). For each filament, the median value of the gradients from all its pixels was adopted to represent its overall properties \citep{Chen2020, Chen24}. In Figure~\ref{fig:velo_par_div}(a), the median magnitudes of $\mid\nabla v_{\perp}\mid$ are plotted against $\mid\nabla v_{\parallel}\mid$ for each filament, showing a strong linear correlation (Pearson’s $r = 0.96$) with a best-fit slope of $0.92 \pm 0.01$. When considering only filaments with aspect ratios greater than 5, the slope increases to 0.96. This near-unity slope indicates that the local velocity gradients perpendicular and parallel to the filament skeletons are comparable. Figure~\ref{fig:velo_par_div}(b) displays the median $\mid\nabla I_{\perp}\mid$ versus $\mid\nabla I_{\parallel}\mid$, which also exhibits a strong correlation ($r = 0.96$) but with a shallower best-fit slope of $0.54 \pm 0.01$. This result suggests that the filaments are undergoing significant compression in the transverse direction. Similarly, Figure~\ref{fig:velo_par_div}(c) presents the median $\mid \text{g}_{\perp}\mid$ versus $\mid \text{g}_{\parallel}\mid$, yielding a linear regression slope of $\sim 0.45$, comparable to that of the intensity gradients. This implies that self-gravity acts predominantly perpendicular to the filament skeletons. Panels (d)–(f) show the corresponding gradient ratios: $\mid\nabla v_{\perp}\mid / \mid\nabla v_{\parallel}\mid$, $\mid\nabla I_{\perp}\mid / \mid\nabla I_{\parallel}\mid$, and $\mid \text{g}_{\perp}\mid / \mid \text{g}_{\parallel}\mid$. The ratio $\mid\nabla v_{\perp}\mid / \mid\nabla v_{\parallel}\mid$ is close to 1, while $\mid\nabla I_{\perp}\mid / \mid\nabla I_{\parallel}\mid$ and $\mid \text{g}_{\perp}\mid / \mid \text{g}_{\parallel}\mid$ are approximately 2. These results consistently reinforce the trends observed in (a)–(c). In summary, these filaments are not dominated by longitudinal flows. Instead, transverse gas motions across these supercritical filaments are also equally important and cannot be neglected.

\begin{figure*}
    \centering
    \includegraphics[width=\linewidth]{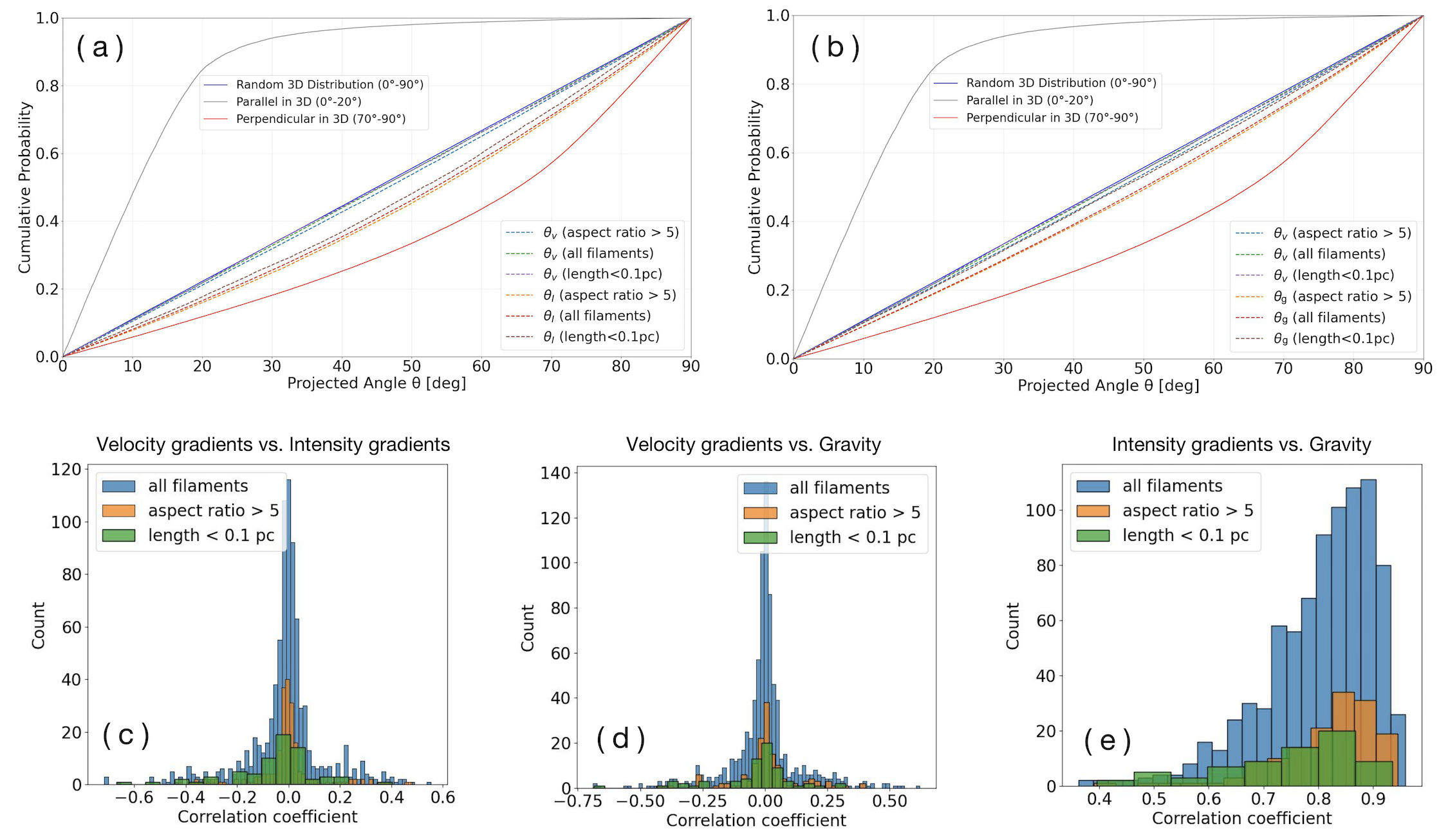}

    \caption{Panels (a) and (b) show the cumulative distribution functions of the relative orientation between the filaments and $\nabla v$, $\nabla I$, and the projected $\theta_{\text{3D}}$, as well as between $\nabla v$, \text{g}, and the projected $\theta_{\text{3D}}$, respectively. Panels (c), (d), and (e) present the distributions of the Pearson correlation coefficients for the correlations between intensity and velocity gradients, velocity gradients and gravity, and intensity gradients and gravity, respectively.}
    \label{fig:cdf}
\end{figure*}

We further investigated the spatial alignment among the gradients, gravitational acceleration, and the filament skeletons by measuring the orientation angles of $\nabla v$, $\nabla I$, and g relative to the filament skeletons in the plane of the sky, denoted as $\theta_{\text{v}}$, $\theta_{\text{I}}$, and $\theta_{\text{g}}$, respectively. Figures~\ref{fig:cdf}(a) and (b) show the cumulative distribution functions (CDFs) of these orientation angles for the full sample, a sub-sample of the 214 longest filaments (with aspect ratios $>5$), and a sub-sample of the shortest 69 filaments (lengths $<0.1$ pc). Projection effects can cause significant differences between the observed two-dimensional angular distributions and their intrinsic three-dimensional configurations \citep{seifried2020}. To account for this, we adopted a Monte Carlo approach \citep{stephens2017}, generating $10^6$ random unit vector pairs with intrinsic angles $\theta_{\text{3D}}$ ranging from $0^\circ$ to $90^\circ$, and projecting them onto the 2D plane to derive the corresponding azimuthal angles $\phi$. The resulting CDFs represent three distinct 3D configurations: (1) preferentially parallel ($\theta_{\text{3D}} = 0^\circ-20^\circ$), (2) random ($\theta_{\text{3D}} = 0^\circ-90^\circ$), and (3) preferentially perpendicular ($\theta_{\text{3D}} = 70^\circ-90^\circ$) \citep{Jiao2024}.

As shown in Figure~\ref{fig:cdf}(a), the distribution of $\theta_{\text{v}}$ is consistent with a random orientation across all samples. In contrast, $\theta_{\text{I}}$ displays a statistically significant preference for perpendicular alignment. This result is consistent with the earlier finding that $\mid\nabla I_{\perp}\mid$ is, on average, nearly twice as large as $\mid\nabla I_{\parallel} \mid$ (Figure~\ref{fig:velo_par_div}(b)). Similarly, Figure~\ref{fig:cdf}(b) reveals that $\theta_{\text{g}}$ follows a trend akin to $\theta_{\text{I}}$, showing a clear preference for perpendicularity. Despite the effects of line-of-sight integration, the random distribution of velocity gradients in two-dimensional projections suggests an underlying three-dimensional isotropy. In summary, the local velocity gradients in these sub-parsec supercritical filaments are randomly oriented relative to their skeletons. Even at scales smaller than 0.1 pc—where gravity has traditionally been considered dominant (e.g., \citep{Bergin2007})—the gas flows traced by these velocity gradients are still likely random. The random nature of velocity gradients observed here is consistent with the random distribution of rotational axes among dense cores embedded within filaments \cite{2020ApJ...894L..20X}, yet stands in clear contrast to the bimodal behavior displayed by magnetic field orientations at dense core scales \cite{2014ApJ...792..116Z}.

Furthermore, we examined the pixel-by-pixel correlations among velocity gradients ($\nabla v$), intensity gradients ($\nabla I$), and gravitational acceleration (\text{g}) across all filaments by computing their Pearson correlation coefficients. The distributions of these coefficients are shown in Figure~\ref{fig:cdf}(c-e). As seen in Figure~\ref{fig:cdf}(c), the distribution of correlation coefficients between $\nabla v$ and $\nabla I$ peaks near zero with small dispersion, indicating no significant correlation between them. Similarly, Figure~\ref{fig:cdf}(d) reveals that velocity gradients show no clear correlation with gravitational acceleration—even in filaments shorter than 0.1 pc. In contrast, intensity gradients and gravitational acceleration are strongly correlated, as demonstrated in Figure~\ref{fig:cdf}(e), confirming that column density aligns closely with the local gravitational field. These results suggest that, from a statistical perspective, the local velocity gradients within these sub-parsec supercritical filaments are unlikely to be governed by gas density distribution or local gravity.

\section{Discussion}\label{sec:4}
\begin{figure*}
    \centering
    \includegraphics[width=\linewidth]{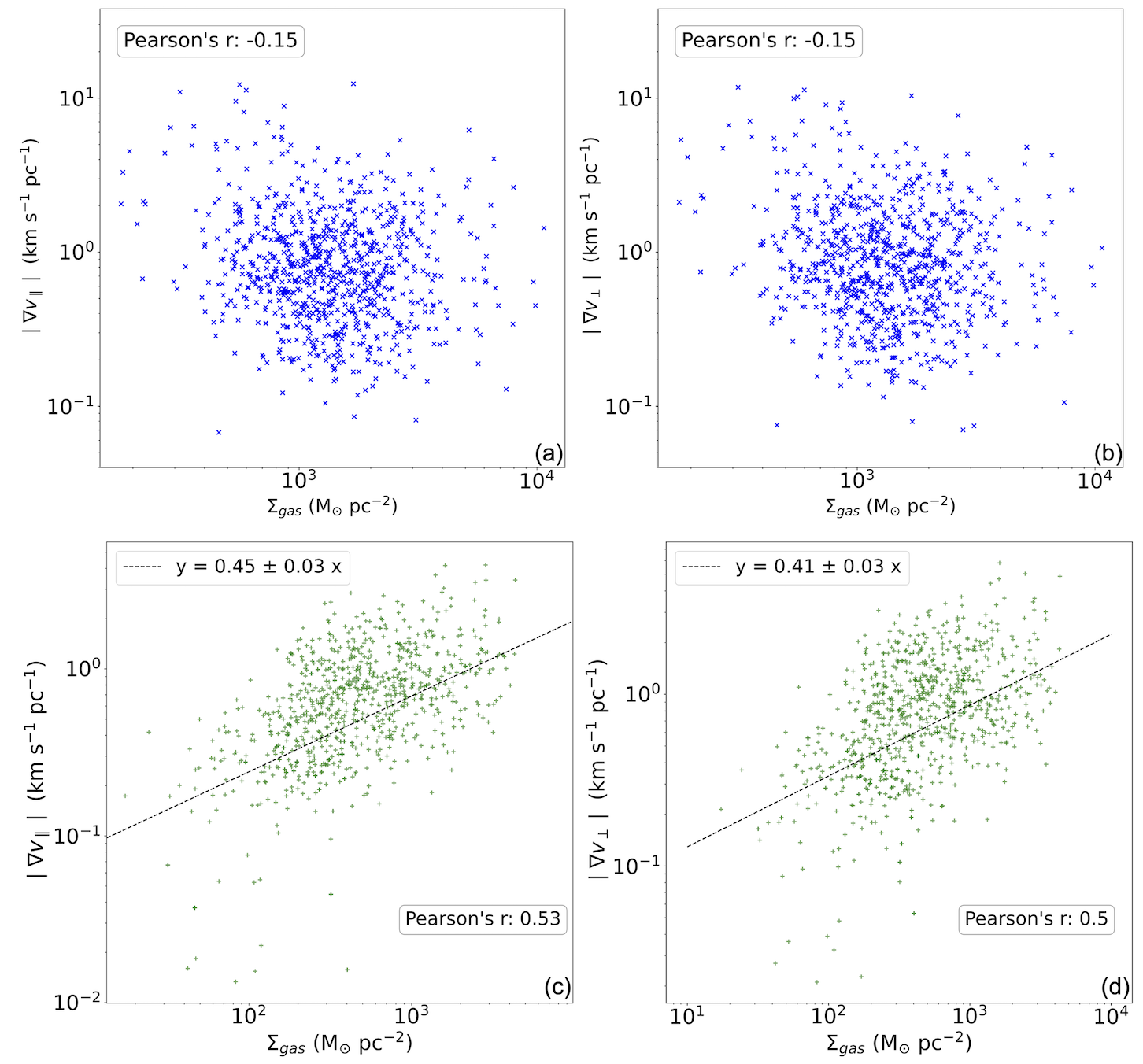}
    
    \caption{The median $\mid \nabla v_{\parallel} \mid$ and $\mid \nabla v_{\perp} \mid$ plotted against surface density. Panels (a) and (b) show the observational data from ATOMS. Panel (c) and (d) show the simulation data. The best-fitting linear regression models for all filaments are shown as black dashed line.}
    \label{fig:gradient_surface}
\end{figure*}

Surface density serves as a proxy for the strength of the gravitational field in molecular clouds, and its relationship with the velocity gradient can provide insight into the interplay between inertia (e.g., turbulence) and gravity. Panels (a) and (b) of Figure~\ref{fig:gradient_surface} show the median magnitudes of $\mid \nabla v_{\parallel} \mid$ and $\mid \nabla v_{\perp} \mid$ as functions of the mean surface density of filaments observed in the ATOMS data. Both quantities yield Pearson correlation coefficients of –0.15, indicating that neither the parallel nor the perpendicular velocity gradients depend significantly on surface density. This suggests that the random local velocity gradients within these supercritical filaments are unlikely to be directly governed by gravity — even though the filaments are gravitationally bound on larger scales. Furthermore, as shown previously in Figure~\ref{fig:cdf}(d), velocity gradients show no significant correlation with gravitational acceleration. If local velocity gradients were dominated by self-gravity, their magnitudes would correlate strongly with gravitational acceleration. However, observations confirm no such correlation, consistent across both the 214 long filaments and the shortest 69 filaments ($<0.1$ pc) . This result further reinforces the conclusion that local velocity gradients are not dominated by self-gravity.

The observed local velocity gradients are likely driven by isotropic turbulence. To test this hypothesis, we compared our observations with numerical simulations of randomly driven, supersonic MHD turbulence \citep{Haugbolle2018} (see Supplementary Materials). Synthetic H$^{13}$CO$^+$ J=1–0 observations were generated from the simulation data at various beam sizes (0.01, 0.02, 0.05, and 0.10 pc). Filaments were then extracted, and their velocity gradients were measured using the same procedure applied to the observational data. The numerical simulation results broadly reproduce the observed phenomena (see Supplementary Materials). As shown in Figure~\ref{fig:dis_vel_gradient}, the distributions of the median velocity gradients from both the ATOMS observational data (Panel a) and the simulation data (Panel b) exhibit notable similarity — both yielding comparable values of $\mid \nabla v_{\parallel} \mid$ and $\mid \nabla v_{\perp} \mid$. This agreement supports the interpretation that gas structure formation on small scales ($\sim$0.1–1 pc) in molecular clouds may be explained by turbulent fragmentation \citep{Haugbolle2018}.

However, we note that in the simulation data, the distribution of $\mid \nabla v_{\perp} \mid$ shows a slight shift toward higher values compared to $\mid \nabla v_{\parallel} \mid$ (Panel b in Figure~\ref{fig:dis_vel_gradient}), a feature that is absent in the observational data. Additionally, as shown in Panels (c) and (d) of Figure~\ref{fig:gradient_surface}, the simulation data exhibit a trend in which velocity gradients increase with surface density, in contrast to the behavior observed in the data. These discrepancies do not appear to be due to gravity, as the gas densities in the simulations are generally lower than those in the observations (Figure~\ref{fig:L_to_M}). Rather, we suggest that turbulence is stronger in the observed massive clumps than in the simulated cloud. The observations target highly active massive star-forming regions, where the velocity dispersion often exceeds the Larson relation \citep{Plume97,Liu16}, whereas the simulations correspond to nearby low-mass star-forming regions with standard Larson normalization — the turbulent forcing in the simulations is calibrated to match typical Larson scaling relations. Furthermore, magnetic fields may play a more dynamically significant role in the simulations during the formation of density structures, as is often observed in nearby clouds \citep{Soler2017}. Overall, the imperfect agreement between simulations and observations calls for further investigation. Future new simulations with varied initial conditions are expected to clarify discrepancies between current simulations and observations.

\begin{figure*}
    \centering
    \includegraphics[width=\textwidth]{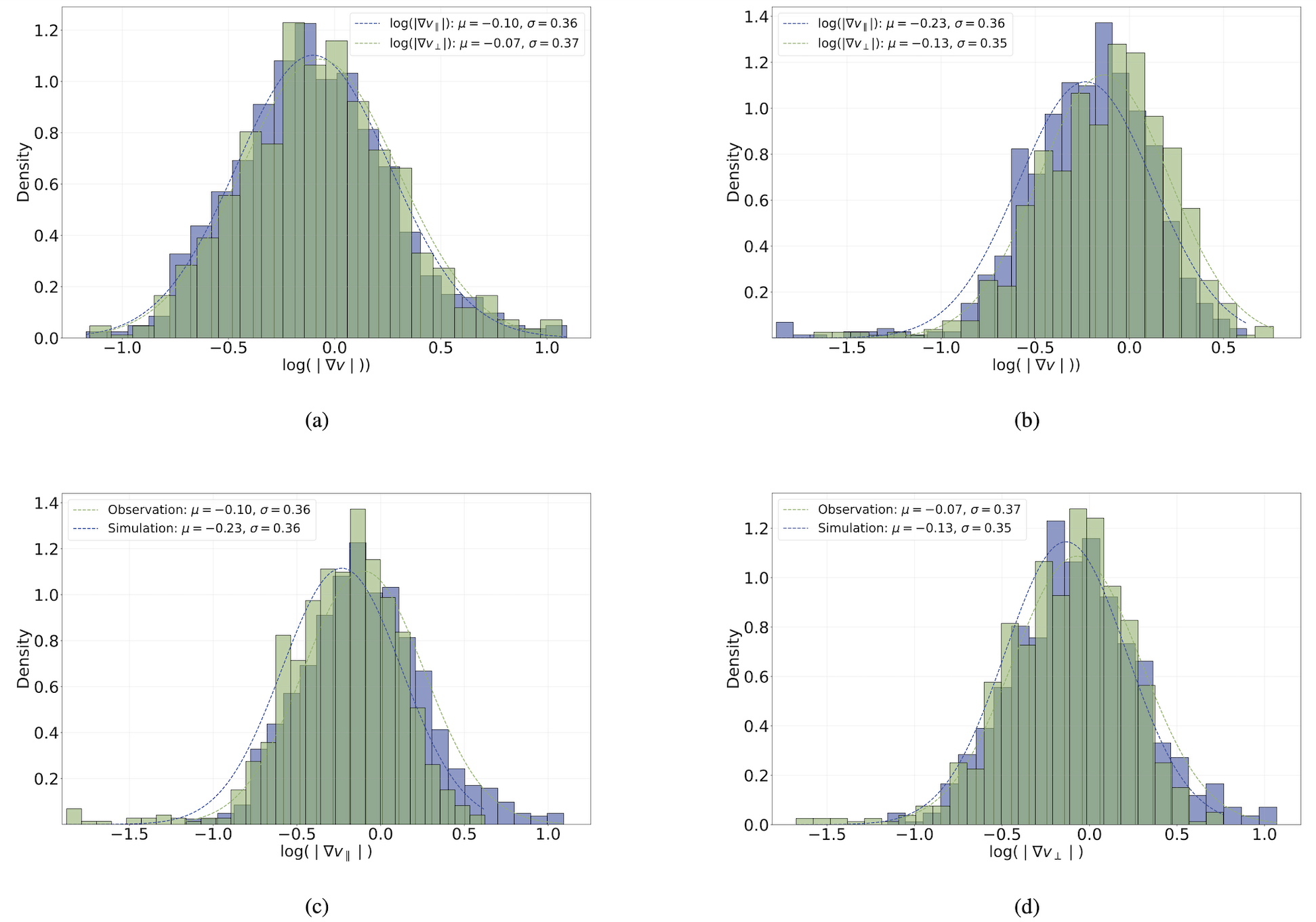}    
    \caption{The distribution of median velocity gradients ($\mid \nabla v_{\parallel} \mid$ and $\mid \nabla v_{\perp} \mid$) in log space. (a) Observational data from ATOMS. (b) Simulation data. (c) The distributions of $\mid \nabla v_{\parallel} \mid$ are compared between observational data (green) and simulation data (blue). (d) The distributions of $\mid \nabla v_{\perp} \mid$ are compared between observational data (green) and simulation data (blue). Dashed lines indicate fitted Gaussian distributions, where $\mu$ and $\sigma$ are the mean and standard deviation, respectively.}
    \label{fig:dis_vel_gradient}
\end{figure*}

\section{Conclusion}

Based on a systematic statistical analysis of the local velocity gradient fields within sub-parsec scale filaments from the large-sample ATOMS survey, this study draws the following main conclusions:

\begin{itemize}
\item{The local velocity gradient components parallel ($\mid\nabla v_{\parallel}\mid$) and perpendicular ($\mid\nabla v_{\perp}\mid$) to the filament skeletons are found to be comparable in magnitude. Their orientations relative to the skeletons are randomly distributed, contrasting sharply with the intensity gradients, which show a strong preferential perpendicular orientation.}

\item{No significant correlation is found between the magnitudes of the local velocity gradients and the filament surface density, nor are the velocity gradients significantly correlated with intensity gradients at the pixel level. This indicates that the local velocity gradients are unlikely gravity-dominated, even within these globally bound structures.}

\item{The isotropic nature of the velocity gradients and their overall similarity to results from numerical simulations of randomly driven turbulence support the interpretation that random turbulence is likely the primary mechanism shaping the velocity fields on small scales ($\sim$0.1--1 pc).}

\item{In conclusion, the results demonstrate that gas motions on sub-parsec scales within formed supercritical filaments in massive clumps could be still predominantly governed by random turbulence, providing new observational constraints for understanding the interplay between gravity and turbulence in the early stages of star formation. This scenario needs to be further tested in future studies.}

\end{itemize}

\section*{Acknowledgements}
This work has been supported by the National Key R\&D Program of China No.\ 2022YFA1603100 and National Science and Technology Major Project 2024ZD1100601. Z.C.\ acknowledges support from the National Natural Science Foundation of China (NSFC), through grants No.\ 12403028, the Basic Research Program of Shanxi Provence (202403021222272). T.L.\ acknowledges support from the National Natural Science Foundation of China (NSFC), through grants No.\ 12073061 and No.\ 12122307, the Tianchi Talent Program of Xinjiang Uygur Autonomous Region, and the international partnership program of the Chinese Academy of Sciences, through grant No.\ 114231KYSB20200009.  MJ acknowledges the support of the Research Council of Finland Grant No. 348342. H.-L. Liu is supported by Yunnan Fundamental Research Project (grant No.\,202301AT070118, 202401AS070121), and by Xingdian Talent Support Plan--Youth Project. G.G. acknowledges support from the ANID BASAL project FB210003. P.P. acknowledges support from the US National Science Foundation under Grant AST 2408023. This research was carried out in part at the Jet Propulsion Laboratory, which is operated by the California Institute of Technology under a contract with the National Aeronautics and Space Administration (80NM0018D0004) DL acknowledges support from NSFC 12588202 and the New Cornerstone Foundation.

\renewcommand{\thesection}{\Alph{section}}
\renewcommand{\thesubsection}{\Alph{section}\arabic{subsection}}
\renewcommand{\thesubsubsection}{\Alph{section}\arabic{subsection}\alph{subsubsection}}
\renewcommand{\thefigure}{\thesection\arabic{figure}}

\setcounter{section}{0}
\section{Supplementary Materials}

\subsection{\label{sec:level1} ALMA observations}
The ATOMS survey, standing for ALMA Three-millimeter Observations of Massive Star-forming regions, has observed 146 active Galactic star-forming regions at band 3 (3 mm) using ALMA \citep{Liu20_1}. These 146 sources were selected from the CS (2-1) survey of UC H{\sc ii} region candidates with bright CS emission $T_{\rm b}$ $>$ 2 K \citep{Bronfman96}, a proxy for moderately high gas density. Most ATOMS sources are gravitationally bound, exhibiting a virial parameter below 2. There are 139 targets located in the first and fourth Galactic Quadrants of the inner Galactic Plane. The rest are distributed in the outer Galaxy. The distances of the sample clumps range from 0.4 kpc to 13.0 kpc with a mean value of 4.5 kpc. The sample includes 27 distant ($d$ $>$7 \,kpc) sources that are either close to the Galactic Center like SgrB2(M) or mini-starbursts like W49A, thus providing extreme environments for star formation studies. This study focuses on the structures and gas kinematics of filaments within these clumps. 

The ALMA observations of the ATOMS survey were conducted from late September to mid November 2019, utilizing both the Atacama Compact Array (ACA; Morita Array, 7-m antennas) and the 12-m array (configured as C43-2 or C43-3) in band 3 (Project ID: 2019.1.00685.S; PI: Tie Liu). The ALMA data were calibrated and imaged using the CASA software \citep{McMullin07}. The 12-m and ACA 7-m array data were jointly imaged using Briggs weighting, setting the robust parameter to 0.5 in the CASA tclean task, for both continuum images and line cubes. In this work, we use H$^{13}$CO$^+$ J=1$-$0 (86.754288 GHz) line data with a spectral resolution of $\sim$0.1 km s$^{-1}$. The typical beam FWHM size and channel rms noise level for H$^{13}$CO$^+$ J=1$-$0 line emission are $\sim$ 2.5$^{\prime\prime}$ and 8 mJy beam$^{-1}$, respectively. The typical maximum recoverable angular scale (MRS) in this survey is about 1 arcmin, comparable to the field of view (FOV) of the 12-m array observations \citep{Liu20_1,Liu2021}. The H$^{13}$CO$^+$ J=1–0 data obtained from the ATOMS Survey are generally optically thin in dense cloud cores \citep{Zhou2022, Zhang2025}. Additionally, the emission of H$^{13}$CO$^+$ J=1$-$0 is predominantly confined to regions smaller than MRS, ensuring minimal missing flux in spectral line profile analyses \citep{Zhou2022,Zhang2025}. 

H$^{13}$CO$^+$ J=1$-$0 has a rather high critical density of 6.2×10$^4$ cm$^{-3}$ at 10 K and an effective excitation density of 3.9×10$^4$ cm$^{-3}$ \citep{Bergin2007,Shirley2015}. Its optical depth is much lower than its main line counterpart HCO$^+$ J=1$-$0, thereby serving as a reliable tracer of dense gas. Moreover, the spatial distribution of the H$^{13}$CO$^+$ J=1$-$0 emission is tightly correlated with the column density of the dense gas revealed by dust emission as seen in Herschel data toward nearby clouds \citep{Shimajiri2017}. In particular, the H$^{13}$CO$^+$ J=1$-$0 emission traces the dense “supercritical” filaments detected by Herschel very well in nearby clouds \citep{Shimajiri2017}. The virial mass estimates derived from the velocity dispersion of H$^{13}$CO$^+$ J=1$-$0 also agree well with the dense gas mass estimates derived from Herschel data for the same sub-regions \citep{Shimajiri2017}. Filaments within the ATOMS sources have been identified from the integrated intensity maps of H$^{13}$CO$^+$ J=1-0 molecular line, and global velocity gradients along the longest filaments have been measured in \cite{Zhou2022}. Building upon that, this work focuses on measuring localized velocity gradients within velocity-coherent filaments (directly extracted in PPV space via H$^{13}$CO$^+$ J=1$-$0 data). 

\subsection{Extraction of filaments}\label{secA1}

Many methods have been developed to identify filamentary structures, including CRISPy, DisPerSE, FilFinder, the local Hessian matrix, getFilaments, and wavelet transform-based algorithms \citep{Sousbie2011,Sousbie2011MNRAS.414..384S,Koch2015,Schisano2014,Chen2020}. The key advantage of wavelet transforms lies in their ability to enhance contrast (particularly for elongated features). FilFinder can uniformly extract hierarchical filamentary structures and performs well even when the image has large intensity variations. Although these methods have been proven to be effective and widely implemented, they inherently lack the capability to verify whether outputs represent single, continuous filamentary structures in three dimensions. The DisPerSE algorithm \citep{Sousbie2011,Sousbie2011MNRAS.414..384S}, while it has recently been used in star formation studies \citep{Arzoumanian2011} and can also run in three dimensions \citep{smith2016}, necessitates two additional user-defined parameters compared to CRISPy \citep{Chen2020}. Therefore, we employ CRISPy for three-dimensional filamentary structure extraction in this study. 

CRISPy has two key parameters: a density threshold and a smoothing bandwidth. The density threshold eliminates noise in the density field, whereas the smoothing bandwidth regulates kernel density estimation for particle-like data distributions. We adopted a 5-$\sigma$ intensity threshold to capture the majority of the emission in the model while avoiding regions near the typical rms noise level of the data. A smoothing length of one pixel was used, corresponding to approximately one-fourth of the sampling width of our data—equivalent to 8 pixels across the FWHM of the beam. This minimal smoothing preserves information better for our 200$\times$200 pixel data. The algorithm identifies emission ridges in PPV space and maps them back to the original grid as "skeletons".

Filamentary structures are identified through their coherent spatial and kinematic properties within the PPV space (see Figure 1). The initial skeleton extraction was subsequently refined through a two-step cleaning process: (1) elimination of short filaments with lengths below three beam-resolved scales (where one scale corresponds to 8 pixels) to remove insignificant substructures, and (2) pruning of minor branches shorter than one beam-resolved scale while retaining more substantial branches exceeding this threshold. Notably, several of these preserved elongated branches exhibit characteristics consistent with hub-filament systems, which are defined as nodes where three or more distinct filaments intersect.

We identified a total of 837 velocity-coherent filamentary structures among the 147 sources in H$^{13}$CO$^+$ data. Among them, 214 filaments have an aspect ratio (length-to-width ratio) greater than 5.

\subsection{Measuring velocity gradients and intensity gradients}\label{subsubsec3.1}

\begin{figure*}
    \centering
    \includegraphics[width=\linewidth]{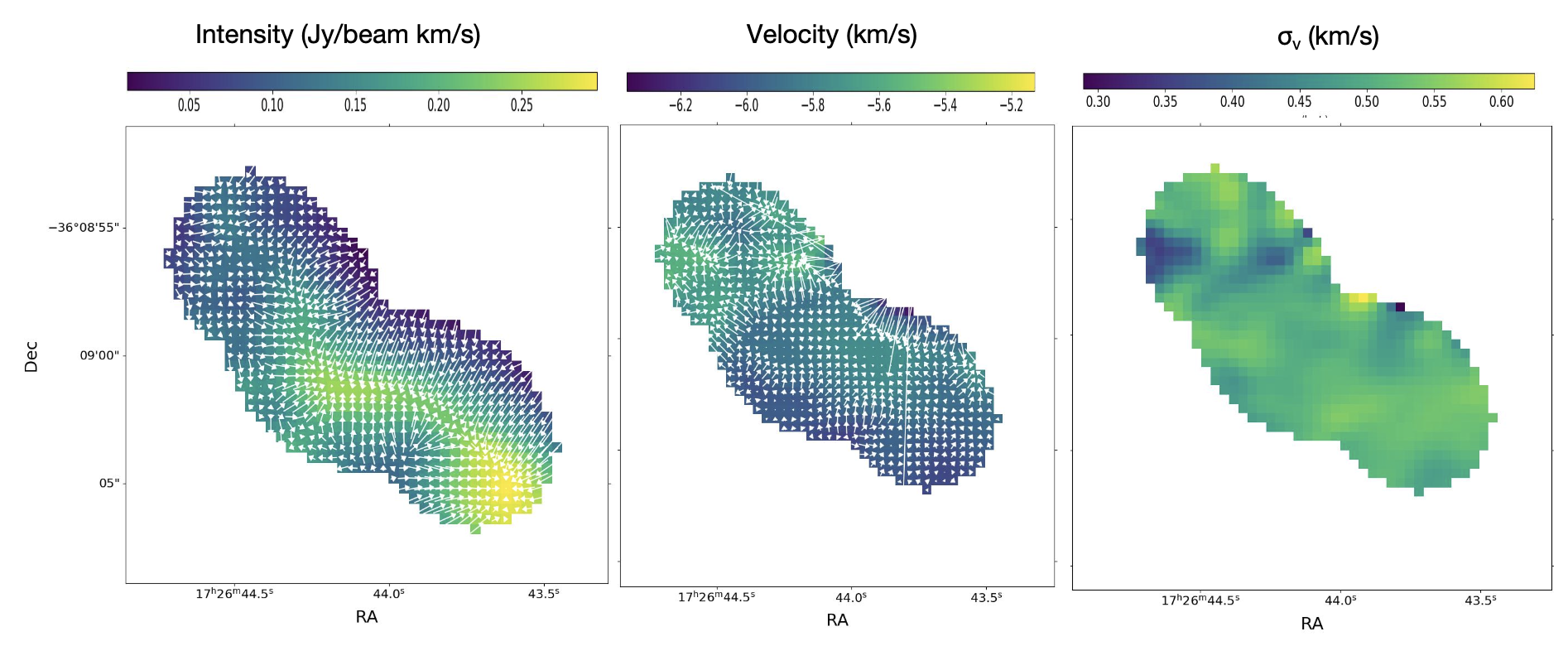}
    \caption{Panels (a), (b), and (c) display the 0th moment, 1st moment, and $\sigma_{v}$ maps, respectively, for a representative filament skeleton in the source I17233-3606. The white arrows represent the gradient direction and magnitude at each pixel position.}
    \label{fig:intensity_velocity_sigma}
\end{figure*}

After filament identification, we projected these filaments onto the 2D position-position space, and then measured their velocity gradients and intensity gradients. The velocity range of the filaments was defined as spanning from the minimum to the maximum velocities of the filament skeleton, extended by $\pm$3 times the spectral resolution ($\sim$0.1 km/s). The integrated intensity, intensity-weighted velocity, and velocity dispersion maps were derived using the zeroth, first, and second moments, respectively. An example of the 0th moment, 1st moment, and $\sigma_{v}$ maps for an exemplary filament skeleton is shown in Figure~\ref{fig:intensity_velocity_sigma}. 

The intensity gradient ($\nabla I$) and velocity gradient ($\nabla v$) are computed from the zeroth and first moment maps. These gradients were calculated pixel by pixel through plane fitting within circular apertures (16-pixel diameter, $\sim$2 beam widths) centered on each pixel. The aperture diameter is deliberately chosen to be twice the beam size to ensure that gradients are calculated over resolved structures. The velocity and intensity gradients of a representative filament are illustrated as white arrows in Figure~\ref{fig:intensity_velocity_sigma}.

\begin{figure*}
    \centering
    \centering
    \includegraphics[width=\textwidth]{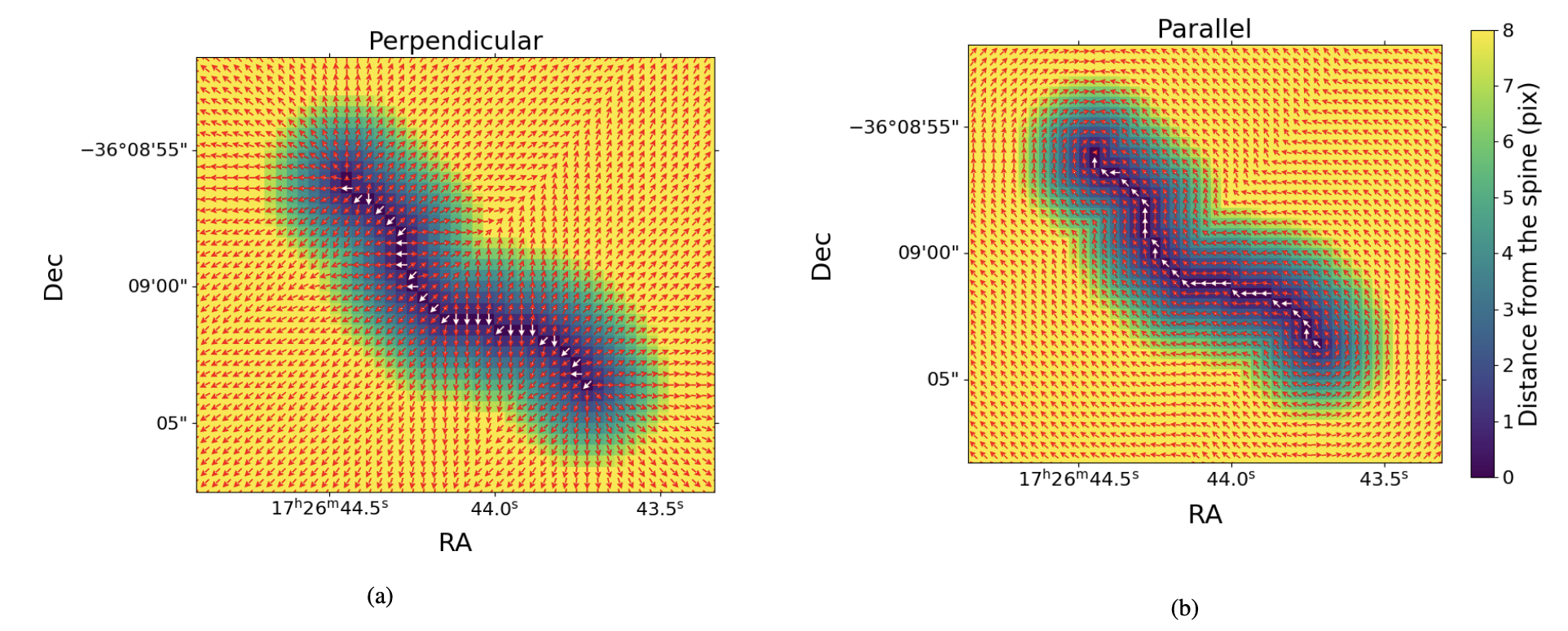}

    \caption{
    Panels (a) and (b) show vector fields that are perpendicular and parallel to the same filament skeleton as shown in Figure~\ref{fig:intensity_velocity_sigma}, respectively. The filament skeleton and the distance between each pixel and the skeleton, from which the vector fields are derived, are shown in the background.  }
    \label{fig:distance-transform}
\end{figure*}

\begin{figure*}
    \centering
    \includegraphics[width=\linewidth]{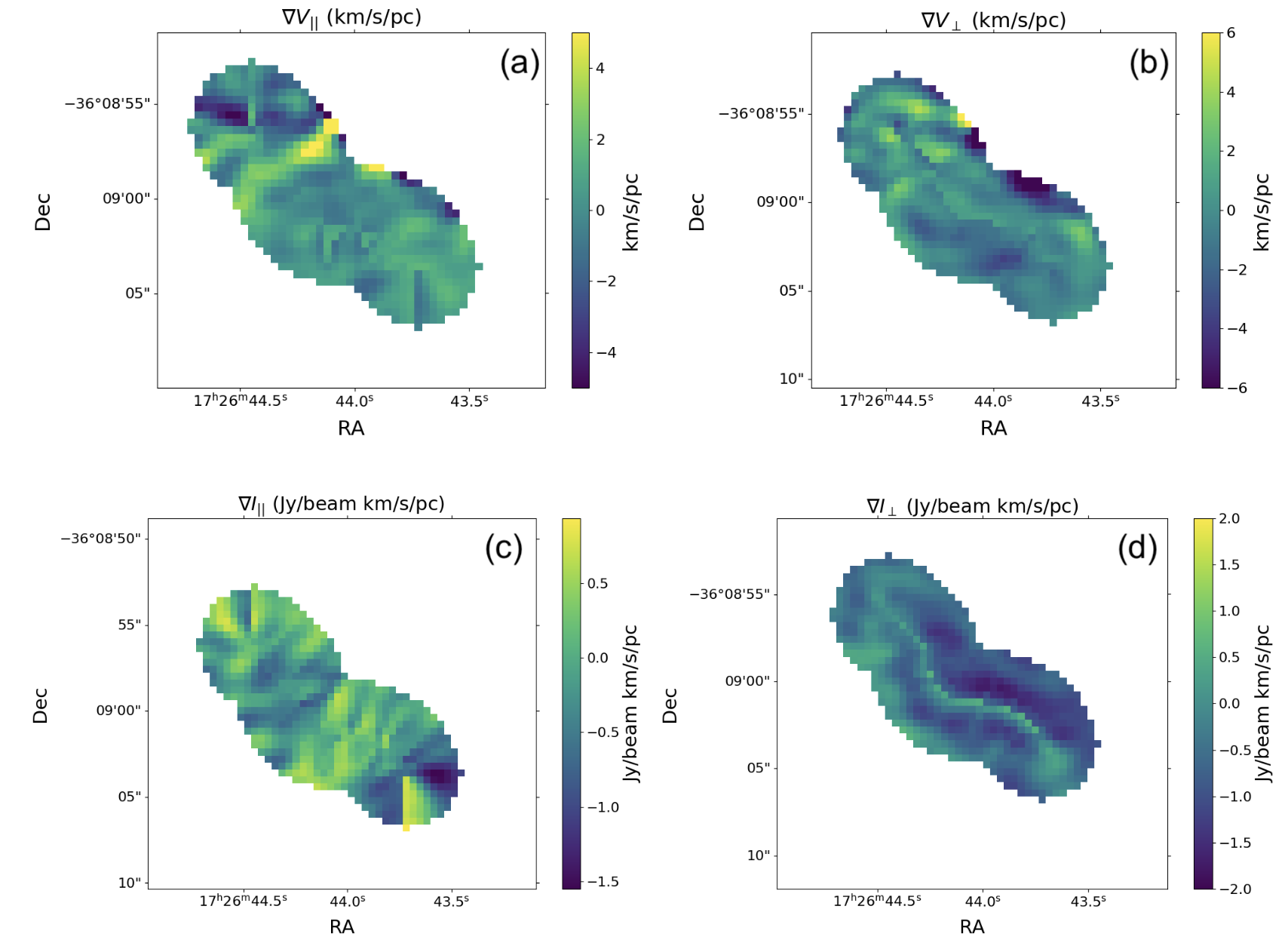}
    \caption{Panels (a) and (b) show the $\nabla v_{\parallel}$ and $\nabla v_{\perp}$ maps of the same filament skeleton as shown in Figure~\ref{fig:intensity_velocity_sigma}, respectively. Panels (c) and (d) show the $\nabla I_{\parallel}$ and $\nabla I_{\perp}$, respectively. }
    \label{fig:int-vel-gradient}
\end{figure*}

\begin{figure*}
    \centering
    \includegraphics[width=\linewidth]{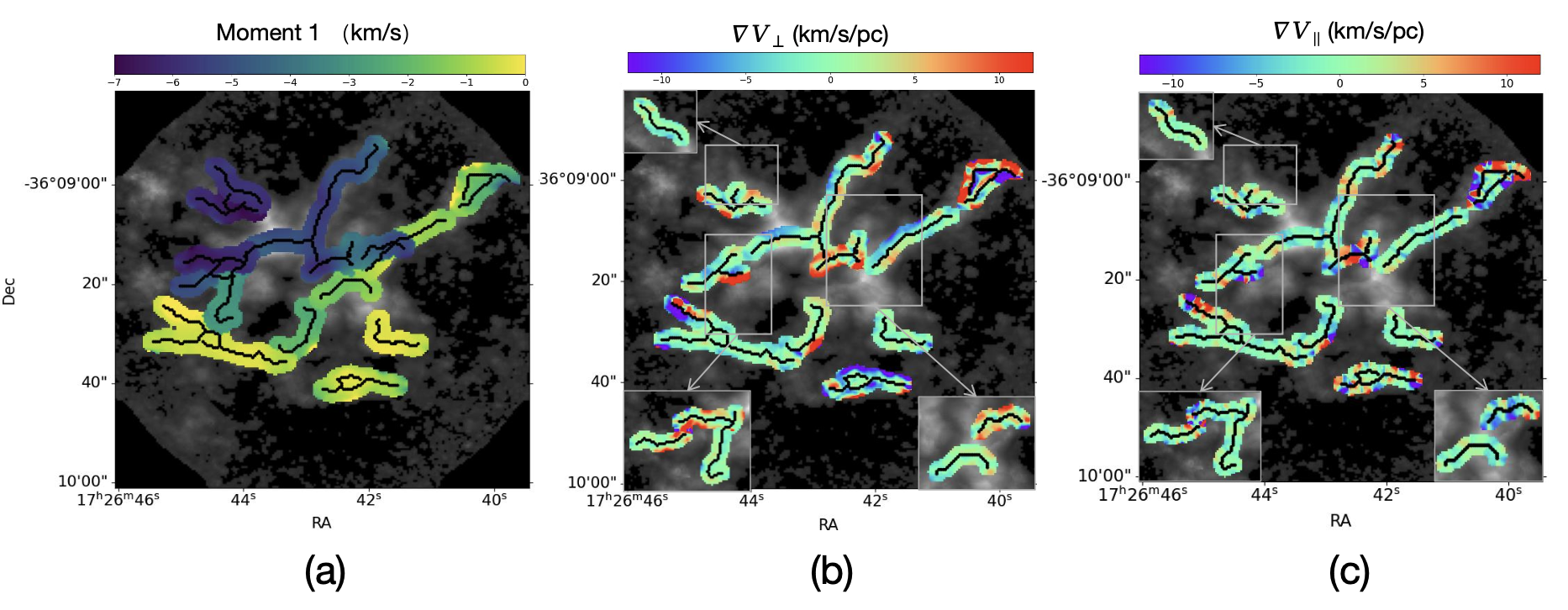}
    \caption{(a) projections of filament skeletons identified in I17233-3606, overlaid on top of the moment 1 maps of selected velocity-coherent filaments (color).  (b) and (c): spatial distribution of the perpendicular and parallel components of $\nabla v$, respectively, relative to their filament skeletons. The call-out boxes in these panels show the same $\nabla v$ components of the additional, overlapping velocity-coherent filaments in the sky. }
    \label{fig:whole-fil}
\end{figure*}

To geometrically analyze gas motions relative to the filament skeleton, we employ the vector field decomposition technique proposed by ref.\cite{Chen2020}. This technique decomposes the velocity gradient field into two orthogonal components: one parallel and one perpendicular to the filament skeleton (Figure~\ref{fig:distance-transform}). The decomposition is achieved by computing dot products between $\nabla v$ and two characteristic vector fields: one constructed from the gradient of pixel distances to the skeleton (perpendicular component), and the other obtained by rotating the former by 90$^\circ$ (parallel component). Figure~\ref{fig:int-vel-gradient} displays the $\nabla v_{\perp}$ and $\nabla v_{\parallel}$, as well as the $\nabla I_{\perp}$ and $\nabla I_{\parallel}$ maps for the same filament. Figure~\ref{fig:whole-fil} displays the velocity and the perpendicular and parallel components of the $\nabla v$ of the filament network in an exemplary source I17233-3606. Images for the remaining sources are available in the supplementary material.

\subsection{Properties of the filaments}

We estimated the total gas masses of the velocity-coherent filaments from H$^{13}$CO$^+$ J=1-0 line emission following a standard procedure \citep{Sanhueza2012,Xu2023}. Under the assumptions of optically thin emission and a unity filling factor, the column density of H$^{13}$CO$^+$ at a pixel $(i,j)$ was calculated using the formulation from ref. \cite{Sanhueza2021}. The calculation is based on the equation:
\begin{align*}
N_{ij} &= \frac{3 k_{B}}{8 \pi^{3} B_{\mathrm{rot}} \mu_{\mathrm{dm}}^{2}} 
         \cdot \frac{T_{\mathrm{ex}}+h B_{\mathrm{rot}} / 3 k_{\mathrm{B}}}{J+1} 
         \cdot \frac{\exp \left(E_{J} / k_{\mathrm{B}}\right)}
              {1-\exp \left(-h \nu / k_{\mathrm{B}} T_{\mathrm{ex}}\right)} \\
       &\quad \times \frac{1}{\left[J\left(T_{\mathrm{ex}}\right)-J\left(T_{\mathrm{bg}}\right)\right]} 
         \int T_{b,ij} \, d\nu ,
\end{align*}

where $k_{\mathrm{B}}$ is the Boltzmann constant, $h$ is the Planck constant, and $T_{\mathrm{ex}}$ is the excitation temperature for which we use the dust temperature as listed in ref.\cite{Liu20_1}. Here, $\nu$ represents the transition frequency (86.754288 GHz), $\mu_{\mathrm{dm}}$ is the permanent dipole moment of the molecule (3.89 Debye), $J$ denotes the rotational quantum number of the lower state, $E_{J}=h B_{\mathrm{rot}} J(J+1)$ is the energy in the level $J$, $B_{\mathrm{rot}}$ is the rotational constant of the molecule (43.377302 GHz), $T_{\mathrm{bg}}$ is the background brightness temperature (2.73 K), and $T_{b,ij}$ is the brightness temperature at Pixel $(i,j)$. 
$J(T)$ is defined as 
\begin{equation}
J(T)=\frac{h \nu}{k_{\mathrm{B}}} \frac{1}{e^{h \nu / k_{\mathrm{B}} T}-1} .
\end{equation} 
The column density is then converted into mass using
\begin{equation}
M_{\text {filament }}=\left(X_{\mathrm{H}^{13} \mathrm{CO}^{+}}\right)^{-1} \mu_{H_2}~m_{\mathrm{H}} A D^{2} \times \sum_{i,j}^{filament} N_{ij}\left(\mathrm{H}^{13} \mathrm{CO}^{+}\right),
\end{equation} 
where $A$ is the angular area of a pixel (e.g. 0.4$^{\prime\prime}$$\times$0.4$^{\prime\prime}$) and $D$ is the distance that is listed in \cite{Liu20_1}. $X_{\mathrm{H}^{13} \mathrm{CO}^{+}}=\left[\mathrm{H}^{13} \mathrm{CO}^{+}] /[ \mathrm{H}_{2}\right]$ is the H$^{13}$CO$^+$ to molecular hydrogen abundance ratio. The mass of a hydrogen atom is denoted as $m_{\mathrm{H}}$. The molecular weight per hydrogen molecule is given as $\mu_{H_2}$ = 2.8\citep{Kauffmann2008}. Ref.\cite{Hoq2013} analyzed MALT90 data from 333 high-mass star-forming regions and reported an abundance of 1.28 $\times$ 10$^{-10}$. Given that our sources are also high-mass star-forming regions, we initially adopted this value. However, $X_{\mathrm{H}^{13}\mathrm{CO}^{+}}$ is the most uncertain factor in the mass calculation. Ref. \cite{Liu2020ApJ} derived a clump-averaged H$^{13}$CO$^+$ abundance of 9 $\times$ 10$^{-12}$ from APEX observations of G34.43+00.24, which is one order of magnitude lower than the previously used value.  
Given the significant impact of this order-of-magnitude difference on mass calculations, and considering that the value from \cite{Hoq2013} is derived from a large sample (333 regions) that better matches our dataset, we treat the value from \cite{Liu2020ApJ} (9 $\times$ 10$^{-12}$) as a potential source of uncertainty and investigate its impact. Accordingly, we recalculated the masses using $X_{\mathrm{H}^{13}\mathrm{CO}^{+}}$ = 9 $\times$ 10$^{-12}$ and marked the corresponding results with gold pentagrams in Figure 2. However, both the highest (1.28 $\times$ 10$^{-10}$) and lowest (9 $\times$ 10$^{-12}$) abundance values yield filament masses that exceed the critical line mass (see below). To calculate the line mass $m$ (mass per unit length, $m=M/L$) of each filament, we further divided each filament’s mass by its skeleton length. 

\begin{figure}
\centering
    \includegraphics[width=\textwidth]{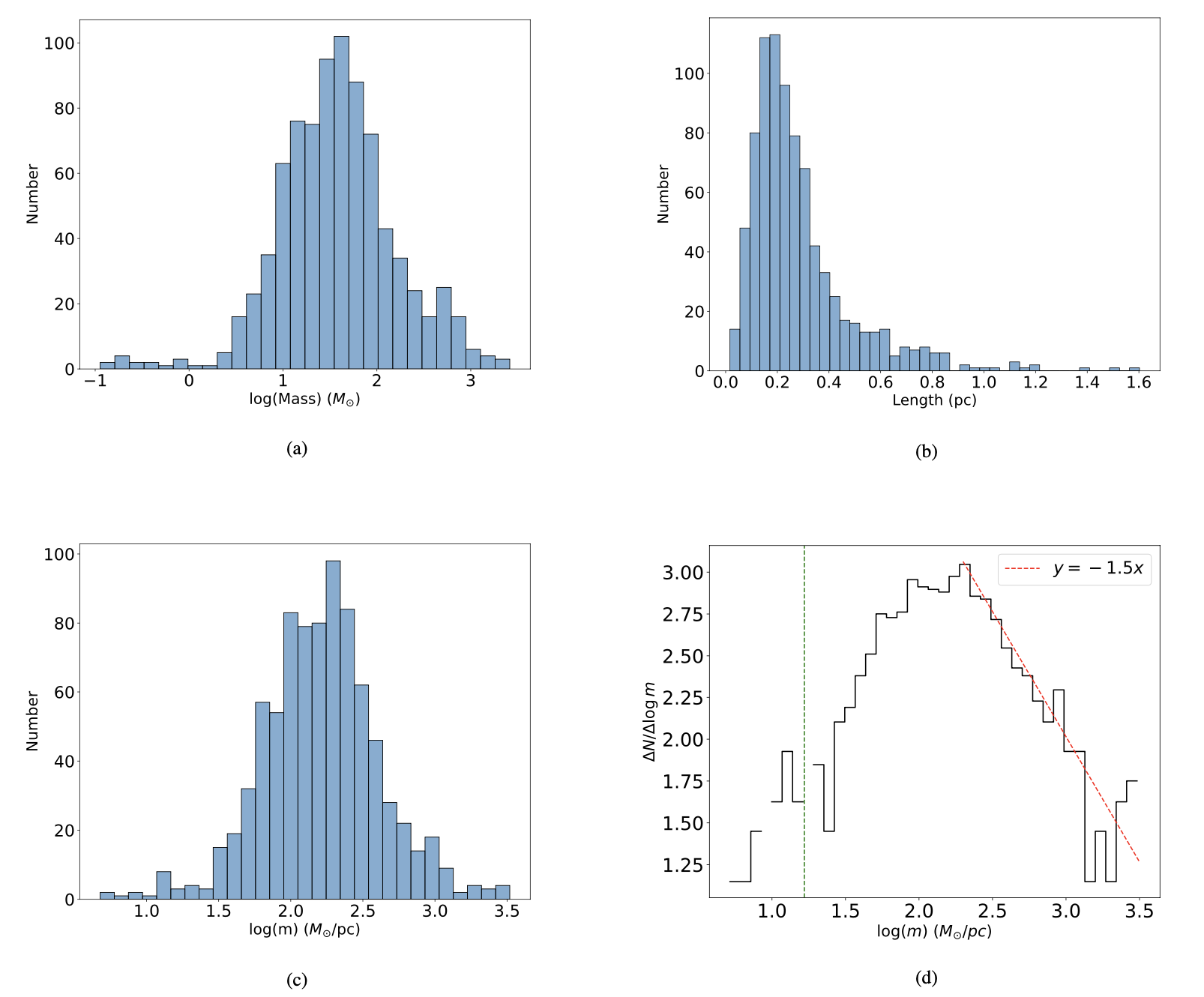}    
    \caption{The distribution of mass (a), length (b) and line mass (c). Panel (d) represents the filament line mass function. The value of $m_\mathrm{crit}$ at 10 K is marked by the green dashed line. The best-fitting linear regression model is shown as the red dashed line.}
    \label{fig:dis_mass_length}
\end{figure}

The derived physical parameters of these filamentary structures (mass, length, and line mass $m$) are presented in Figure~\ref{fig:dis_mass_length} (panels a to c), providing a comprehensive statistical overview of their properties. Specifically, the mass spans 0.1-2.6$\times$10$^{3}$ $M_{\odot}$(mean = 106 $M_{\odot}$, median = 37 $M_{\odot}$), the length ranges from 0.02 to 1.6 pc (mean = 0.29 pc, median = 0.23 pc), and the line mass $m$ varies from 4.8 to 3.3$\times$10$^{3}$ $M_{\odot}$ pc$^{-1}$ (mean = 256 $M_{\odot}$ pc$^{-1}$, median = 161 $M_{\odot}$ pc$^{-1}$). In summary, this diverse sample of filamentary structures is well-suited for statistical studies of sub-parsec-scale gas formation and evolution.

Ref.\cite{Hacar23} provides an overview of the mass and length of various filamentary structures. The molecular filaments exhibit a continuous distribution in terms of mass and length, encompassing all nearby filaments and fibers and extending to the longest structures in the Galactic Plane and the Giant Filaments. The approximate scaling relation follows $L \propto M^{0.5}$. Figure 2 illustrates the mass--length relationship of our data. The majority of filaments detected at scales $L \gtrsim 0.1$ pc exhibit high line mass $m$ (up to $\gtrsim$ 10 $M_{\odot}\,\mathrm{pc}^{-1}$. We fitted the L-M relation and found that $L \propto M^{0.4\pm 0.01}$. This shallower slope compared to Ref. \cite{Hacar23} may arise because not all mass is assigned to smaller-scale filaments (e.g., due to density-dependent observational tracers), leading to a reduced slope \cite{Hacar23}. Other studies also indicate that the HFS system may exhibit some deviations from the $L \propto M^{0.5}$ relation \cite{Shen2024}.

The filament line mass distribution function presented in Figure~\ref{fig:dis_mass_length}(d) exhibits a well-defined power-law behavior at the high-mass end, characterized by $\Delta N/\Delta \log M_{\rm line} \propto M_{\rm line}^{-1.5 \pm 0.1}$. This result agrees well with the pioneering work of \cite{Andre2019}, who reported a similar power-law index of -1.6 for supercritical-mass filaments. The preferred theoretical model that explains the remarkably robust power-law slope at the high-mass end \citep{Salpeter1955} is the "gravo-turbulent fragmentation" scenario \citep{Hennebelle2008, Padoan2002}, where supersonic turbulence naturally produces the power-law tail.

\subsection{The critical line mass}\label{sce4.1}

Line mass serves as an indicator of filament growth, while critical line mass ($m_\mathrm{crit}$) provides a measure of filament stability \citep{Chen2020}. Filaments with $m$ $>$ $m_\mathrm{crit}$ (supercritical) become radially unstable and inevitably collapse under self-gravity, whereas those with $m$ $<$ $m_\mathrm{crit}$ (subcritical) can maintain hydrostatic equilibrium. This critical line mass plays a similar role to the isothermal Jeans mass in early studies of molecular clouds \citep{Klessen2000}. It is based on several assumptions: hydrostatic equilibrium, isolation, and isothermality. The calculation of $m_\mathrm{crit}$ is derived from ref.\cite{Stod1963,Ostriker1964}:
\begin{equation}
m_\mathrm{crit}(T)=\frac{2c_\mathrm{s}^2}{G}\sim16.6\left(\frac{T}{10\mathrm{~K}}\right)M_\odot\mathrm{pc}^{-1}.
\end{equation}
Here, the isothermal sound speed is calculated as $c_s=(k_\mathrm{b}T/\mu_p\mathrm{m_H})^{1/2}$, where $T$ is the isothermal gas temperature, and G, $k_b$, $\mu_p$, and $m_H$ represent the gravitational constant, Boltzmann constant, mean interstellar molecular weight, and hydrogen atomic mass, respectively. Assuming a mean molecular weight per free particle of $\mu_p$ = 2.33 \citep{Kauffmann2008}, a 10 K \citep{Chen24,Rosolowsky2008,Friesen2017} filament would have $m_{crit}$ = 16.6 $M_\odot~\mathrm{pc}^{-1}$.

Observations have shown that core formation likely occurs over a broader range of $m$ values \citep{Arzoumanian13, Arzoumanian2019, Konyves2015, Konyves2020}. For instance, the empirical core formation efficiency rises sharply within a factor of 2 of $m_\mathrm{crit}$ (8–33 $M_{\odot}$pc$^{-1}$) before plateauing near 2$m_\mathrm{crit}$ \citep{Konyves2020}. Furthermore, cores in models can form in subcritical filaments with $m >$ 0.5 $m_\mathrm{crit}$ due to compressive instabilities. Accordingly, following the convention of ref. \cite{Chen24}, we define a thermally transcritical regime as 0.5$m_\mathrm{crit}$$\leq$ $m$ $\leq$2$m_\mathrm{crit}$. By this definition, 28 (3.3\%) and 805 (96.2\%) of the velocity-coherent filaments in our sample are trans- and supercritical, respectively, assuming a gas temperature of 10 K. Four (0.5\%) filaments are subcritical.

As shown in Figure 2 , the masses of the filaments we identified in star-forming regions mostly significantly exceed the expected $m_\mathrm{crit}$, indicating that they are thermally supercritical. This implies that they are likely undergoing contraction or fragmentation due to insufficient gravitational support.

\subsection{Turbulent support of filaments}

Turbulent motions within filaments can also provide supports against gravity. The effective sound speed ($C_{\rm s,eff}$) including turbulent support is,
\begin{align*}
C_{s, \mathrm{eff}}=\left[\left(C_{s}\right)^{2}+\left(\sigma_{\mathrm{NT}}\right)^{2}\right]^{1 / 2},
\end{align*}
where $C_{s}$ is the thermal sound speed and $\sigma_{\mathrm{NT}}$ is the non-thermal one-dimensional velocity dispersion. For $T$ = 10 K, the sound speed is $0.19 \mathrm{~km} \mathrm{~s}^{-1}$. The non-thermal one dimensional velocity dispersion  $\sigma_{\mathrm{NT}}$ can be calculated as follows:
\begin{align*}
\sigma_{\mathrm{NT}}=\sqrt{\sigma_{ H^{13}CO^{+}}^{2}-\frac{k T}{m_{H^{13}CO^{+}}}}
\end{align*}

The variation of $C_{\rm s,eff}$ with respect to the line mass $m$ has been investigated by ref.\cite{Arzoumanian13}, who found that $C_{\rm s,eff} \propto m^{0.36}$. Ref. \cite{Hacar23} reported a slightly steeper scaling ($C_{\rm s,eff} \propto m^{0.5}$) for filaments with $m >$ 100 $M_{\odot}$pc$^{-1}$. Ref.\cite{Arzoumanian13} interpreted the observed correlation between $C_{\rm s,eff}$ and $m$ as a consequence of accretion/contraction-driven turbulence, where turbulence increases as the filament grows through accretion. 

\begin{figure*}
    \centering
    \includegraphics[width=\linewidth]{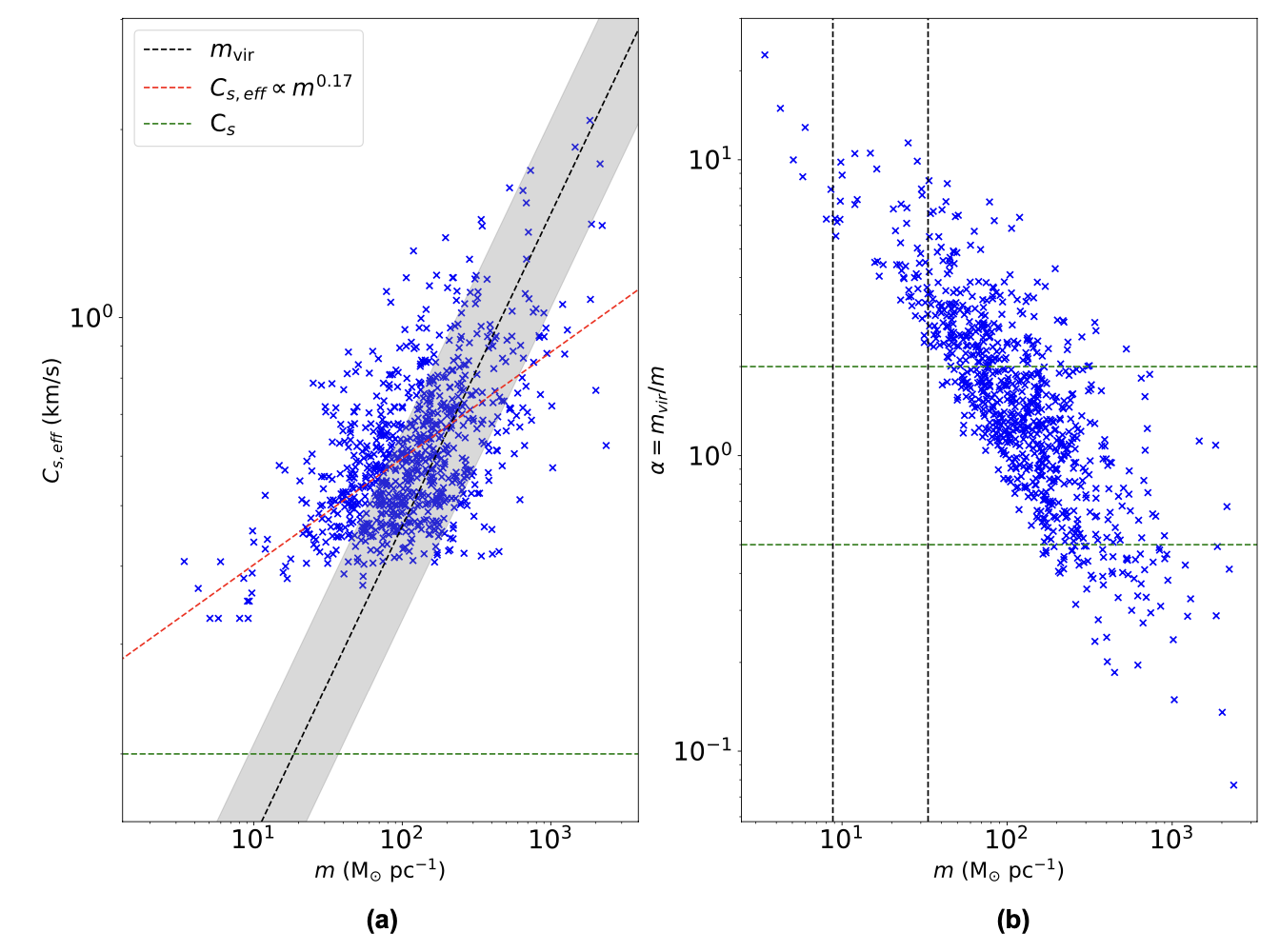}
    \caption{(a) The $C_{\rm s,eff}$ of each filament plotted against their respective $m$. The value of $m_\mathrm{crit}(C_{\rm s,eff})$ is marked by the black dashed line, with the shaded region representing 0.5 $m_\mathrm{crit}(C_{\rm s,eff})$ $\leq$ $m$ $\leq$ 2 $m_\mathrm{crit}(C_{\rm s,eff})$. The red dashed line represents $C_{\rm s,eff} \propto m^{0.17}$. The horizontal green dotted line shows the $c_{\rm s}$ value of gas at 10 K with a nonthermal component that is sonic. (b) The virial parameter the intrinsic mass per unit length. The black dashed lines represent 0.5$m_{crit}$ and 2$m_{crit}$, respectively. The green lines represent $\alpha=0.5$ and $\alpha=2$.}
    \label{fig:mvir_sigmav}
\end{figure*}

Figure~\ref{fig:mvir_sigmav} (a) illustrates the median value $C_{\rm s,eff}$ of each filament plotted against their respective $m$. Our data yield a Pearson’s correlation coefficient (Pearson’s $r$) of 0.55, suggesting a weak linear correlation between $C_{\rm s,eff}$ and $m$. We observed that $C_{\rm s,eff} \propto m^{0.17}$, with exponent smaller than the values reported in the aforementioned studies. This discrepancy likely arises because, at smaller scales, factors such as stellar feedback can significantly influence turbulent motions. This discrepancy likely stems from differences in the samples: the study by \cite{Chen24} had a much smaller sample size and a narrower line mass range (concentrated around 10–100 $M_{\odot}$ pc$^{-1}$). Notably, Figure~\ref{fig:mvir_sigmav} (a) reveals an upward trend across the 10-100 $M_{\odot}$pc$^{-1}$ range. 

If the contribution of non-thermal motions to filament stability is considered, $c_s$ can be replaced with $C_{s, \mathrm{eff}}$, yielding the turbulent critical line mass:
\begin{equation}
m_\mathrm{crit}(C_{\rm s,eff})=\frac{2C_{\rm s,eff}^2}{G}.
\end{equation}

We note that $C_{\rm s,eff}$ encompasses both the thermal and nonthermal motions. To more specifically estimate the stability of the individual cores, we derived the virial parameter $\alpha=m_\mathrm{crit}(C_{\rm s,eff})/m$ for the whole sample of filaments. The black dashed lines denote 0.5$m_{crit}$ and 2$m_{crit}$, while the green lines correspond to $\alpha=0.5$ and $\alpha=2$. Our dataset exhibits a Pearson’s correlation coefficient of -0.35, which indicates decreasing trend for $\alpha$ as a function of $m$. Among all filaments, 133 (16\%) are supercritical, 543 (65\%) are transcritical, and 161 (19\%) are subcritical. 
These results suggest that a large fraction of filaments in our sample are unlikely to be in equilibrium even if nonthermal motions can provide significant additional support against self-gravity. Unless significant magnetic support is also present, these filaments are likely actively accreting from their surroundings to prevent total radial collapse, consistent with the proposal of ref. \cite{Arzoumanian13}. Additionally, thermally trans- and supervirial filaments may rely on continuous accretion to survive and form cores \citep{Chen24}. The presence of some subvirial filaments in our sample indicates that nonthermal motions can provide support against self-gravity.

\subsection{Gravitational acceleration}

Gravity is the key force driving star formation in interstellar molecular clouds. Ref.\cite{He2023} used numerical simulations to verify whether the gravitational acceleration maps derived from 2D surface density can accurately represent the true 3D gravitational field. In their study, the gravitational potential was computed by solving the Poisson equation. In the 3D case, Poisson’s equation can be solved efficiently in the Fourier space ($k$ space). In the case of 2D, assuming 3D density is distributed in a thin plate of half-thickness H, the potential is:
\begin{equation}
    \Phi_{k, 2 \mathrm{D}}=-\frac{2 \pi G \Sigma_{k}}{\left|k_{2 \mathrm{D}}\right|\left(1+\left|k_{2 \mathrm{D}} H\right|\right)}  
\end{equation}
$\Phi_{k}$ is the gravitational potential in the $k$ space, $\Sigma_{k}$ is the surface density in the $k$ space for the 2D case. $G$ is the gravitational constant, $k_{2 \mathrm{D}}=\sqrt{k_{x}^{2}+k_{y}^{2}}$ in the 2D application. To derive the gravitational potential, one can transform the density distribution into the $k$ space, compute $\Phi_{k}$, and back to the real space to get $\Phi$. This method offers a practical tool for studying the relationship between molecular cloud morphology and gravitational collapse, particularly in observational contexts where the 3D density distribution is not directly available.  

\begin{figure*}
    \centering
    \includegraphics[width=\linewidth]{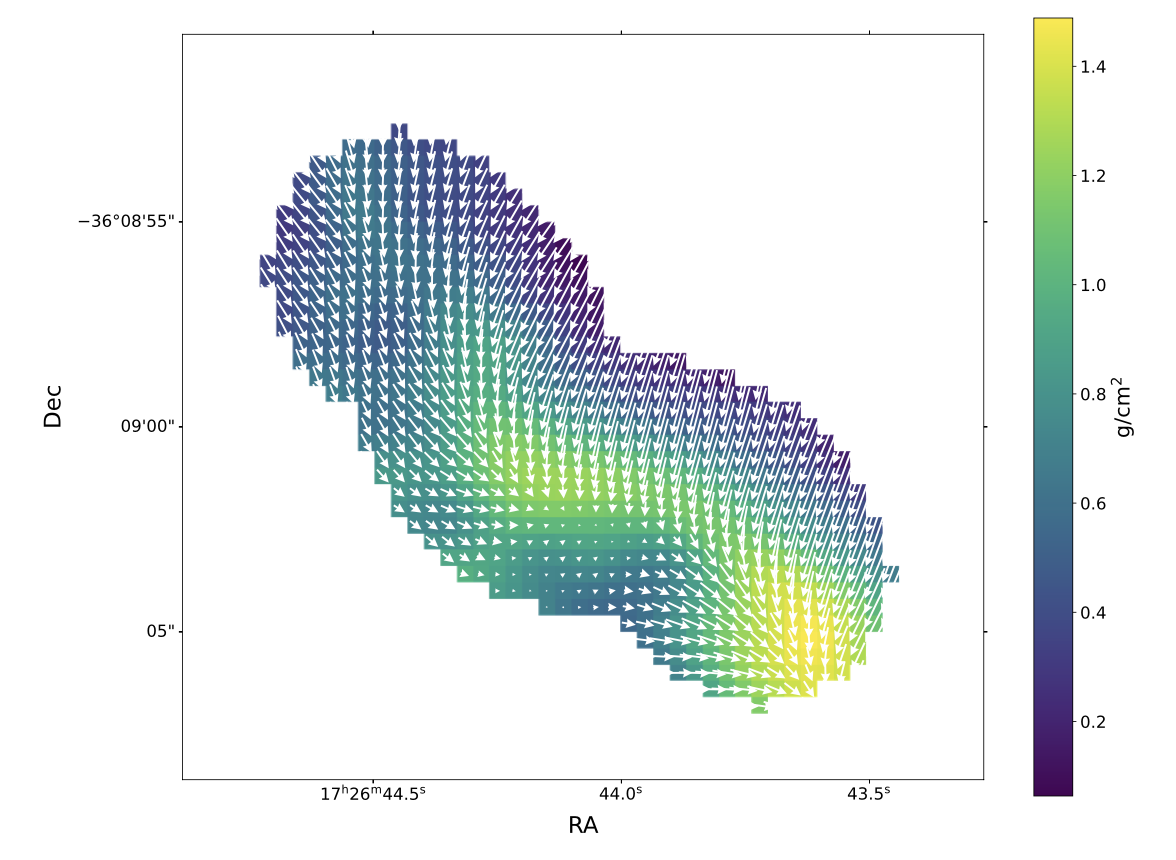}
    \caption{The surface density of the same filament skeleton as shown in Figure~\ref{fig:intensity_velocity_sigma}. The white arrows represent the Gravitational acceleration gradient direction and magnitude at each pixel position.} 
    \label{fig:acc_gra1}
\end{figure*}

\begin{figure*}
    \centering
    \includegraphics[width=\linewidth]{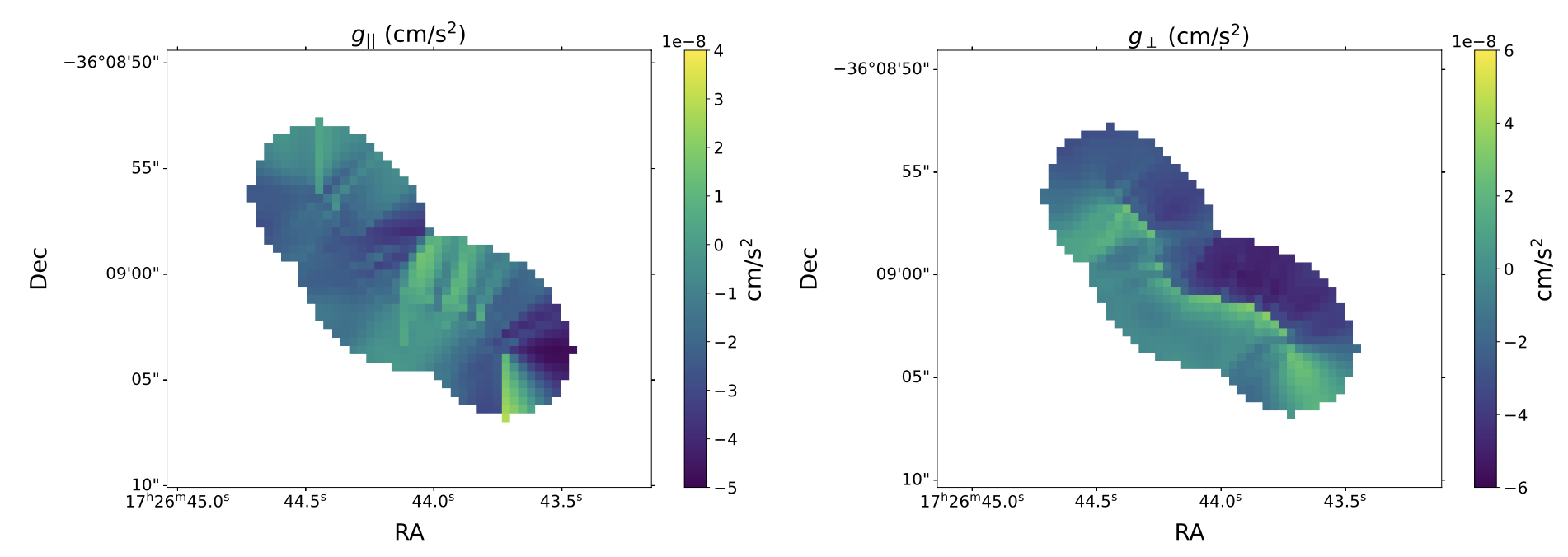}
    \caption{The $\text{g}_{\parallel}$ and $\text{g}_{\perp}$ maps of the same filament skeleton as shown in Figure~\ref{fig:intensity_velocity_sigma}.} 
    \label{fig:acc_gra2}
\end{figure*}

Therefore, we apply this method to compute the gradient of the gravitational potential ($\Phi$) from the 2D surface density, thereby reflecting variations in the gravitational field. For further details, see \cite{He2023}. Figure~\ref{fig:acc_gra1} shows the surface density of the same filament skeleton as in Figure~\ref{fig:intensity_velocity_sigma}, with the gravitational acceleration ($\text{g}=\nabla \Phi$) indicated by white arrows. The $\text{g}_{{\perp}}$ and $\text{g}_{{\parallel}}$ maps of the same filament skeleton, as shown in Figure~\ref{fig:intensity_velocity_sigma}, are presented in Figure~\ref{fig:acc_gra2}. It can be seen that the $\text{g}_{{\perp}}$ and $\text{g}_{{\parallel}}$ maps exhibit very similar morphological features to the $\nabla I_{\parallel}$ and $\nabla I_{\perp}$ maps.

\subsection{\label{sec:level2} Gradients in simulated data}

We conducted a comparative study with simulation data from the magnetohydrodynamic (MHD) cloud simulations of \cite{Haugbolle2018}. Synthetic H$^{13}$CO$^{+}$ observations were generated from the simulation data via non-LTE radiative transfer calculations using the LOC program \citep{Juvela_2020_LOC}. The H$^{13}$CO$^{+}$ modeling used a maximum fractional abundance of [H$^{13}$CO$^{+}$]/[H$_2$] = $1.0 \times 10^{-11}$, and the abundances were further scaled by a factor $n({\rm H}_2)^{2.45}/(3.0\times 10^8+n({\rm H}_2)^{2.45})$ to account for the decrease of molecular abundances at lower densities \citep{Glover2010}. As the MHD simulation did not provide accurate kinetic temperatures, we followed the procedures outlined in \cite{Juvela2022} and adopted $T_{\rm kin} = T_{\rm dust}$, with dust temperatures $T_{\rm dust}$ derived from separate continuum radiative transfer calculations. The resulting kinetic temperatures are generally above 10\,K, with a median values of 15.6\,K. The highest values locally exceeded 40\,K, reflecting the presence of embedded sources inherent to the MHD simulation output. 

The calculated H$^{13}$CO$^{+}$ maps cover the 4$\times$4\,pc and 1.56$\times$1.56\,pc projected size of the MHD model.
The FWHM resolution of the synthetic maps was set to 0.01, 0.02, 0.05, and 0.10\,pc, with pixel
sizes fixed to 1/4 of the FWHM value. The velocity resolution was set to 0.1\,km\,s$^{-1}$. In
addition to the four different beam sizes, spectral line maps were calculated for three orthogonal
view directions. We use directly the maps obtained from the radiative transfer calculations,
without further simulating the effects of interferometric observations.

\begin{figure*}
\centering
    \centering
    \includegraphics[width=\textwidth]{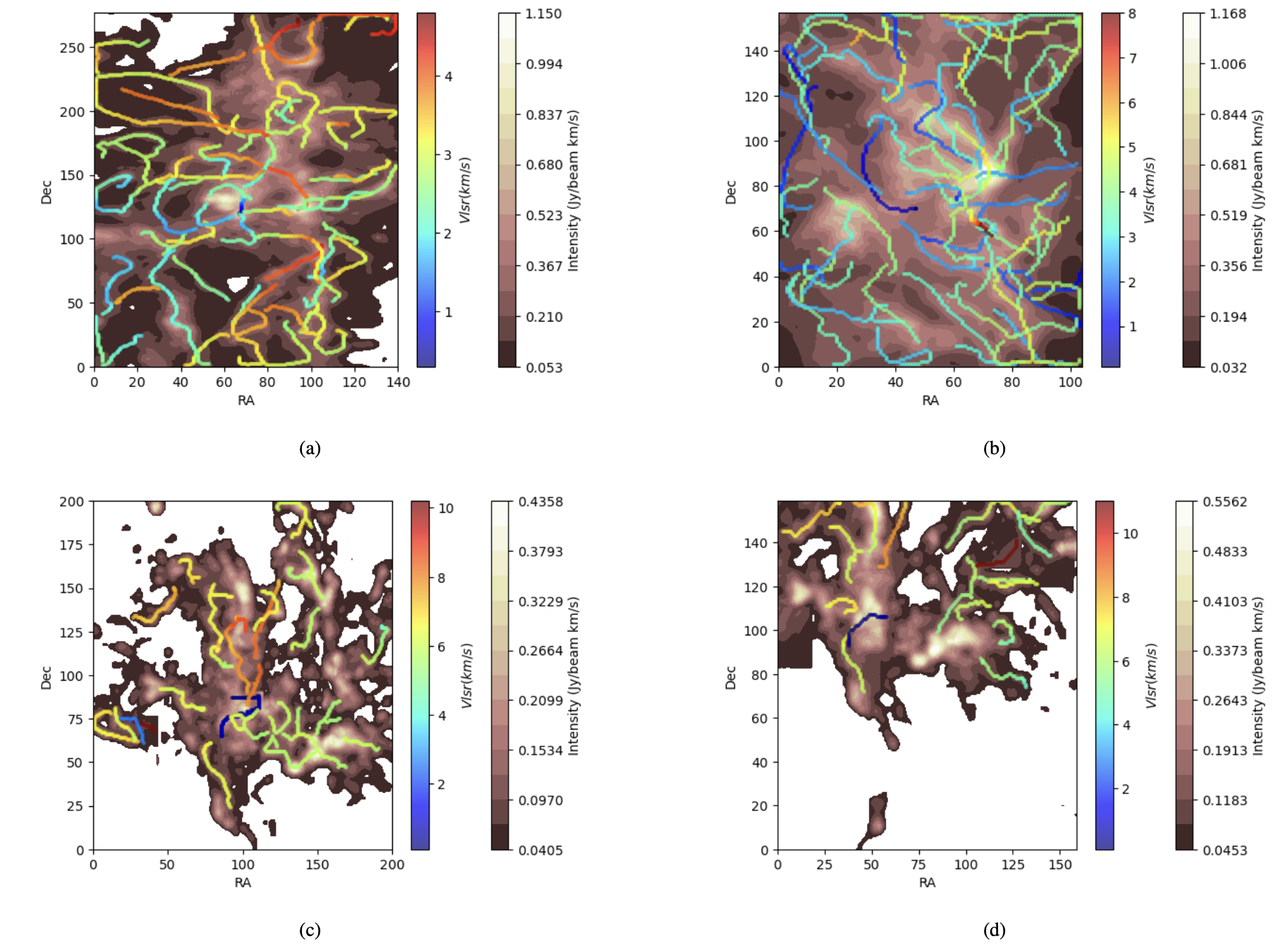}
    \caption{The projection of the three-dimensional simulation data onto the position-position plane. The images starting from the (a) panels are the simulated data at resolutions of 0.01 pc, 0.02 pc, 0.05 pc, and 0.1 pc, respectively. The colored points indicate the skeletons of the filamentary structures, and different colors represent different velocities of the skeletons.}
    \label{fig:extract_sim_fil}
\end{figure*}

Based on the simulation data described above, we identified a total of 817 filamentary structures (Figure \ref{fig:extract_sim_fil}) and calculated their mass, gradients, and other parameters. The gray plus signs in Figure 2 represent the L-M relationship of the simulation data. It shows that the simulation data have lower mass compared to the observational data. According to the previous definition, among the velocity-coherent filaments in simulation data, 411 (50\%) are trans-critical, 333 (41\%) are supercritical, and 73 (9\%) are subcritical. 

\begin{figure*}
    \centering
    \includegraphics[width=\linewidth]{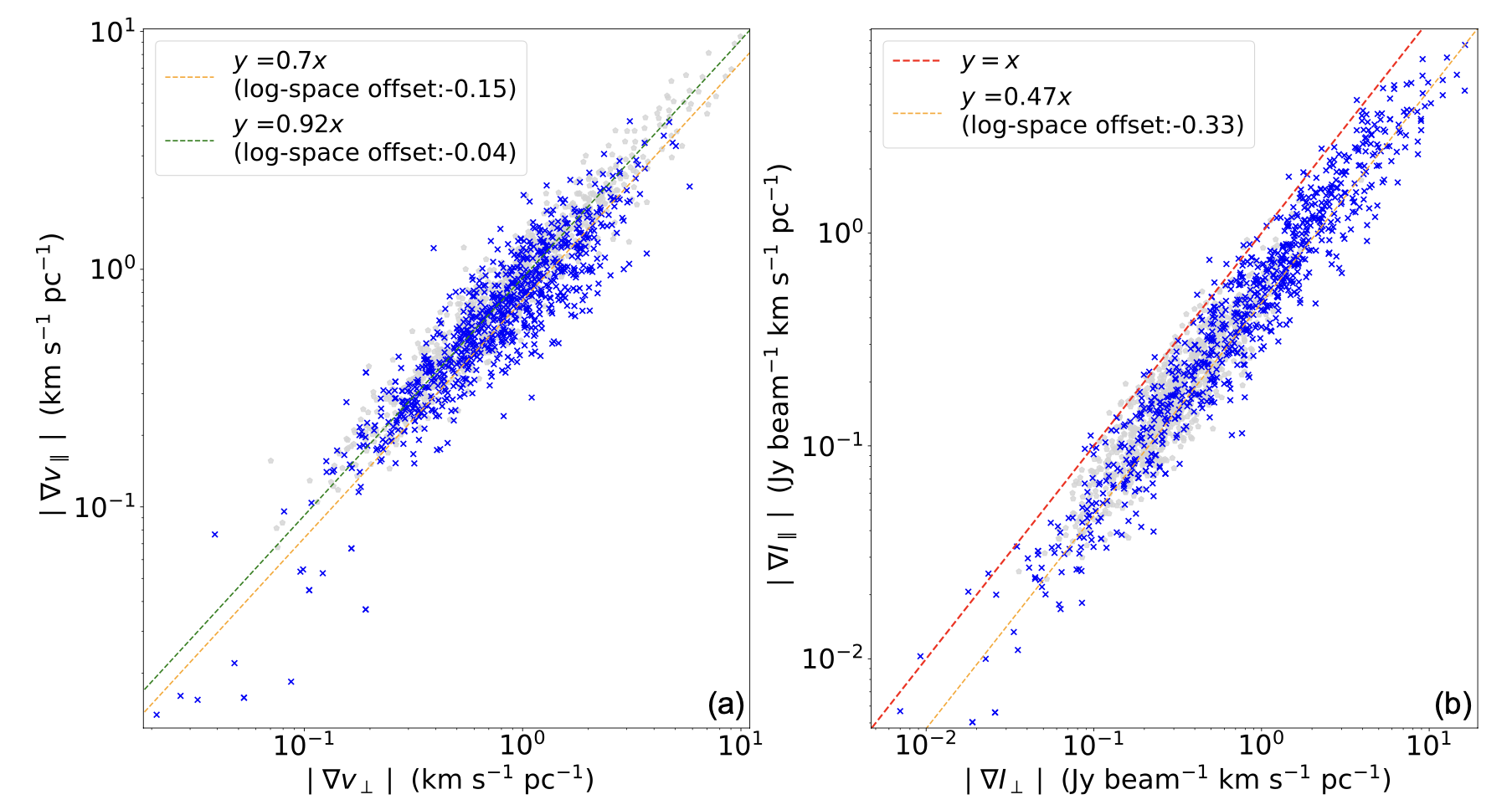}
    \caption{Gray pentagons denote filaments derived from observational data, and blue crosses represent those obtained from simulations. (a) The median $\mid \nabla v_{\perp} \mid$  plotted against their $\mid\nabla v_{\parallel} \mid$ counterparts. (b) The median $\mid\nabla I_{\perp}\mid$ plotted against their $\mid\nabla I_{\parallel}\mid$ counterparts. The best-fitting linear regression models of simulated data are represented by the orange dashed line. The red dashed line represents the y=x line. The best-fitting linear regression model derived from observation data is shown as a green dashed line.} 
    \label{fig:sim_gra}
\end{figure*}

The histogram of velocity gradients shown in Figure 6 (b), reveals that the distribution of $\mid\nabla v_{\perp}\mid$ exhibits a small shift trend compared to $\mid\nabla v_{\parallel}\mid$. Figure~\ref{fig:sim_gra} (a) presents the correlation between the median magnitudes of $\mid\nabla v_{\perp}\mid$ and $\mid\nabla v_{\parallel}\mid$, showing a Pearson correlation coefficient of r = 0.86 and a best-fit linear regression slope of 0.7$\pm$0.02. Figure~\ref{fig:sim_gra} (b) displays the correlation between the median magnitudes of $\mid\nabla I_{\perp}\mid$ and $\mid\nabla I_{\parallel}\mid$ (r = 0.86, slope = 0.47$\pm$0.01).  
The best-fitting linear regression model for observation data is shown as a green dashed line (slope = 0.92). However, the velocity gradient distribution, despite some differences, follows a broadly similar trend to the observational data. The intensity gradient distribution of the simulated data shows strong consistency with observational data. 

\begin{figure*}
\centering
    \centering
    \includegraphics[width=0.7\textwidth]{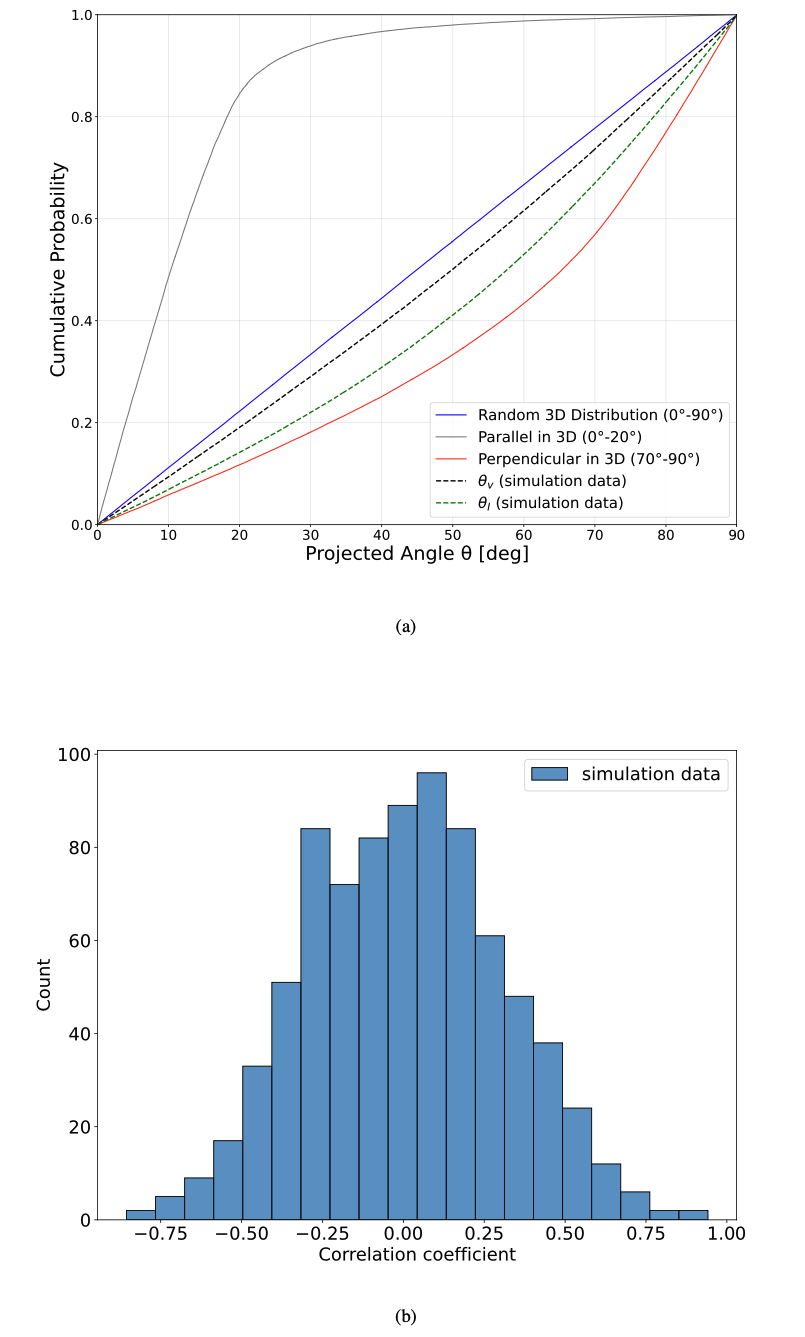}
    
    \caption{Statistical analysis results from simulation data. Panel (a): Cumulative distribution functions of relative orientation between $\nabla v$ and filaments, as well as between $\nabla I$ and filaments and the projected $\theta_{\text{3D}}$ for simulation data. Panel (b): Distribition of the Pearson correlation coefficients for the correlations between the intensity gradients and velocity gradients for simulation data.}
    \label{fig:sim-cdf}
\end{figure*}

We also examined the spatial alignment of the simulation data. Figure~\ref{fig:sim-cdf}(a) presents the cumulative distribution functions of the relative orientations for the entire simulation sample. Unlike in observational data, the velocity gradients tend to be slightly perpendicular to the filament skeletons in simulation, while the intensity gradients are much closer to perpendicular. 

We further investigated the pixel-by-pixel correlations between intensity gradients and velocity gradients across the simulation filaments and calculated their Pearson correlation coefficients. Figure~\ref{fig:sim-cdf}(b) shows the distribution of these coefficients: 73\% of the filaments have an absolute correlation coefficient less than 0.2, and 5.4\% exhibit correlations as high as 0.5 between intensity and velocity gradients.

\subsection{Uncertainties in measuring gradients}

To rigorously exclude the possibility that the observed linear relation between $\nabla v_{\perp}$ and $\nabla v_{\parallel}$ arises from numerical artifacts associated with the step size used in the gradient computation, we recomputed both gradient components employing an alternative step of 2 beams (equivalent to 16 pixels).

\begin{figure*}
    \centering
    \includegraphics[width=\linewidth]{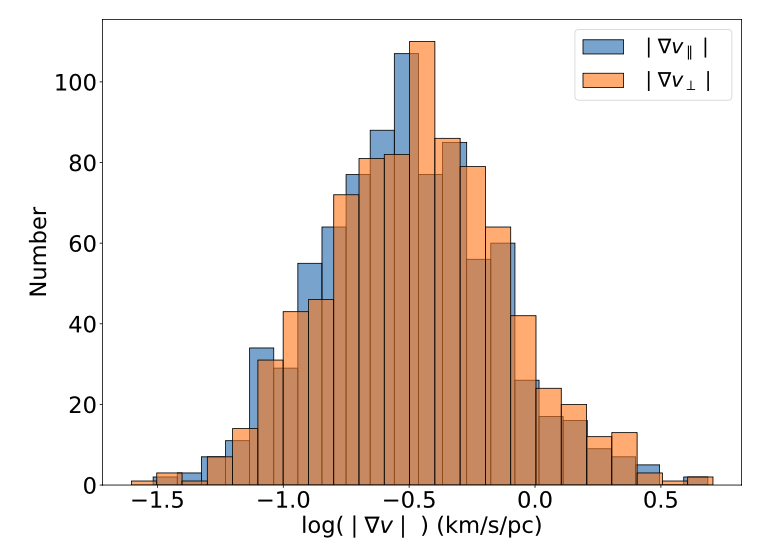}
    \caption{Velocity gradient distribution obtained with an alternative computational parameter. The distribution was derived using a step size of 2 beams (equivalent to 16 pixels) for observational data, compared to the conventional 1-beam step size (8 pixels) used in our previous analysis (Figure 6 (a)).} 
    \label{fig:2beam_dis}
\end{figure*}

\begin{figure*}
    \centering
    \includegraphics[width=\linewidth]{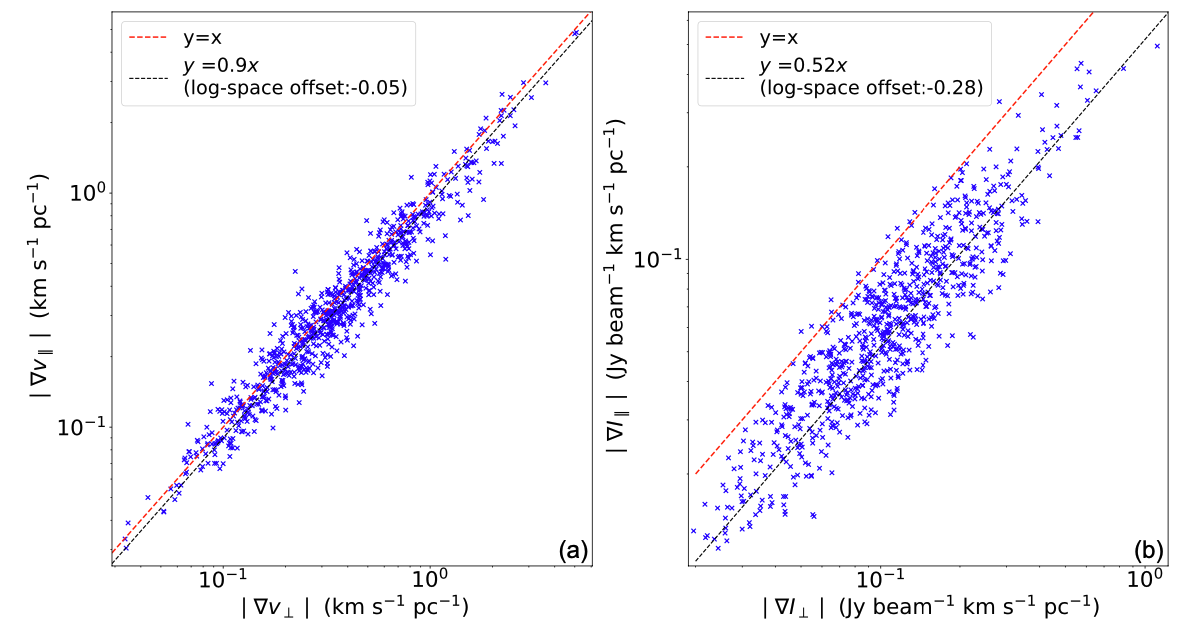}
    \caption{The gradients was derived using an alternative step size of 2 beams (equivalent to 16 pixels) for observation data. (a) The median $\mid \nabla v_{\perp} \mid$  plotted against their $\mid\nabla v_{\parallel} \mid$ counterparts. (b) The median $\mid\nabla I_{\perp}\mid$ plotted against their $\mid\nabla I_{\parallel}\mid$ counterparts. The best-fitting linear regression models of simulated data are represented by the orange dashed line. The red dashed line represents the y=x line.} 
    \label{fig:2beam_gra}
\end{figure*}

Figure~\ref{fig:2beam_dis} shows the distributions of the two $\mid\nabla v\mid$ components from the new calculation. Figure~\ref{fig:2beam_gra} illustrates the correlation between the median magnitudes of the gradients. The new calculations yield fundamentally consistent results with the previous calculations. The robust preservation of the linear relation between $\mid\nabla v_{\perp}\mid$ and $\mid\nabla v_{\parallel}\mid$ conclusively confirms this relationship as an intrinsic physical characteristic of filaments, rather than a numerical artifact of the data analysis.

While the linear relationship between $\nabla v_{\perp}$ and $\nabla v_{\parallel}$ has been conclusively established, the overall velocity gradient decreases when the step size is 2 beams compared to one beam (Figure 6 (a)). This is expected because a larger step size results in a relatively flatter (smaller) velocity gradient. To better interpret this effect, we analyzed the velocity gradient distributions at different step sizes (half beam, one beam, and two beams) using simulation data. 

\begin{figure*}
\centering
    \includegraphics[width=\textwidth]{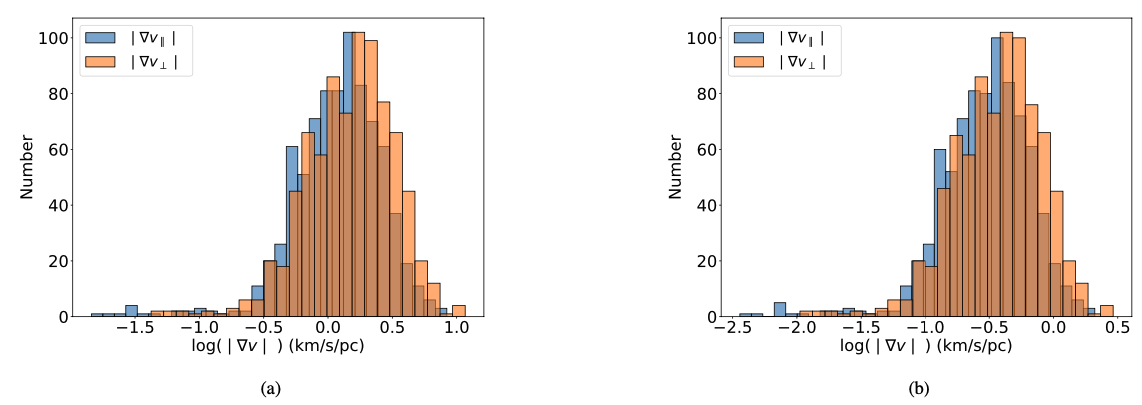}

    \caption{The distribution of median velocity gradients in simulations. (a) half-beam step size. (b) two beams step size.}
    \label{fig:sim_step_half_two}
\end{figure*}

Figures~\ref{fig:sim_step_half_two}(a) and (b) show the distributions of median velocity gradients for step sizes of half beam and two beams, respectively. The distribution of median velocity gradients for the one beam step size is displayed in Figure 6 (b). These figures demonstrate that larger step sizes correspond to smaller overall gradients, without significantly altering the overall distribution shapes. 

\bibliographystyle{plainnat}
\bibliography{sample}

\begin{thebibliography}{96}
\providecommand{\natexlab}[1]{#1}
\providecommand{\url}[1]{\texttt{#1}}
\expandafter\ifx\csname urlstyle\endcsname\relax
  \providecommand{\doi}[1]{doi: #1}\else
  \providecommand{\doi}{doi: \begingroup \urlstyle{rm}\Url}\fi

\bibitem[{Andr{\'e}} et~al.(2014){Andr{\'e}}, {Di Francesco}, {Ward-Thompson},
  {Inutsuka}, {Pudritz}, and {Pineda}]{2014prpl.conf...27A}
P.~{Andr{\'e}}, J.~{Di Francesco}, D.~{Ward-Thompson}, S.-I. {Inutsuka}, R.~E.
  {Pudritz}, and J.~E. {Pineda}.
\newblock {From Filamentary Networks to Dense Cores in Molecular Clouds: Toward
  a New Paradigm for Star Formation}.
\newblock In Henrik {Beuther}, Ralf~S. {Klessen}, Cornelis~P. {Dullemond}, and
  Thomas {Henning}, editors, \emph{Protostars and Planets VI}, pages 27--51,
  January 2014.
\newblock \doi{10.2458/azu_uapress_9780816531240-ch002}.

\bibitem[{Andr{\'e}} et~al.(2019){Andr{\'e}}, {Arzoumanian}, {K{\"o}nyves},
  {Shimajiri}, and {Palmeirim}]{Andre2019}
Ph. {Andr{\'e}}, D.~{Arzoumanian}, V.~{K{\"o}nyves}, Y.~{Shimajiri}, and
  P.~{Palmeirim}.
\newblock {The role of molecular filaments in the origin of the prestellar core
  mass function and stellar initial mass function}.
\newblock \emph{Astronomy \& Astrophysics}, 629:\penalty0 L4, September 2019.
\newblock \doi{10.1051/0004-6361/201935915}.

\bibitem[{Arzoumanian} et~al.(2011){Arzoumanian}, {Andr{\'e}}, {Didelon},
  {K{\"o}nyves}, {Schneider}, {Men'shchikov}, {Sousbie}, {Zavagno}, {Bontemps},
  {di Francesco}, {Griffin}, {Hennemann}, {Hill}, {Kirk}, {Martin}, {Minier},
  {Molinari}, {Motte}, {Peretto}, {Pezzuto}, {Spinoglio}, {Ward-Thompson},
  {White}, and {Wilson}]{Arzoumanian2011}
D.~{Arzoumanian}, Ph. {Andr{\'e}}, P.~{Didelon}, V.~{K{\"o}nyves},
  N.~{Schneider}, A.~{Men'shchikov}, T.~{Sousbie}, A.~{Zavagno}, S.~{Bontemps},
  J.~{di Francesco}, M.~{Griffin}, M.~{Hennemann}, T.~{Hill}, J.~{Kirk},
  P.~{Martin}, V.~{Minier}, S.~{Molinari}, F.~{Motte}, N.~{Peretto},
  S.~{Pezzuto}, L.~{Spinoglio}, D.~{Ward-Thompson}, G.~{White}, and C.~D.
  {Wilson}.
\newblock {Characterizing interstellar filaments with Herschel in IC 5146}.
\newblock \emph{Astronomy \& Astrophysics}, 529:\penalty0 L6, May 2011.
\newblock \doi{10.1051/0004-6361/201116596}.

\bibitem[{Arzoumanian} et~al.(2013){Arzoumanian}, {Andr{\'e}}, {Peretto}, and
  {K{\"o}nyves}]{Arzoumanian13}
D.~{Arzoumanian}, Ph. {Andr{\'e}}, N.~{Peretto}, and V.~{K{\"o}nyves}.
\newblock {Formation and evolution of interstellar filaments. Hints from
  velocity dispersion measurements}.
\newblock \emph{Astronomy \& Astrophysics}, 553:\penalty0 A119, May 2013.
\newblock \doi{10.1051/0004-6361/201220822}.

\bibitem[{Arzoumanian} et~al.(2019){Arzoumanian}, {Andr{\'e}}, {K{\"o}nyves},
  {Palmeirim}, {Roy}, {Schneider}, {Benedettini}, {Didelon}, {Di Francesco},
  {Kirk}, and {Ladjelate}]{Arzoumanian2019}
D.~{Arzoumanian}, Ph. {Andr{\'e}}, V.~{K{\"o}nyves}, P.~{Palmeirim}, A.~{Roy},
  N.~{Schneider}, M.~{Benedettini}, P.~{Didelon}, J.~{Di Francesco}, J.~{Kirk},
  and B.~{Ladjelate}.
\newblock {Characterizing the properties of nearby molecular filaments observed
  with Herschel}.
\newblock \emph{Astronomy \& Astrophysics}, 621:\penalty0 A42, January 2019.
\newblock \doi{10.1051/0004-6361/201832725}.

\bibitem[{Arzoumanian} et~al.(2018){Arzoumanian}, {Shimajiri}, {Inutsuka},
  {Inoue}, and {Tachihara}]{Arzoumanian2018}
Doris {Arzoumanian}, Yoshito {Shimajiri}, Shu-ichiro {Inutsuka}, Tsuyoshi
  {Inoue}, and Kengo {Tachihara}.
\newblock {Molecular filament formation and filament-cloud interaction: Hints
  from Nobeyama 45 m telescope observations}.
\newblock \emph{Publications of the Astronomical Society of Japan}, 70\penalty0
  (5):\penalty0 96, October 2018.
\newblock \doi{10.1093/pasj/psy095}.

\bibitem[{Baug} et~al.(2018){Baug}, {Dewangan}, {Ojha}, {Tachihara}, {Pandey},
  {Sharma}, {Tamura}, {Ninan}, and {Ghosh}]{Baug2018}
T.~{Baug}, L.~K. {Dewangan}, D.~K. {Ojha}, Kengo {Tachihara}, A.~K. {Pandey},
  Saurabh {Sharma}, M.~{Tamura}, J.~P. {Ninan}, and S.~K. {Ghosh}.
\newblock {Star Formation in the Sh 2-53 Region Influenced by Accreting
  Molecular Filaments}.
\newblock \emph{The Astrophysical Journal}, 852\penalty0 (2):\penalty0 119,
  January 2018.
\newblock \doi{10.3847/1538-4357/aaa429}.

\bibitem[{Bergin} and {Tafalla}(2007)]{Bergin2007}
Edwin~A. {Bergin} and Mario {Tafalla}.
\newblock {Cold Dark Clouds: The Initial Conditions for Star Formation}.
\newblock \emph{Annual Review of Astronomy \& Astrophysics}, 45\penalty0
  (1):\penalty0 339--396, September 2007.
\newblock \doi{10.1146/annurev.astro.45.071206.100404}.

\bibitem[{Beuther} et~al.(2015){Beuther}, {Ragan}, {Johnston}, {Henning},
  {Hacar}, and {Kainulainen}]{2015A&A...584A..67B}
H.~{Beuther}, S.~E. {Ragan}, K.~{Johnston}, Th. {Henning}, A.~{Hacar}, and
  J.~T. {Kainulainen}.
\newblock {Filament fragmentation in high-mass star formation}.
\newblock \emph{Astronomy \& Astrophysics}, 584:\penalty0 A67, December 2015.
\newblock \doi{10.1051/0004-6361/201527108}.

\bibitem[{Beuther} et~al.(2025){Beuther}, {Olguin}, {Sanhueza}, {Cunningham},
  and {Ginsburg}]{2025A&A...695A..51B}
H.~{Beuther}, F.~A. {Olguin}, P.~{Sanhueza}, N.~{Cunningham}, and
  A.~{Ginsburg}.
\newblock {Hierarchical accretion flow from the G351 infrared dark filament to
  its central cores}.
\newblock \emph{Astronomy \& Astrophysics}, 695:\penalty0 A51, March 2025.
\newblock \doi{10.1051/0004-6361/202452754}.

\bibitem[{Bronfman} et~al.(1996){Bronfman}, {Nyman}, and {May}]{Bronfman96}
L.~{Bronfman}, L.~A. {Nyman}, and J.~{May}.
\newblock {A CS(2-1) survey of IRAS point sources with color characteristics of
  ultra-compact HII regions.}
\newblock \emph{Astronomy \& Astrophysics, Supplement}, 115:\penalty0 81,
  January 1996.

\bibitem[{Chen} et~al.(2019){Chen}, {Zhang}, {Wright}, {Busquet}, {Lin}, {Liu},
  {Olguin}, {Sanhueza}, {Nakamura}, {Palau}, {Ohashi}, {Tatematsu}, and
  {Liao}]{2019ApJ...875...24C}
Huei-Ru~Vivien {Chen}, Qizhou {Zhang}, M.~C.~H. {Wright}, Gemma {Busquet},
  Yuxin {Lin}, Hauyu~Baobab {Liu}, F.~A. {Olguin}, Patricio {Sanhueza},
  Fumitaka {Nakamura}, Aina {Palau}, Satoshi {Ohashi}, Ken'ichi {Tatematsu},
  and Li-Wen {Liao}.
\newblock {Filamentary Accretion Flows in the Infrared Dark Cloud G14.225-0.506
  Revealed by ALMA}.
\newblock \emph{The Astrophysical Journal}, 875\penalty0 (1):\penalty0 24,
  April 2019.
\newblock \doi{10.3847/1538-4357/ab0f3e}.

\bibitem[{Chen} et~al.(2020{\natexlab{a}}){Chen}, {Di Francesco}, {Rosolowsky},
  {Keown}, {Pineda}, {Friesen}, {Caselli}, {Chen}, {Matzner}, {Offner},
  {Punanova}, {Redaelli}, {Scibelli}, and {Shirley}]{Chen2020}
Michael Chun-Yuan {Chen}, James {Di Francesco}, Erik {Rosolowsky}, Jared
  {Keown}, Jaime~E. {Pineda}, Rachel~K. {Friesen}, Paola {Caselli}, How-Huan
  {Chen}, Christopher~D. {Matzner}, Stella~S. {Offner}, Anna {Punanova}, Elena
  {Redaelli}, Samantha {Scibelli}, and Yancy {Shirley}.
\newblock {Velocity-coherent Filaments in NGC 1333: Evidence for Accretion
  Flow?}
\newblock \emph{The Astrophysical Journal}, 891\penalty0 (1):\penalty0 84,
  March 2020{\natexlab{a}}.
\newblock \doi{10.3847/1538-4357/ab7378}.

\bibitem[{Chen} et~al.(2020{\natexlab{b}}){Chen}, {Di Francesco}, {Rosolowsky},
  {Keown}, {Pineda}, {Friesen}, {Caselli}, {Chen}, {Matzner}, {Offner},
  {Punanova}, {Redaelli}, {Scibelli}, and {Shirley}]{Chen2020ApJ...891...84C}
Michael Chun-Yuan {Chen}, James {Di Francesco}, Erik {Rosolowsky}, Jared
  {Keown}, Jaime~E. {Pineda}, Rachel~K. {Friesen}, Paola {Caselli}, How-Huan
  {Chen}, Christopher~D. {Matzner}, Stella~S. {Offner}, Anna {Punanova}, Elena
  {Redaelli}, Samantha {Scibelli}, and Yancy {Shirley}.
\newblock {Velocity-coherent Filaments in NGC 1333: Evidence for Accretion
  Flow?}
\newblock \emph{The Astrophysical Journal}, 891\penalty0 (1):\penalty0 84,
  March 2020{\natexlab{b}}.
\newblock \doi{10.3847/1538-4357/ab7378}.

\bibitem[{Chen} et~al.(2024){Chen}, {Di Francesco}, {Friesen}, {Pineda},
  {Caselli}, {Ginsburg}, {Kirk}, {Punanova}, and {The GAS
  Collaboration}]{Chen24}
Michael Chun-Yuan {Chen}, James {Di Francesco}, Rachel~K. {Friesen}, Jaime~E.
  {Pineda}, Paola {Caselli}, Adam {Ginsburg}, Helen {Kirk}, Anna {Punanova},
  and {The GAS Collaboration}.
\newblock {Filament Accretion and Fragmentation in the Perseus Molecular
  Cloud}.
\newblock \emph{The Astrophysical Journal}, 977\penalty0 (1):\penalty0 135,
  December 2024.
\newblock \doi{10.3847/1538-4357/ad88e8}.

\bibitem[{Dhabal} et~al.(2018){Dhabal}, {Mundy}, {Rizzo}, {Storm}, and
  {Teuben}]{Dhabal2018}
Arnab {Dhabal}, Lee~G. {Mundy}, Maxime~J. {Rizzo}, Shaye {Storm}, and Peter
  {Teuben}.
\newblock {Morphology and Kinematics of Filaments in the Serpens and Perseus
  Molecular Clouds}.
\newblock \emph{The Astrophysical Journal}, 853\penalty0 (2):\penalty0 169,
  February 2018.
\newblock \doi{10.3847/1538-4357/aaa76b}.

\bibitem[{Dutta} et~al.(2018){Dutta}, {Mondal}, {Samal}, and {Jose}]{Dutta2018}
Somnath {Dutta}, Soumen {Mondal}, Manash~R. {Samal}, and Jessy {Jose}.
\newblock {The Planck Cold Clump G108.37-01.06: A Site of Complex Interplay
  between H II Regions, Young Clusters, and Filaments}.
\newblock \emph{The Astrophysical Journal}, 864\penalty0 (2):\penalty0 154,
  September 2018.
\newblock \doi{10.3847/1538-4357/aadb3e}.

\bibitem[{Feng} et~al.(2024){Feng}, {Smith}, {Hacar}, {Clark}, and
  {Seifried}]{Feng2024}
Jiancheng {Feng}, Rowan~J. {Smith}, Alvaro {Hacar}, Susan~E. {Clark}, and
  Daniel {Seifried}.
\newblock {On the evolution of the observed mass-to-length relationship for
  star-forming filaments}.
\newblock \emph{Monthly Notices of the Royal Astronomical Society},
  528\penalty0 (4):\penalty0 6370--6387, March 2024.
\newblock \doi{10.1093/mnras/stae407}.

\bibitem[{Fern{\'a}ndez-L{\'o}pez} et~al.(2014){Fern{\'a}ndez-L{\'o}pez},
  {Arce}, {Looney}, {Mundy}, {Storm}, {Teuben}, {Lee}, {Segura-Cox}, {Isella},
  {Tobin}, {Rosolowsky}, {Plunkett}, {Kwon}, {Kauffmann}, {Ostriker}, {Tassis},
  {Shirley}, and {Pound}]{Fern2014}
M.~{Fern{\'a}ndez-L{\'o}pez}, H.~G. {Arce}, L.~{Looney}, L.~G. {Mundy},
  S.~{Storm}, P.~J. {Teuben}, K.~{Lee}, D.~{Segura-Cox}, A.~{Isella}, J.~J.
  {Tobin}, E.~{Rosolowsky}, A.~{Plunkett}, W.~{Kwon}, J.~{Kauffmann},
  E.~{Ostriker}, K.~{Tassis}, Y.~L. {Shirley}, and M.~{Pound}.
\newblock {CARMA Large Area Star Formation Survey: Observational Analysis of
  Filaments in the Serpens South Molecular Cloud}.
\newblock \emph{The Astrophysical Journal Letters}, 790\penalty0 (2):\penalty0
  L19, August 2014.
\newblock \doi{10.1088/2041-8205/790/2/L19}.

\bibitem[{Friesen} et~al.(2013){Friesen}, {Medeiros}, {Schnee}, {Bourke}, {di
  Francesco}, {Gutermuth}, and {Myers}]{Friesen2013}
R.~K. {Friesen}, L.~{Medeiros}, S.~{Schnee}, T.~L. {Bourke}, J.~{di Francesco},
  R.~{Gutermuth}, and P.~C. {Myers}.
\newblock {Abundant cyanopolyynes as a probe of infall in the Serpens South
  cluster-forming region}.
\newblock \emph{Monthly Notices of the Royal Astronomical Society},
  436\penalty0 (2):\penalty0 1513--1529, December 2013.
\newblock \doi{10.1093/mnras/stt1671}.

\bibitem[{Friesen} et~al.(2017){Friesen}, {Pineda}, {co-PIs}, {Rosolowsky},
  {Alves}, {Chac{\'o}n-Tanarro}, {How-Huan Chen}, {Chun-Yuan Chen}, {Di
  Francesco}, {Keown}, {Kirk}, {Punanova}, {Seo}, {Shirley}, {Ginsburg},
  {Hall}, {Offner}, {Singh}, {Arce}, {Caselli}, {Goodman}, {Martin}, {Matzner},
  {Myers}, {Redaelli}, and {GAS Collaboration}]{Friesen2017}
Rachel~K. {Friesen}, Jaime~E. {Pineda}, {co-PIs}, Erik {Rosolowsky}, Felipe
  {Alves}, Ana {Chac{\'o}n-Tanarro}, Hope {How-Huan Chen}, Michael {Chun-Yuan
  Chen}, James {Di Francesco}, Jared {Keown}, Helen {Kirk}, Anna {Punanova},
  Youngmin {Seo}, Yancy {Shirley}, Adam {Ginsburg}, Christine {Hall}, Stella
  S.~R. {Offner}, Ayushi {Singh}, H{\'e}ctor~G. {Arce}, Paola {Caselli},
  Alyssa~A. {Goodman}, Peter~G. {Martin}, Christopher {Matzner}, Philip~C.
  {Myers}, Elena {Redaelli}, and {GAS Collaboration}.
\newblock {The Green Bank Ammonia Survey: First Results of NH$_{3}$ Mapping of
  the Gould Belt}.
\newblock \emph{The Astrophysical Journal}, 843\penalty0 (1):\penalty0 63, July
  2017.
\newblock \doi{10.3847/1538-4357/aa6d58}.

\bibitem[{Ge} and {Wang}(2022)]{Ge2022}
Yifei {Ge} and Ke~{Wang}.
\newblock {A Census of 163 Large-scale ({\ensuremath{\geq}}10 pc),
  Velocity-coherent Filaments in the Inner Galactic Plane: Physical Properties,
  Dense-gas Fraction, and Association with Spiral Arms}.
\newblock \emph{The Astrophysical Journal Supplement Series}, 259\penalty0
  (2):\penalty0 36, April 2022.
\newblock \doi{10.3847/1538-4365/ac4a76}.

\bibitem[{Glover} et~al.(2010){Glover}, {Federrath}, {Mac Low}, and
  {Klessen}]{Glover2010}
S.~C.~O. {Glover}, C.~{Federrath}, M.~M. {Mac Low}, and R.~S. {Klessen}.
\newblock {Modelling CO formation in the turbulent interstellar medium}.
\newblock \emph{Monthly Notices of the Royal Astronomical Society},
  404\penalty0 (1):\penalty0 2--29, May 2010.
\newblock \doi{10.1111/j.1365-2966.2009.15718.x}.

\bibitem[{Gong} et~al.(2018){Gong}, {Li}, {Mao}, {Henkel}, {Menten}, {Fang},
  {Wang}, and {Sun}]{Gong2018}
Y.~{Gong}, G.~X. {Li}, R.~Q. {Mao}, C.~{Henkel}, K.~M. {Menten}, M.~{Fang},
  M.~{Wang}, and J.~X. {Sun}.
\newblock {The Serpens filament at the onset of slightly supercritical
  collapse}.
\newblock \emph{Astronomy \& Astrophysics}, 620:\penalty0 A62, November 2018.
\newblock \doi{10.1051/0004-6361/201833583}.

\bibitem[{Hacar} et~al.(2023){Hacar}, {Clark}, {Heitsch}, {Kainulainen},
  {Panopoulou}, {Seifried}, and {Smith}]{Hacar23}
A.~{Hacar}, S.~E. {Clark}, F.~{Heitsch}, J.~{Kainulainen}, G.~V. {Panopoulou},
  D.~{Seifried}, and R.~{Smith}.
\newblock {Initial Conditions for Star Formation: a Physical Description of the
  Filamentary ISM}.
\newblock In S.~{Inutsuka}, Y.~{Aikawa}, T.~{Muto}, K.~{Tomida}, and
  M.~{Tamura}, editors, \emph{Protostars and Planets VII}, volume 534 of
  \emph{Astronomical Society of the Pacific Conference Series}, page 153, July
  2023.
\newblock \doi{10.48550/arXiv.2203.09562}.

\bibitem[{Hacar} et~al.(2025){Hacar}, {Konietzka}, {Seifried}, {Clark},
  {Socci}, {Bonanomi}, {Burkert}, {Schisano}, {Kainulainen}, and
  {Smith}]{Hacar2025}
A.~{Hacar}, R.~{Konietzka}, D.~{Seifried}, S.~E. {Clark}, A.~{Socci},
  F.~{Bonanomi}, A.~{Burkert}, E.~{Schisano}, J.~{Kainulainen}, and R.~{Smith}.
\newblock {Emergence of high-mass stars in complex fiber networks (EMERGE): V.
  From filaments to spheroids: the origin of the hub-filament systems}.
\newblock \emph{Astronomy \& Astrophysics}, 694:\penalty0 A69, February 2025.
\newblock \doi{10.1051/0004-6361/202450779}.

\bibitem[{Haugb{\o}lle} et~al.(2018){Haugb{\o}lle}, {Padoan}, and
  {Nordlund}]{Haugbolle2018}
Troels {Haugb{\o}lle}, Paolo {Padoan}, and {\r{A}}ke {Nordlund}.
\newblock {The Stellar IMF from Isothermal MHD Turbulence}.
\newblock \emph{The Astrophysical Journal}, 854\penalty0 (1):\penalty0 35,
  February 2018.
\newblock \doi{10.3847/1538-4357/aaa432}.

\bibitem[{He} et~al.(2023){He}, {Li}, and {Burkert}]{He2023}
Zhen-Zhen {He}, Guang-Xing {Li}, and Andreas {Burkert}.
\newblock {Mapping gravity in stellar nurseries - establishing the
  effectiveness of 2D acceleration maps}.
\newblock \emph{Monthly Notices of the Royal Astronomical Society},
  526\penalty0 (1):\penalty0 L20--L25, November 2023.
\newblock \doi{10.1093/mnrasl/slad104}.

\bibitem[{Hennebelle} and {Chabrier}(2008)]{Hennebelle2008}
Patrick {Hennebelle} and Gilles {Chabrier}.
\newblock {Analytical Theory for the Initial Mass Function: CO Clumps and
  Prestellar Cores}.
\newblock \emph{The Astrophysical Journal}, 684\penalty0 (1):\penalty0
  395--410, September 2008.
\newblock \doi{10.1086/589916}.

\bibitem[{Hopkins}(2012)]{Hopkins2012}
Philip~F. {Hopkins}.
\newblock {The stellar initial mass function, core mass function and the
  last-crossing distribution}.
\newblock \emph{Monthly Notices of the Royal Astronomical Society},
  423\penalty0 (3):\penalty0 2037--2044, July 2012.
\newblock \doi{10.1111/j.1365-2966.2012.20731.x}.

\bibitem[{Hoq} et~al.(2013){Hoq}, {Jackson}, {Foster}, {Sanhueza},
  {Guzm{\'a}n}, {Whitaker}, {Claysmith}, {Rathborne}, {Vasyunina}, and
  {Vasyunin}]{Hoq2013}
Sadia {Hoq}, James~M. {Jackson}, Jonathan~B. {Foster}, Patricio {Sanhueza},
  Andr{\'e}s {Guzm{\'a}n}, J.~Scott {Whitaker}, Christopher {Claysmith},
  Jill~M. {Rathborne}, Tatiana {Vasyunina}, and Anton {Vasyunin}.
\newblock {Chemical Evolution in High-mass Star-forming Regions: Results from
  the MALT90 Survey}.
\newblock \emph{The Astrophysical Journal}, 777\penalty0 (2):\penalty0 157,
  November 2013.
\newblock \doi{10.1088/0004-637X/777/2/157}.

\bibitem[{Hsieh} et~al.(2021){Hsieh}, {Arce}, {Mardones}, {Kong}, and
  {Plunkett}]{2021ApJ...908...92H}
Cheng-Han {Hsieh}, H{\'e}ctor~G. {Arce}, Diego {Mardones}, Shuo {Kong}, and
  Adele {Plunkett}.
\newblock {Rotating Filament in Orion B: Do Cores Inherit Their Angular
  Momentum from Their Parent Filament?}
\newblock \emph{The Astrophysical Journal}, 908\penalty0 (1):\penalty0 92,
  February 2021.
\newblock \doi{10.3847/1538-4357/abd034}.

\bibitem[{Inutsuka}(2001)]{Inutsuka2001}
Shu-ichiro {Inutsuka}.
\newblock {The Mass Function of Molecular Cloud Cores}.
\newblock \emph{The Astrophysical Journal Letters}, 559\penalty0 (2):\penalty0
  L149--L152, October 2001.
\newblock \doi{10.1086/323786}.

\bibitem[{Inutsuka} and {Miyama}(1992)]{Inutsuka1992}
Shu-Ichiro {Inutsuka} and Shoken~M. {Miyama}.
\newblock {Self-similar Solutions and the Stability of Collapsing Isothermal
  Filaments}.
\newblock \emph{The Astrophysical Journal}, 388:\penalty0 392, April 1992.
\newblock \doi{10.1086/171162}.

\bibitem[{Inutsuka} and {Miyama}(1997)]{Inutsuka1997}
Shu-ichiro {Inutsuka} and Shoken~M. {Miyama}.
\newblock {A Production Mechanism for Clusters of Dense Cores}.
\newblock \emph{The Astrophysical Journal}, 480\penalty0 (2):\penalty0
  681--693, May 1997.
\newblock \doi{10.1086/303982}.

\bibitem[{Jappsen} et~al.(2005){Jappsen}, {Klessen}, {Larson}, {Li}, and {Mac
  Low}]{Jappsen2005}
A.~K. {Jappsen}, R.~S. {Klessen}, R.~B. {Larson}, Y.~{Li}, and M.~M. {Mac Low}.
\newblock {The stellar mass spectrum from non-isothermal gravoturbulent
  fragmentation}.
\newblock \emph{Astronomy \& Astrophysics}, 435\penalty0 (2):\penalty0
  611--623, May 2005.
\newblock \doi{10.1051/0004-6361:20042178}.

\bibitem[{Jiao} et~al.(2024){Jiao}, {Wang}, {Xu}, {Wang}, and
  {Beuther}]{Jiao2024}
Wenyu {Jiao}, Ke~{Wang}, Fengwei {Xu}, Chao {Wang}, and Henrik {Beuther}.
\newblock {Relative alignment between gas structures and magnetic field in
  Orion A at different scales using different molecular gas tracers}.
\newblock \emph{Astronomy \& Astrophysics}, 686:\penalty0 A202, June 2024.
\newblock \doi{10.1051/0004-6361/202449182}.

\bibitem[{Juvela} et~al.(2022){Juvela}, {Mannfors}, {Liu}, and
  {T{\'o}th}]{Juvela2022}
M.~{Juvela}, E.~{Mannfors}, T.~{Liu}, and L.~V. {T{\'o}th}.
\newblock {Synthetic Next Generation Very Large Array line observations of a
  massive star-forming cloud}.
\newblock \emph{Astronomy \& Astrophysics}, 666:\penalty0 A74, October 2022.
\newblock \doi{10.1051/0004-6361/202244026}.

\bibitem[{Juvela}(2020)]{Juvela_2020_LOC}
Mika {Juvela}.
\newblock {LOC program for line radiative transfer}.
\newblock \emph{Astronomy \& Astrophysics}, 644:\penalty0 A151, December 2020.
\newblock \doi{10.1051/0004-6361/202039456}.

\bibitem[{Kauffmann} et~al.(2008){Kauffmann}, {Bertoldi}, {Bourke}, {Evans},
  and {Lee}]{Kauffmann2008}
J.~{Kauffmann}, F.~{Bertoldi}, T.~L. {Bourke}, N.~J. {Evans}, II, and C.~W.
  {Lee}.
\newblock {MAMBO mapping of Spitzer c2d small clouds and cores}.
\newblock \emph{Astronomy \& Astrophysics}, 487\penalty0 (3):\penalty0
  993--1017, September 2008.
\newblock \doi{10.1051/0004-6361:200809481}.

\bibitem[{Kirk} et~al.(2013{\natexlab{a}}){Kirk}, {Myers}, {Bourke},
  {Gutermuth}, {Hedden}, and {Wilson}]{2013ApJ...766..115K}
Helen {Kirk}, Philip~C. {Myers}, Tyler~L. {Bourke}, Robert~A. {Gutermuth},
  Abigail {Hedden}, and Grant~W. {Wilson}.
\newblock {Filamentary Accretion Flows in the Embedded Serpens South
  Protocluster}.
\newblock \emph{The Astrophysical Journal}, 766\penalty0 (2):\penalty0 115,
  April 2013{\natexlab{a}}.
\newblock \doi{10.1088/0004-637X/766/2/115}.

\bibitem[{Kirk} et~al.(2013{\natexlab{b}}){Kirk}, {Myers}, {Bourke},
  {Gutermuth}, {Hedden}, and {Wilson}]{Kirk2013}
Helen {Kirk}, Philip~C. {Myers}, Tyler~L. {Bourke}, Robert~A. {Gutermuth},
  Abigail {Hedden}, and Grant~W. {Wilson}.
\newblock {Filamentary Accretion Flows in the Embedded Serpens South
  Protocluster}.
\newblock \emph{The Astrophysical Journal}, 766\penalty0 (2):\penalty0 115,
  April 2013{\natexlab{b}}.
\newblock \doi{10.1088/0004-637X/766/2/115}.

\bibitem[{Klessen} et~al.(2000){Klessen}, {Heitsch}, and {Mac
  Low}]{Klessen2000}
Ralf~S. {Klessen}, Fabian {Heitsch}, and Mordecai-Mark {Mac Low}.
\newblock {Gravitational Collapse in Turbulent Molecular Clouds. I.
  Gasdynamical Turbulence}.
\newblock \emph{The Astrophysical Journal}, 535\penalty0 (2):\penalty0
  887--906, June 2000.
\newblock \doi{10.1086/308891}.

\bibitem[{Koch} and {Rosolowsky}(2015)]{Koch2015}
Eric~W. {Koch} and Erik~W. {Rosolowsky}.
\newblock {Filament identification through mathematical morphology}.
\newblock \emph{Monthly Notices of the Royal Astronomical Society},
  452\penalty0 (4):\penalty0 3435--3450, October 2015.
\newblock \doi{10.1093/mnras/stv1521}.

\bibitem[{K{\"o}nyves} et~al.(2015{\natexlab{a}}){K{\"o}nyves}, {Andr{\'e}},
  {Men'shchikov}, {Palmeirim}, {Arzoumanian}, {Schneider}, {Roy}, {Didelon},
  {Maury}, {Shimajiri}, {Di Francesco}, {Bontemps}, {Peretto}, {Benedettini},
  {Bernard}, {Elia}, {Griffin}, {Hill}, {Kirk}, {Ladjelate}, {Marsh}, {Martin},
  {Motte}, {Nguy{\^e}n Luong}, {Pezzuto}, {Roussel}, {Rygl}, {Sadavoy},
  {Schisano}, {Spinoglio}, {Ward-Thompson}, and {White}]{Knyves2015}
V.~{K{\"o}nyves}, Ph. {Andr{\'e}}, A.~{Men'shchikov}, P.~{Palmeirim},
  D.~{Arzoumanian}, N.~{Schneider}, A.~{Roy}, P.~{Didelon}, A.~{Maury},
  Y.~{Shimajiri}, J.~{Di Francesco}, S.~{Bontemps}, N.~{Peretto},
  M.~{Benedettini}, J.~Ph. {Bernard}, D.~{Elia}, M.~J. {Griffin}, T.~{Hill},
  J.~{Kirk}, B.~{Ladjelate}, K.~{Marsh}, P.~G. {Martin}, F.~{Motte},
  Q.~{Nguy{\^e}n Luong}, S.~{Pezzuto}, H.~{Roussel}, K.~L.~J. {Rygl}, S.~I.
  {Sadavoy}, E.~{Schisano}, L.~{Spinoglio}, D.~{Ward-Thompson}, and G.~J.
  {White}.
\newblock {A census of dense cores in the Aquila cloud complex: SPIRE/PACS
  observations from the Herschel Gould Belt survey}.
\newblock \emph{Astronomy \& Astrophysics}, 584:\penalty0 A91, December
  2015{\natexlab{a}}.
\newblock \doi{10.1051/0004-6361/201525861}.

\bibitem[{K{\"o}nyves} et~al.(2015{\natexlab{b}}){K{\"o}nyves}, {Andr{\'e}},
  {Men'shchikov}, {Palmeirim}, {Arzoumanian}, {Schneider}, {Roy}, {Didelon},
  {Maury}, {Shimajiri}, {Di Francesco}, {Bontemps}, {Peretto}, {Benedettini},
  {Bernard}, {Elia}, {Griffin}, {Hill}, {Kirk}, {Ladjelate}, {Marsh}, {Martin},
  {Motte}, {Nguy{\^e}n Luong}, {Pezzuto}, {Roussel}, {Rygl}, {Sadavoy},
  {Schisano}, {Spinoglio}, {Ward-Thompson}, and {White}]{Konyves2015}
V.~{K{\"o}nyves}, Ph. {Andr{\'e}}, A.~{Men'shchikov}, P.~{Palmeirim},
  D.~{Arzoumanian}, N.~{Schneider}, A.~{Roy}, P.~{Didelon}, A.~{Maury},
  Y.~{Shimajiri}, J.~{Di Francesco}, S.~{Bontemps}, N.~{Peretto},
  M.~{Benedettini}, J.~Ph. {Bernard}, D.~{Elia}, M.~J. {Griffin}, T.~{Hill},
  J.~{Kirk}, B.~{Ladjelate}, K.~{Marsh}, P.~G. {Martin}, F.~{Motte},
  Q.~{Nguy{\^e}n Luong}, S.~{Pezzuto}, H.~{Roussel}, K.~L.~J. {Rygl}, S.~I.
  {Sadavoy}, E.~{Schisano}, L.~{Spinoglio}, D.~{Ward-Thompson}, and G.~J.
  {White}.
\newblock {A census of dense cores in the Aquila cloud complex: SPIRE/PACS
  observations from the Herschel Gould Belt survey}.
\newblock \emph{Astronomy \& Astrophysics}, 584:\penalty0 A91, December
  2015{\natexlab{b}}.
\newblock \doi{10.1051/0004-6361/201525861}.

\bibitem[{K{\"o}nyves} et~al.(2020){K{\"o}nyves}, {Andr{\'e}}, {Arzoumanian},
  {Schneider}, {Men'shchikov}, {Bontemps}, {Ladjelate}, {Didelon}, {Pezzuto},
  {Benedettini}, {Bracco}, {Di Francesco}, {Goodwin}, {Rygl}, {Shimajiri},
  {Spinoglio}, {Ward-Thompson}, and {White}]{Konyves2020}
V.~{K{\"o}nyves}, Ph. {Andr{\'e}}, D.~{Arzoumanian}, N.~{Schneider},
  A.~{Men'shchikov}, S.~{Bontemps}, B.~{Ladjelate}, P.~{Didelon}, S.~{Pezzuto},
  M.~{Benedettini}, A.~{Bracco}, J.~{Di Francesco}, S.~{Goodwin}, K.~L.~J.
  {Rygl}, Y.~{Shimajiri}, L.~{Spinoglio}, D.~{Ward-Thompson}, and G.~J.
  {White}.
\newblock {Properties of the dense core population in Orion B as seen by the
  Herschel Gould Belt survey}.
\newblock \emph{Astronomy \& Astrophysics}, 635:\penalty0 A34, March 2020.
\newblock \doi{10.1051/0004-6361/201834753}.

\bibitem[{Liu} et~al.(2020{\natexlab{a}}){Liu}, {Sanhueza}, {Liu}, {Zavagno},
  {Tang}, {Wu}, and {Zhang}]{Liu2020ApJ}
Hong-Li {Liu}, Patricio {Sanhueza}, Tie {Liu}, Annie {Zavagno}, Xin-Di {Tang},
  Yuefang {Wu}, and Siju {Zhang}.
\newblock {Chemistry of Protostellar Clumps in the High-mass, Star-forming
  Filamentary Infrared Dark Cloud G034.43+00.24}.
\newblock \emph{The Astrophysical Journal}, 901\penalty0 (1):\penalty0 31,
  September 2020{\natexlab{a}}.
\newblock \doi{10.3847/1538-4357/abadfe}.

\bibitem[{Liu} et~al.(2021){Liu}, {Liu}, {Evans}, {Wang}, {Garay}, {Qin}, {Li},
  {Stutz}, {Goldsmith}, {Liu}, {Tej}, {Zhang}, {Juvela}, {Li}, {Wang},
  {Bronfman}, {Ren}, {Wu}, {Kim}, {Lee}, {Tatematsu}, {Cunningham}, {Liu},
  {Wu}, {Hirota}, {Lee}, {Li}, {Kang}, {Mardones}, {Ristorcelli}, {Zhang},
  {Luo}, {Toth}, {Yi}, {Yun}, {Peng}, {Li}, {Zhu}, {Shen}, {Baug}, {Dewangan},
  {Chakali}, {Liu}, {Xu}, {Wang}, {Zhang}, {Li}, {Zhang}, {Zhou}, {Tang},
  {Xue}, {Issac}, {Soam}, and {{\'A}lvarez-Guti{\'e}rrez}]{Liu2021}
Hong-Li {Liu}, Tie {Liu}, Neal~J. {Evans}, II, Ke~{Wang}, Guido {Garay},
  Sheng-Li {Qin}, Shanghuo {Li}, Amelia {Stutz}, Paul~F. {Goldsmith},
  Sheng-Yuan {Liu}, Anandmayee {Tej}, Qizhou {Zhang}, Mika {Juvela}, Di~{Li},
  Jun-Zhi {Wang}, Leonardo {Bronfman}, Zhiyuan {Ren}, Yue-Fang {Wu}, Kee-Tae
  {Kim}, Chang~Won {Lee}, Ken'ichi {Tatematsu}, Maria~R. {Cunningham},
  Xun-Chuan {Liu}, Jing-Wen {Wu}, Tomoya {Hirota}, Jeong-Eun {Lee}, Pak-Shing
  {Li}, Sung-Ju {Kang}, Diego {Mardones}, Isabelle {Ristorcelli}, Yong {Zhang},
  Qiu-Yi {Luo}, L.~Viktor {Toth}, Hee-weon {Yi}, Hyeong-Sik {Yun}, Ya-Ping
  {Peng}, Juan {Li}, Feng-Yao {Zhu}, Zhi-Qiang {Shen}, Tapas {Baug}, L.~K.
  {Dewangan}, Eswaraiah {Chakali}, Rong {Liu}, Feng-Wei {Xu}, Yu~{Wang}, Chao
  {Zhang}, Jinzeng {Li}, Chao {Zhang}, Jianwen {Zhou}, Mengyao {Tang}, Qiaowei
  {Xue}, Namitha {Issac}, Archana {Soam}, and Rodrigo~H.
  {{\'A}lvarez-Guti{\'e}rrez}.
\newblock {ATOMS: ALMA three-millimeter observations of massive star-forming
  regions - III. Catalogues of candidate hot molecular cores and hyper/ultra
  compact H II regions}.
\newblock \emph{Monthly Notices of the Royal Astronomical Society},
  505\penalty0 (2):\penalty0 2801--2818, August 2021.
\newblock \doi{10.1093/mnras/stab1352}.

\bibitem[{Liu} et~al.(2016){Liu}, {Kim}, {Yoo}, {Liu}, {Tatematsu}, {Qin},
  {Zhang}, {Wu}, {Wang}, {Goldsmith}, {Juvela}, {Lee}, {T{\'o}th}, {Mardones},
  {Garay}, {Bronfman}, {Cunningham}, {Li}, {Lo}, {Ristorcelli}, and
  {Schnee}]{Liu16}
Tie {Liu}, Kee-Tae {Kim}, Hyunju {Yoo}, Sheng-yuan {Liu}, Ken'ichi {Tatematsu},
  Sheng-Li {Qin}, Qizhou {Zhang}, Yuefang {Wu}, Ke~{Wang}, Paul~F. {Goldsmith},
  Mika {Juvela}, Jeong-Eun {Lee}, L.~Viktor {T{\'o}th}, Diego {Mardones}, Guido
  {Garay}, Leonardo {Bronfman}, Maria~R. {Cunningham}, Di~{Li}, Nadia {Lo},
  Isabelle {Ristorcelli}, and Scott {Schnee}.
\newblock {Star Formation Laws in Both Galactic Massive Clumps and External
  Galaxies: Extensive Study with Dust Coninuum, HCN (4-3), and CS (7-6)}.
\newblock \emph{The Astrophysical Journal}, 829\penalty0 (2):\penalty0 59,
  October 2016.
\newblock \doi{10.3847/0004-637X/829/2/59}.

\bibitem[{Liu} et~al.(2020{\natexlab{b}}){Liu}, {Evans}, {Kim}, {Goldsmith},
  {Liu}, {Zhang}, {Tatematsu}, {Wang}, {Juvela}, {Bronfman}, {Cunningham},
  {Garay}, {Hirota}, {Lee}, {Kang}, {Li}, {Li}, {Mardones}, {Qin},
  {Ristorcelli}, {Tej}, {Toth}, {Wu}, {Wu}, {Yi}, {Yun}, {Liu}, {Peng}, {Li},
  {Li}, {Lee}, {Shen}, {Baug}, {Wang}, {Zhang}, {Issac}, {Zhu}, {Luo}, {Soam},
  {Liu}, {Xu}, {Wang}, {Zhang}, {Ren}, and {Zhang}]{Liu20_1}
Tie {Liu}, Neal~J. {Evans}, Kee-Tae {Kim}, Paul~F. {Goldsmith}, Sheng-Yuan
  {Liu}, Qizhou {Zhang}, Ken'ichi {Tatematsu}, Ke~{Wang}, Mika {Juvela},
  Leonardo {Bronfman}, Maria~R. {Cunningham}, Guido {Garay}, Tomoya {Hirota},
  Jeong-Eun {Lee}, Sung-Ju {Kang}, Di~{Li}, Pak-Shing {Li}, Diego {Mardones},
  Sheng-Li {Qin}, Isabelle {Ristorcelli}, Anandmayee {Tej}, L.~Viktor {Toth},
  Jing-Wen {Wu}, Yue-Fang {Wu}, Hee-weon {Yi}, Hyeong-Sik {Yun}, Hong-Li {Liu},
  Ya-Ping {Peng}, Juan {Li}, Shang-Huo {Li}, Chang~Won {Lee}, Zhi-Qiang {Shen},
  Tapas {Baug}, Jun-Zhi {Wang}, Yong {Zhang}, Namitha {Issac}, Feng-Yao {Zhu},
  Qiu-Yi {Luo}, Archana {Soam}, Xun-Chuan {Liu}, Feng-Wei {Xu}, Yu~{Wang}, Chao
  {Zhang}, Zhiyuan {Ren}, and Chao {Zhang}.
\newblock {ATOMS: ALMA Three-millimeter Observations of Massive Star-forming
  regions - I. Survey description and a first look at G9.62+0.19}.
\newblock \emph{Monthly Notices of the Royal Astronomical Society},
  496\penalty0 (3):\penalty0 2790--2820, June 2020{\natexlab{b}}.
\newblock \doi{10.1093/mnras/staa1577}.

\bibitem[{Lu} et~al.(2018){Lu}, {Zhang}, {Liu}, {Sanhueza}, {Tatematsu},
  {Feng}, {Smith}, {Myers}, {Sridharan}, and {Gu}]{Lu2018}
Xing {Lu}, Qizhou {Zhang}, Hauyu~Baobab {Liu}, Patricio {Sanhueza}, Ken'ichi
  {Tatematsu}, Siyi {Feng}, Howard~A. {Smith}, Philip~C. {Myers}, T.~K.
  {Sridharan}, and Qiusheng {Gu}.
\newblock {Filamentary Fragmentation and Accretion in High-mass Star-forming
  Molecular Clouds}.
\newblock \emph{The Astrophysical Journal}, 855\penalty0 (1):\penalty0 9, March
  2018.
\newblock \doi{10.3847/1538-4357/aaad11}.

\bibitem[{McMullin} et~al.(2007){McMullin}, {Waters}, {Schiebel}, {Young}, and
  {Golap}]{McMullin07}
J.~P. {McMullin}, B.~{Waters}, D.~{Schiebel}, W.~{Young}, and K.~{Golap}.
\newblock {CASA Architecture and Applications}.
\newblock In R.~A. {Shaw}, F.~{Hill}, and D.~J. {Bell}, editors,
  \emph{Astronomical Data Analysis Software and Systems XVI}, volume 376 of
  \emph{Astronomical Society of the Pacific Conference Series}, page 127,
  October 2007.

\bibitem[{Men'shchikov} et~al.(2010){Men'shchikov}, {Andr{\'e}}, {Didelon},
  {K{\"o}nyves}, {Schneider}, {Motte}, {Bontemps}, {Arzoumanian}, {Attard},
  {Abergel}, {Baluteau}, {Bernard}, {Cambr{\'e}sy}, {Cox}, {di Francesco}, {di
  Giorgio}, {Griffin}, {Hargrave}, {Huang}, {Kirk}, {Li}, {Martin}, {Minier},
  {Miville-Desch{\^e}nes}, {Molinari}, {Olofsson}, {Pezzuto}, {Roussel},
  {Russeil}, {Saraceno}, {Sauvage}, {Sibthorpe}, {Spinoglio}, {Testi},
  {Ward-Thompson}, {White}, {Wilson}, {Woodcraft}, and {Zavagno}]{Men2010}
A.~{Men'shchikov}, Ph. {Andr{\'e}}, P.~{Didelon}, V.~{K{\"o}nyves},
  N.~{Schneider}, F.~{Motte}, S.~{Bontemps}, D.~{Arzoumanian}, M.~{Attard},
  A.~{Abergel}, J.~P. {Baluteau}, J.~Ph. {Bernard}, L.~{Cambr{\'e}sy},
  P.~{Cox}, J.~{di Francesco}, A.~M. {di Giorgio}, M.~{Griffin}, P.~{Hargrave},
  M.~{Huang}, J.~{Kirk}, J.~Z. {Li}, P.~{Martin}, V.~{Minier}, M.~A.
  {Miville-Desch{\^e}nes}, S.~{Molinari}, G.~{Olofsson}, S.~{Pezzuto},
  H.~{Roussel}, D.~{Russeil}, P.~{Saraceno}, M.~{Sauvage}, B.~{Sibthorpe},
  L.~{Spinoglio}, L.~{Testi}, D.~{Ward-Thompson}, G.~{White}, C.~D. {Wilson},
  A.~{Woodcraft}, and A.~{Zavagno}.
\newblock {Filamentary structures and compact objects in the Aquila and Polaris
  clouds observed by Herschel}.
\newblock \emph{Astronomy \& Astrophysics}, 518:\penalty0 L103, July 2010.
\newblock \doi{10.1051/0004-6361/201014668}.

\bibitem[{Myers}(2009)]{Myers2009}
Philip~C. {Myers}.
\newblock {Filamentary Structure of Star-forming Complexes}.
\newblock \emph{The Astrophysical Journal}, 700\penalty0 (2):\penalty0
  1609--1625, August 2009.
\newblock \doi{10.1088/0004-637X/700/2/1609}.

\bibitem[{Ostriker}(1964)]{Ostriker1964}
J.~{Ostriker}.
\newblock {The Equilibrium of Polytropic and Isothermal Cylinders.}
\newblock \emph{The Astrophysical Journal}, 140:\penalty0 1056, October 1964.
\newblock \doi{10.1086/148005}.

\bibitem[{Padoan}(1995)]{Padoan1995}
Paolo {Padoan}.
\newblock {Supersonic turbulent flows and the fragmentation of a cold medium}.
\newblock \emph{Monthly Notices of the Royal Astronomical Society},
  277\penalty0 (2):\penalty0 377--388, November 1995.
\newblock \doi{10.1093/mnras/277.2.377}.

\bibitem[{Padoan} and {Nordlund}(1999)]{Padoan1999}
Paolo {Padoan} and {\r{A}}ke {Nordlund}.
\newblock {A Super-Alfv{\'e}nic Model of Dark Clouds}.
\newblock \emph{The Astrophysical Journal}, 526\penalty0 (1):\penalty0
  279--294, November 1999.
\newblock \doi{10.1086/307956}.

\bibitem[{Padoan} and {Nordlund}(2002)]{Padoan2002}
Paolo {Padoan} and {\r{A}}ke {Nordlund}.
\newblock {The Stellar Initial Mass Function from Turbulent Fragmentation}.
\newblock \emph{The Astrophysical Journal}, 576\penalty0 (2):\penalty0
  870--879, September 2002.
\newblock \doi{10.1086/341790}.

\bibitem[{Padoan} et~al.(2020){Padoan}, {Pan}, {Juvela}, {Haugb{\o}lle}, and
  {Nordlund}]{Padoan2020}
Paolo {Padoan}, Liubin {Pan}, Mika {Juvela}, Troels {Haugb{\o}lle}, and
  {\r{A}}ke {Nordlund}.
\newblock {The Origin of Massive Stars: The Inertial-inflow Model}.
\newblock \emph{The Astrophysical Journal}, 900\penalty0 (1):\penalty0 82,
  September 2020.
\newblock \doi{10.3847/1538-4357/abaa47}.

\bibitem[{Palmeirim} et~al.(2013){Palmeirim}, {Andr{\'e}}, {Kirk},
  {Ward-Thompson}, {Arzoumanian}, {K{\"o}nyves}, {Didelon}, {Schneider},
  {Benedettini}, {Bontemps}, {Di Francesco}, {Elia}, {Griffin}, {Hennemann},
  {Hill}, {Martin}, {Men'shchikov}, {Molinari}, {Motte}, {Nguyen Luong},
  {Nutter}, {Peretto}, {Pezzuto}, {Roy}, {Rygl}, {Spinoglio}, and
  {White}]{Palmeirim2013}
P.~{Palmeirim}, Ph. {Andr{\'e}}, J.~{Kirk}, D.~{Ward-Thompson},
  D.~{Arzoumanian}, V.~{K{\"o}nyves}, P.~{Didelon}, N.~{Schneider},
  M.~{Benedettini}, S.~{Bontemps}, J.~{Di Francesco}, D.~{Elia}, M.~{Griffin},
  M.~{Hennemann}, T.~{Hill}, P.~G. {Martin}, A.~{Men'shchikov}, S.~{Molinari},
  F.~{Motte}, Q.~{Nguyen Luong}, D.~{Nutter}, N.~{Peretto}, S.~{Pezzuto},
  A.~{Roy}, K.~L.~J. {Rygl}, L.~{Spinoglio}, and G.~L. {White}.
\newblock {Herschel view of the Taurus B211/3 filament and striations: evidence
  of filamentary growth?}
\newblock \emph{Astronomy \& Astrophysics}, 550:\penalty0 A38, February 2013.
\newblock \doi{10.1051/0004-6361/201220500}.

\bibitem[{Peretto} et~al.(2014){Peretto}, {Fuller}, {Andr{\'e}}, {Arzoumanian},
  {Rivilla}, {Bardeau}, {Duarte Puertas}, {Guzman Fernandez}, {Lenfestey},
  {Li}, {Olguin}, {R{\"o}ck}, {de Villiers}, and {Williams}]{Peretto2014}
N.~{Peretto}, G.~A. {Fuller}, Ph. {Andr{\'e}}, D.~{Arzoumanian}, V.~M.
  {Rivilla}, S.~{Bardeau}, S.~{Duarte Puertas}, J.~P. {Guzman Fernandez},
  C.~{Lenfestey}, G.~X. {Li}, F.~A. {Olguin}, B.~R. {R{\"o}ck}, H.~{de
  Villiers}, and J.~{Williams}.
\newblock {SDC13 infrared dark clouds: Longitudinally collapsing filaments?}
\newblock \emph{Astronomy \& Astrophysics}, 561:\penalty0 A83, January 2014.
\newblock \doi{10.1051/0004-6361/201322172}.

\bibitem[{Plume} et~al.(1997){Plume}, {Jaffe}, {Evans}, {Mart{\'\i}n-Pintado},
  and {G{\'o}mez-Gonz{\'a}lez}]{Plume97}
Ren{\'e} {Plume}, D.~T. {Jaffe}, Neal~J. {Evans}, II, J.~{Mart{\'\i}n-Pintado},
  and J.~{G{\'o}mez-Gonz{\'a}lez}.
\newblock {Dense Gas and Star Formation: Characteristics of Cloud Cores
  Associated with Water Masers}.
\newblock \emph{The Astrophysical Journal}, 476\penalty0 (2):\penalty0
  730--749, February 1997.
\newblock \doi{10.1086/303654}.

\bibitem[{Rayner} et~al.(2017){Rayner}, {Griffin}, {Schneider}, {Motte},
  {K{\"o}nyves}, {Andr{\'e}}, {Di Francesco}, {Didelon}, {Pattle},
  {Ward-Thompson}, {Anderson}, {Benedettini}, {Bernard}, {Bontemps}, {Elia},
  {Fuente}, {Hennemann}, {Hill}, {Kirk}, {Marsh}, {Men'shchikov}, {Nguyen
  Luong}, {Peretto}, {Pezzuto}, {Rivera-Ingraham}, {Roy}, {Rygl},
  {S{\'a}nchez-Monge}, {Spinoglio}, {Tig{\'e}}, {Trevi{\~n}o-Morales}, and
  {White}]{Rayner2017}
T.~S.~M. {Rayner}, M.~J. {Griffin}, N.~{Schneider}, F.~{Motte},
  V.~{K{\"o}nyves}, P.~{Andr{\'e}}, J.~{Di Francesco}, P.~{Didelon},
  K.~{Pattle}, D.~{Ward-Thompson}, L.~D. {Anderson}, M.~{Benedettini}, J.~P.
  {Bernard}, S.~{Bontemps}, D.~{Elia}, A.~{Fuente}, M.~{Hennemann}, T.~{Hill},
  J.~{Kirk}, K.~{Marsh}, A.~{Men'shchikov}, Q.~{Nguyen Luong}, N.~{Peretto},
  S.~{Pezzuto}, A.~{Rivera-Ingraham}, A.~{Roy}, K.~{Rygl},
  {\'A}.~{S{\'a}nchez-Monge}, L.~{Spinoglio}, J.~{Tig{\'e}}, S.~P.
  {Trevi{\~n}o-Morales}, and G.~J. {White}.
\newblock {Far-infrared observations of a massive cluster forming in the
  Monoceros R2 filament hub}.
\newblock \emph{Astronomy \& Astrophysics}, 607:\penalty0 A22, October 2017.
\newblock \doi{10.1051/0004-6361/201630039}.

\bibitem[{Rosolowsky} et~al.(2008){Rosolowsky}, {Pineda}, {Foster}, {Borkin},
  {Kauffmann}, {Caselli}, {Myers}, and {Goodman}]{Rosolowsky2008}
E.~W. {Rosolowsky}, J.~E. {Pineda}, J.~B. {Foster}, M.~A. {Borkin},
  J.~{Kauffmann}, P.~{Caselli}, P.~C. {Myers}, and A.~A. {Goodman}.
\newblock {An Ammonia Spectral Atlas of Dense Cores in Perseus}.
\newblock \emph{The Astrophysical Journal Supplement Series}, 175\penalty0
  (2):\penalty0 509--521, April 2008.
\newblock \doi{10.1086/524299}.

\bibitem[{Salpeter}(1955)]{Salpeter1955}
Edwin~E. {Salpeter}.
\newblock {The Luminosity Function and Stellar Evolution.}
\newblock \emph{The Astrophysical Journal}, 121:\penalty0 161, January 1955.
\newblock \doi{10.1086/145971}.

\bibitem[{Sanhueza} et~al.(2012){Sanhueza}, {Jackson}, {Foster}, {Garay},
  {Silva}, and {Finn}]{Sanhueza2012}
Patricio {Sanhueza}, James~M. {Jackson}, Jonathan~B. {Foster}, Guido {Garay},
  Andrea {Silva}, and Susanna~C. {Finn}.
\newblock {Chemistry in Infrared Dark Cloud Clumps: A Molecular Line Survey at
  3 mm}.
\newblock \emph{The Astrophysical Journal}, 756\penalty0 (1):\penalty0 60,
  September 2012.
\newblock \doi{10.1088/0004-637X/756/1/60}.

\bibitem[{Sanhueza} et~al.(2021){Sanhueza}, {Girart}, {Padovani}, {Galli},
  {Hull}, {Zhang}, {Cortes}, {Stephens}, {Fern{\'a}ndez-L{\'o}pez}, {Jackson},
  {Frau}, {Kock}, {Wu}, {Zapata}, {Olguin}, {Lu}, {Silva}, {Tang}, {Sakai},
  {Guzm{\'a}n}, {Tatematsu}, {Nakamura}, and {Chen}]{Sanhueza2021}
Patricio {Sanhueza}, Josep~Miquel {Girart}, Marco {Padovani}, Daniele {Galli},
  Charles L.~H. {Hull}, Qizhou {Zhang}, Paulo {Cortes}, Ian~W. {Stephens},
  Manuel {Fern{\'a}ndez-L{\'o}pez}, James~M. {Jackson}, Pau {Frau}, Patrick~M.
  {Kock}, Benjamin {Wu}, Luis~A. {Zapata}, Fernando {Olguin}, Xing {Lu}, Andrea
  {Silva}, Ya-Wen {Tang}, Takeshi {Sakai}, Andr{\'e}s~E. {Guzm{\'a}n}, Ken'ichi
  {Tatematsu}, Fumitaka {Nakamura}, and Huei-Ru~Vivien {Chen}.
\newblock {Gravity-driven Magnetic Field at 1000 au Scales in High-mass Star
  Formation}.
\newblock \emph{The Astrophysical Journal Letters}, 915\penalty0 (1):\penalty0
  L10, July 2021.
\newblock \doi{10.3847/2041-8213/ac081c}.

\bibitem[{Schisano} et~al.(2014){Schisano}, {Rygl}, {Molinari}, {Busquet},
  {Elia}, {Pestalozzi}, {Polychroni}, {Billot}, {Carey}, {Paladini},
  {Noriega-Crespo}, {Moore}, {Plume}, {Glover}, and
  {V{\'a}zquez-Semadeni}]{Schisano2014}
E.~{Schisano}, K.~L.~J. {Rygl}, S.~{Molinari}, G.~{Busquet}, D.~{Elia},
  M.~{Pestalozzi}, D.~{Polychroni}, N.~{Billot}, S.~{Carey}, R.~{Paladini},
  A.~{Noriega-Crespo}, T.~J.~T. {Moore}, R.~{Plume}, S.~C.~O. {Glover}, and
  E.~{V{\'a}zquez-Semadeni}.
\newblock {The Identification of Filaments on Far-infrared and Submillimiter
  Images: Morphology, Physical Conditions and Relation with Star Formation of
  Filamentary Structure}.
\newblock \emph{The Astrophysical Journal}, 791\penalty0 (1):\penalty0 27,
  August 2014.
\newblock \doi{10.1088/0004-637X/791/1/27}.

\bibitem[{Schneider} et~al.(2010){Schneider}, {Csengeri}, {Bontemps}, {Motte},
  {Simon}, {Hennebelle}, {Federrath}, and {Klessen}]{Schneider2010}
N.~{Schneider}, T.~{Csengeri}, S.~{Bontemps}, F.~{Motte}, R.~{Simon},
  P.~{Hennebelle}, C.~{Federrath}, and R.~{Klessen}.
\newblock {Dynamic star formation in the massive DR21 filament}.
\newblock \emph{Astronomy \& Astrophysics}, 520:\penalty0 A49, September 2010.
\newblock \doi{10.1051/0004-6361/201014481}.

\bibitem[{Seifried} et~al.(2020){Seifried}, {Walch}, {Weis}, {Reissl}, {Soler},
  {Klessen}, and {Joshi}]{seifried2020}
D.~{Seifried}, S.~{Walch}, M.~{Weis}, S.~{Reissl}, J.~D. {Soler}, R.~S.
  {Klessen}, and P.~R. {Joshi}.
\newblock {From parallel to perpendicular - On the orientation of magnetic
  fields in molecular clouds}.
\newblock \emph{Monthly Notices of the Royal Astronomical Society},
  497\penalty0 (4):\penalty0 4196--4212, October 2020.
\newblock \doi{10.1093/mnras/staa2231}.

\bibitem[{Shen} et~al.(2024){Shen}, {Liu}, {Ren}, {Tej}, {Li}, {Liu}, {Fuller},
  {Xie}, {Jiao}, {Yang}, {Koch}, {Xu}, {Sanhueza}, {Diep}, {Peretto}, {Yadav},
  {Kramer}, {Sugiyama}, {Rawlings}, {Lee}, {Tatematsu}, {Harsono}, {Eden},
  {Kwon}, {Tsai}, {White}, {Kim}, {Liu}, {Wang}, {Zhang}, {Jiao}, {Yang},
  {Das}, {Wu}, and {Wang}]{Shen2024}
Xianjin {Shen}, Hong-Li {Liu}, Zhiyuan {Ren}, Anandmayee {Tej}, Di~{Li},
  Hauyu~Baobab {Liu}, Gary~A. {Fuller}, Jinjin {Xie}, Sihan {Jiao}, Aiyuan
  {Yang}, Patrick~M. {Koch}, Fengwei {Xu}, Patricio {Sanhueza}, Pham~Ngoc
  {Diep}, Nicolas {Peretto}, R.~K. {Yadav}, Busaba~H. {Kramer}, Koichiro
  {Sugiyama}, Mark~G. {Rawlings}, Chang~Won {Lee}, Ken'ichi {Tatematsu}, Daniel
  {Harsono}, David {Eden}, Woojin {Kwon}, Chao-Wei {Tsai}, Glenn~J. {White},
  Kee-Tae {Kim}, Tie {Liu}, Ke~{Wang}, Siju {Zhang}, Wenyu {Jiao}, Dongting
  {Yang}, Swagat~R. {Das}, Jingwen {Wu}, and Chen {Wang}.
\newblock {JCMT 850 {\ensuremath{\mu}}m Continuum Observations of Density
  Structures in the G35 Molecular Complex}.
\newblock \emph{The Astrophysical Journal}, 974\penalty0 (2):\penalty0 239,
  October 2024.
\newblock \doi{10.3847/1538-4357/ad6a5f}.

\bibitem[{Shimajiri} et~al.(2017){Shimajiri}, {Andr{\'e}}, {Braine},
  {K{\"o}nyves}, {Schneider}, {Bontemps}, {Ladjelate}, {Roy}, {Gao}, and
  {Chen}]{Shimajiri2017}
Y.~{Shimajiri}, Ph. {Andr{\'e}}, J.~{Braine}, V.~{K{\"o}nyves}, N.~{Schneider},
  S.~{Bontemps}, B.~{Ladjelate}, A.~{Roy}, Y.~{Gao}, and H.~{Chen}.
\newblock {Testing the universality of the star-formation efficiency in dense
  molecular gas}.
\newblock \emph{Astronomy \& Astrophysics}, 604:\penalty0 A74, August 2017.
\newblock \doi{10.1051/0004-6361/201730633}.

\bibitem[{Shimajiri} et~al.(2019){Shimajiri}, {Andr{\'e}}, {Palmeirim},
  {Arzoumanian}, {Bracco}, {K{\"o}nyves}, {Ntormousi}, and
  {Ladjelate}]{Shimajiri2019}
Y.~{Shimajiri}, Ph. {Andr{\'e}}, P.~{Palmeirim}, D.~{Arzoumanian}, A.~{Bracco},
  V.~{K{\"o}nyves}, E.~{Ntormousi}, and B.~{Ladjelate}.
\newblock {Probing accretion of ambient cloud material into the Taurus
  B211/B213 filament}.
\newblock \emph{Astronomy \& Astrophysics}, 623:\penalty0 A16, March 2019.
\newblock \doi{10.1051/0004-6361/201834399}.

\bibitem[{Shirley}(2015)]{Shirley2015}
Yancy~L. {Shirley}.
\newblock {The Critical Density and the Effective Excitation Density of
  Commonly Observed Molecular Dense Gas Tracers}.
\newblock \emph{Publications of the Astronomical Society of the Pacific},
  127\penalty0 (949):\penalty0 299, March 2015.
\newblock \doi{10.1086/680342}.

\bibitem[{Smith} et~al.(2016){Smith}, {Glover}, {Klessen}, and
  {Fuller}]{smith2016}
Rowan~J. {Smith}, Simon C.~O. {Glover}, Ralf~S. {Klessen}, and Gary~A.
  {Fuller}.
\newblock {On the nature of star-forming filaments - II. Subfilaments and
  velocities}.
\newblock \emph{Monthly Notices of the Royal Astronomical Society},
  455\penalty0 (4):\penalty0 3640--3655, February 2016.
\newblock \doi{10.1093/mnras/stv2559}.

\bibitem[{Soler} et~al.(2017){Soler}, {Ade}, {Angil{\`e}}, {Ashton}, {Benton},
  {Devlin}, {Dober}, {Fissel}, {Fukui}, {Galitzki}, {Gandilo}, {Hennebelle},
  {Klein}, {Li}, {Korotkov}, {Martin}, {Matthews}, {Moncelsi}, {Netterfield},
  {Novak}, {Pascale}, {Poidevin}, {Santos}, {Savini}, {Scott}, {Shariff},
  {Thomas}, {Tucker}, {Tucker}, and {Ward-Thompson}]{Soler2017}
J.~D. {Soler}, P.~A.~R. {Ade}, F.~E. {Angil{\`e}}, P.~{Ashton}, S.~J. {Benton},
  M.~J. {Devlin}, B.~{Dober}, L.~M. {Fissel}, Y.~{Fukui}, N.~{Galitzki}, N.~N.
  {Gandilo}, P.~{Hennebelle}, J.~{Klein}, Z.~Y. {Li}, A.~L. {Korotkov}, P.~G.
  {Martin}, T.~G. {Matthews}, L.~{Moncelsi}, C.~B. {Netterfield}, G.~{Novak},
  E.~{Pascale}, F.~{Poidevin}, F.~P. {Santos}, G.~{Savini}, D.~{Scott}, J.~A.
  {Shariff}, N.~E. {Thomas}, C.~E. {Tucker}, G.~S. {Tucker}, and
  D.~{Ward-Thompson}.
\newblock {The relation between the column density structures and the magnetic
  field orientation in the Vela C molecular complex}.
\newblock \emph{Astronomy \& Astrophysics}, 603:\penalty0 A64, July 2017.
\newblock \doi{10.1051/0004-6361/201730608}.

\bibitem[{Sousbie}(2011)]{Sousbie2011}
T.~{Sousbie}.
\newblock {The persistent cosmic web and its filamentary structure - I. Theory
  and implementation}.
\newblock \emph{Monthly Notices of the Royal Astronomical Society},
  414\penalty0 (1):\penalty0 350--383, June 2011.
\newblock \doi{10.1111/j.1365-2966.2011.18394.x}.

\bibitem[{Sousbie} et~al.(2011){Sousbie}, {Pichon}, and
  {Kawahara}]{Sousbie2011MNRAS.414..384S}
T.~{Sousbie}, C.~{Pichon}, and H.~{Kawahara}.
\newblock {The persistent cosmic web and its filamentary structure - II.
  Illustrations}.
\newblock \emph{Monthly Notices of the Royal Astronomical Society},
  414\penalty0 (1):\penalty0 384--403, June 2011.
\newblock \doi{10.1111/j.1365-2966.2011.18395.x}.

\bibitem[{Stephens} et~al.(2017){Stephens}, {Dunham}, {Myers}, {Pokhrel},
  {Sadavoy}, {Vorobyov}, {Tobin}, {Pineda}, {Offner}, {Lee}, {Kristensen},
  {J{\o}rgensen}, {Goodman}, {Bourke}, {Arce}, and {Plunkett}]{stephens2017}
Ian~W. {Stephens}, Michael~M. {Dunham}, Philip~C. {Myers}, Riwaj {Pokhrel},
  Sarah~I. {Sadavoy}, Eduard~I. {Vorobyov}, John~J. {Tobin}, Jaime~E. {Pineda},
  Stella S.~R. {Offner}, Katherine~I. {Lee}, Lars~E. {Kristensen}, Jes~K.
  {J{\o}rgensen}, Alyssa~A. {Goodman}, Tyler~L. {Bourke}, H{\'e}ctor~G. {Arce},
  and Adele~L. {Plunkett}.
\newblock {Alignment between Protostellar Outflows and Filamentary Structure}.
\newblock \emph{The Astrophysical Journal}, 846\penalty0 (1):\penalty0 16,
  September 2017.
\newblock \doi{10.3847/1538-4357/aa8262}.

\bibitem[{Stod{\'o}lkiewicz}(1963)]{Stod1963}
J.~S. {Stod{\'o}lkiewicz}.
\newblock {On the Gravitational Instability of Some Magneto-Hydrodynamical
  Systems of Astrophysical Interest. Part III.}
\newblock \emph{Acta Astronomica}, 13:\penalty0 30--54, January 1963.

\bibitem[{Sugitani} et~al.(2011){Sugitani}, {Nakamura}, {Watanabe}, {Tamura},
  {Nishiyama}, {Nagayama}, {Kandori}, {Nagata}, {Sato}, {Gutermuth}, {Wilson},
  and {Kawabe}]{Sugitani2011}
K.~{Sugitani}, F.~{Nakamura}, M.~{Watanabe}, M.~{Tamura}, S.~{Nishiyama},
  T.~{Nagayama}, R.~{Kandori}, T.~{Nagata}, S.~{Sato}, R.~A. {Gutermuth}, G.~W.
  {Wilson}, and R.~{Kawabe}.
\newblock {Near-infrared-imaging Polarimetry Toward Serpens South: Revealing
  the Importance of the Magnetic Field}.
\newblock \emph{The Astrophysical Journal}, 734\penalty0 (1):\penalty0 63, June
  2011.
\newblock \doi{10.1088/0004-637X/734/1/63}.

\bibitem[{Tackenberg} et~al.(2014){Tackenberg}, {Beuther}, {Henning}, {Linz},
  {Sakai}, {Ragan}, {Krause}, {Nielbock}, {Hennemann}, {Pitann}, and
  {Schmiedeke}]{Tackenberg2014}
J.~{Tackenberg}, H.~{Beuther}, Th. {Henning}, H.~{Linz}, T.~{Sakai}, S.~E.
  {Ragan}, O.~{Krause}, M.~{Nielbock}, M.~{Hennemann}, J.~{Pitann}, and
  A.~{Schmiedeke}.
\newblock {Kinematic structure of massive star-forming regions. I. Accretion
  along filaments}.
\newblock \emph{Astronomy \& Astrophysics}, 565:\penalty0 A101, May 2014.
\newblock \doi{10.1051/0004-6361/201321555}.

\bibitem[{Tang} et~al.(2019{\natexlab{a}}){Tang}, {Koch}, {Peretto}, {Novak},
  {Duarte-Cabral}, {Chapman}, {Hsieh}, and {Yen}]{2019ApJ...878...10T}
Ya-Wen {Tang}, Patrick~M. {Koch}, Nicolas {Peretto}, Giles {Novak}, Ana
  {Duarte-Cabral}, Nicholas~L. {Chapman}, Pei-Ying {Hsieh}, and Hsi-Wei {Yen}.
\newblock {Gravity, Magnetic Field, and Turbulence: Relative Importance and
  Impact on Fragmentation in the Infrared Dark Cloud G34.43+00.24}.
\newblock \emph{The Astrophysical Journal}, 878\penalty0 (1):\penalty0 10, June
  2019{\natexlab{a}}.
\newblock \doi{10.3847/1538-4357/ab1484}.

\bibitem[{Tang} et~al.(2019{\natexlab{b}}){Tang}, {Koch}, {Peretto}, {Novak},
  {Duarte-Cabral}, {Chapman}, {Hsieh}, and {Yen}]{Tang2019}
Ya-Wen {Tang}, Patrick~M. {Koch}, Nicolas {Peretto}, Giles {Novak}, Ana
  {Duarte-Cabral}, Nicholas~L. {Chapman}, Pei-Ying {Hsieh}, and Hsi-Wei {Yen}.
\newblock {Gravity, Magnetic Field, and Turbulence: Relative Importance and
  Impact on Fragmentation in the Infrared Dark Cloud G34.43+00.24}.
\newblock \emph{The Astrophysical Journal}, 878\penalty0 (1):\penalty0 10, June
  2019{\natexlab{b}}.
\newblock \doi{10.3847/1538-4357/ab1484}.

\bibitem[{Trevi{\~n}o-Morales} et~al.(2019){Trevi{\~n}o-Morales}, {Fuente},
  {S{\'a}nchez-Monge}, {Kainulainen}, {Didelon}, {Suri}, {Schneider},
  {Ballesteros-Paredes}, {Lee}, {Hennebelle}, {Pilleri},
  {Gonz{\'a}lez-Garc{\'\i}a}, {Kramer}, {Garc{\'\i}a-Burillo}, {Luna},
  {Goicoechea}, {Tremblin}, and {Geen}]{Trevi2019}
S.~P. {Trevi{\~n}o-Morales}, A.~{Fuente}, {\'A}.~{S{\'a}nchez-Monge},
  J.~{Kainulainen}, P.~{Didelon}, S.~{Suri}, N.~{Schneider},
  J.~{Ballesteros-Paredes}, Y.~N. {Lee}, P.~{Hennebelle}, P.~{Pilleri},
  M.~{Gonz{\'a}lez-Garc{\'\i}a}, C.~{Kramer}, S.~{Garc{\'\i}a-Burillo},
  A.~{Luna}, J.~R. {Goicoechea}, P.~{Tremblin}, and S.~{Geen}.
\newblock {Dynamics of cluster-forming hub-filament systems. The case of the
  high-mass star-forming complex Monoceros R2}.
\newblock \emph{Astronomy \& Astrophysics}, 629:\penalty0 A81, September 2019.
\newblock \doi{10.1051/0004-6361/201935260}.

\bibitem[{Urquhart} et~al.(2018){Urquhart}, {K{\"o}nig}, {Giannetti},
  {Leurini}, {Moore}, {Eden}, {Pillai}, {Thompson}, {Braiding}, {Burton},
  {Csengeri}, {Dempsey}, {Figura}, {Froebrich}, {Menten}, {Schuller}, {Smith},
  and {Wyrowski}]{Urquhart18}
J.~S. {Urquhart}, C.~{K{\"o}nig}, A.~{Giannetti}, S.~{Leurini}, T.~J.~T.
  {Moore}, D.~J. {Eden}, T.~{Pillai}, M.~A. {Thompson}, C.~{Braiding}, M.~G.
  {Burton}, T.~{Csengeri}, J.~T. {Dempsey}, C.~{Figura}, D.~{Froebrich}, K.~M.
  {Menten}, F.~{Schuller}, M.~D. {Smith}, and F.~{Wyrowski}.
\newblock {ATLASGAL - properties of a complete sample of Galactic clumps}.
\newblock \emph{Monthly Notices of the Royal Astronomical Society},
  473\penalty0 (1):\penalty0 1059--1102, January 2018.
\newblock \doi{10.1093/mnras/stx2258}.

\bibitem[{Vazquez-Semadeni}(1994)]{Vazquez1994}
Enrique {Vazquez-Semadeni}.
\newblock {Hierarchical Structure in Nearly Pressureless Flows as a Consequence
  of Self-similar Statistics}.
\newblock \emph{The Astrophysical Journal}, 423:\penalty0 681, March 1994.
\newblock \doi{10.1086/173847}.

\bibitem[{V{\'a}zquez-Semadeni} et~al.(2024){V{\'a}zquez-Semadeni}, {Palau},
  {G{\'o}mez}, {Arroyo-Ch{\'a}vez}, {Alig}, {Ballesteros-Paredes}, {Camacho},
  {Traficante}, {Gonz{\'a}lez-Samaniego}, {Zamora-Avil{\'e}s}, and
  {Burkert}]{Enrique2024}
Enrique {V{\'a}zquez-Semadeni}, Aina {Palau}, Gilberto~C. {G{\'o}mez}, Griselda
  {Arroyo-Ch{\'a}vez}, Christian {Alig}, Javier {Ballesteros-Paredes}, Vianey
  {Camacho}, Alessio {Traficante}, Alejandro {Gonz{\'a}lez-Samaniego}, Manuel
  {Zamora-Avil{\'e}s}, and Andreas {Burkert}.
\newblock {The Turbulent Support (TS) and Global Hierarchical Collapse (GHC)
  models for molecular clouds compared. Differences, convergence, and myths}.
\newblock \emph{arXiv e-prints}, art. arXiv:2408.10406, August 2024.
\newblock \doi{10.48550/arXiv.2408.10406}.

\bibitem[{Wang} et~al.(2020){Wang}, {Koch}, {Galv{\'a}n-Madrid}, {Lai}, {Liu},
  {Lin}, and {Pattle}]{Wang2020}
Jia-Wei {Wang}, Patrick~M. {Koch}, Roberto {Galv{\'a}n-Madrid}, Shih-Ping
  {Lai}, Hauyu~Baobab {Liu}, Sheng-Jun {Lin}, and Kate {Pattle}.
\newblock {Formation of the Hub-Filament System G33.92+0.11: Local Interplay
  between Gravity, Velocity, and Magnetic Field}.
\newblock \emph{The Astrophysical Journal}, 905\penalty0 (2):\penalty0 158,
  December 2020.
\newblock \doi{10.3847/1538-4357/abc74e}.

\bibitem[{Wang} et~al.(2024){Wang}, {Koch}, {Clarke}, {Fuller}, {Peretto},
  {Tang}, {Yen}, {Lai}, {Ohashi}, {Arzoumanian}, {Johnstone}, {Furuya},
  {Inutsuka}, {Lee}, {Ward-Thompson}, {Le Gouellec}, {Liu}, {Fanciullo},
  {Hwang}, {Pattle}, {Poidevin}, {Tahani}, {Onaka}, {Rawlings}, {Chung}, {Liu},
  {Lyo}, {Priestley}, {Hoang}, {Tamura}, {Berry}, {Bastien}, {Ching},
  {Coud{\'e}}, {Kwon}, {Chen}, {Eswaraiah}, {Soam}, {Hasegawa}, {Qiu},
  {Bourke}, {Byun}, {Chen}, {Chen}, {Chen}, {Cho}, {Choi}, {Choi}, {Choi},
  {Chrysostomou}, {Dai}, {Di Francesco}, {Diep}, {Doi}, {Duan}, {Duan}, {Eden},
  {Fiege}, {Fissel}, {Franzmann}, {Friberg}, {Friesen}, {Gledhill}, {Graves},
  {Greaves}, {Griffin}, {Gu}, {Han}, {Hayashi}, {Houde}, {Inoue}, {Iwasaki},
  {Jeong}, {K{\"o}nyves}, {Kang}, {Kang}, {Karoly}, {Kataoka}, {Kawabata},
  {Khan}, {Kim}, {Kim}, {Kim}, {Kim}, {Kim}, {Kim}, {Kim}, {Kirchschlager},
  {Kirk}, {Kobayashi}, {Kusune}, {Kwon}, {Lacaille}, {Law}, {Lee}, {Lee},
  {Lee}, {Lee}, {Li}, {Li}, {Li}, {Li}, {Lin}, {Liu}, {Liu}, {Lu}, {Mairs},
  {Matsumura}, {Matthews}, {Moriarty-Schieven}, {Nagata}, {Nakamura},
  {Nakanishi}, {Ngoc}, {Park}, {Parsons}, {Pyo}, {Qian}, {Rao}, {Rawlings},
  {Retter}, {Richer}, {Rigby}, {Sadavoy}, {Saito}, {Savini}, {Seta}, {Sharma},
  {Shimajiri}, {Shinnaga}, {Tang}, {Thuong}, {Tomisaka}, {Tram}, {Tsukamoto},
  {Viti}, {Wang}, {Whitworth}, {Wu}, {Xie}, {Yang}, {Yoo}, {Yuan}, {Yun},
  {Zenko}, {Zhang}, {Zhang}, {Zhang}, {Zhou}, {Zhu}, {de Looze}, {Andr{\'e}},
  {Dowell}, {Eyres}, {Falle}, {Robitaille}, and {van Loo}]{Wang2024}
Jia-Wei {Wang}, Patrick~M. {Koch}, Seamus~D. {Clarke}, Gary {Fuller}, Nicolas
  {Peretto}, Ya-Wen {Tang}, Hsi-Wei {Yen}, Shih-Ping {Lai}, Nagayoshi {Ohashi},
  Doris {Arzoumanian}, Doug {Johnstone}, Ray {Furuya}, Shu-ichiro {Inutsuka},
  Chang~Won {Lee}, Derek {Ward-Thompson}, Valentin J.~M. {Le Gouellec}, Hong-Li
  {Liu}, Lapo {Fanciullo}, Jihye {Hwang}, Kate {Pattle}, Fr{\'e}d{\'e}rick
  {Poidevin}, Mehrnoosh {Tahani}, Takashi {Onaka}, Mark~G. {Rawlings}, Eun~Jung
  {Chung}, Junhao {Liu}, A.~Ran {Lyo}, Felix {Priestley}, Thiem {Hoang},
  Motohide {Tamura}, David {Berry}, Pierre {Bastien}, Tao-Chung {Ching}, Simon
  {Coud{\'e}}, Woojin {Kwon}, Mike {Chen}, Chakali {Eswaraiah}, Archana {Soam},
  Tetsuo {Hasegawa}, Keping {Qiu}, Tyler~L. {Bourke}, Do-Young {Byun}, Zhiwei
  {Chen}, Huei-Ru~Vivien {Chen}, Wen~Ping {Chen}, Jungyeon {Cho}, Minho {Choi},
  Yunhee {Choi}, Youngwoo {Choi}, Antonio {Chrysostomou}, Sophia {Dai}, James
  {Di Francesco}, Pham~Ngoc {Diep}, Yasuo {Doi}, Yan {Duan}, Hao-Yuan {Duan},
  David {Eden}, Jason {Fiege}, Laura~M. {Fissel}, Erica {Franzmann}, Per
  {Friberg}, Rachel {Friesen}, Tim {Gledhill}, Sarah {Graves}, Jane {Greaves},
  Matt {Griffin}, Qilao {Gu}, Ilseung {Han}, Saeko {Hayashi}, Martin {Houde},
  Tsuyoshi {Inoue}, Kazunari {Iwasaki}, Il-Gyo {Jeong}, Vera {K{\"o}nyves},
  Ji-hyun {Kang}, Miju {Kang}, Janik {Karoly}, Akimasa {Kataoka}, Koji
  {Kawabata}, Zacariyya {Khan}, Mi-Ryang {Kim}, Kee-Tae {Kim}, Kyoung~Hee
  {Kim}, Shinyoung {Kim}, Jongsoo {Kim}, Hyosung {Kim}, Gwanjeong {Kim},
  Florian {Kirchschlager}, Jason {Kirk}, Masato I.~N. {Kobayashi}, Takayoshi
  {Kusune}, Jungmi {Kwon}, Kevin {Lacaille}, Chi-Yan {Law}, Sang-Sung {Lee},
  Hyeseung {Lee}, Jeong-Eun {Lee}, Chin-Fei {Lee}, Dalei {Li}, Hua-bai {Li},
  Guangxing {Li}, Di~{Li}, Sheng-Jun {Lin}, Tie {Liu}, Sheng-Yuan {Liu}, Xing
  {Lu}, Steve {Mairs}, Masafumi {Matsumura}, Brenda {Matthews}, Gerald
  {Moriarty-Schieven}, Tetsuya {Nagata}, Fumitaka {Nakamura}, Hiroyuki
  {Nakanishi}, Nguyen~Bich {Ngoc}, Geumsook {Park}, Harriet {Parsons}, Tae-Soo
  {Pyo}, Lei {Qian}, Ramprasad {Rao}, Jonathan {Rawlings}, Brendan {Retter},
  John {Richer}, Andrew {Rigby}, Sarah {Sadavoy}, Hiro {Saito}, Giorgio
  {Savini}, Masumichi {Seta}, Ekta {Sharma}, Yoshito {Shimajiri}, Hiroko
  {Shinnaga}, Xindi {Tang}, Hoang~Duc {Thuong}, Kohji {Tomisaka}, Le~Ngoc
  {Tram}, Yusuke {Tsukamoto}, Serena {Viti}, Hongchi {Wang}, Anthony
  {Whitworth}, Jintai {Wu}, Jinjin {Xie}, Meng-Zhe {Yang}, Hyunju {Yoo},
  Jinghua {Yuan}, Hyeong-Sik {Yun}, Tetsuya {Zenko}, Chuan-Peng {Zhang}, Yapeng
  {Zhang}, Guoyin {Zhang}, Jianjun {Zhou}, Lei {Zhu}, Ilse {de Looze}, Philippe
  {Andr{\'e}}, C.~Darren {Dowell}, Stewart {Eyres}, Sam {Falle},
  Jean-Fran{\c{c}}ois {Robitaille}, and Sven {van Loo}.
\newblock {Filamentary Network and Magnetic Field Structures Revealed with
  BISTRO in the High-mass Star-forming Region NGC 2264: Global Properties and
  Local Magnetogravitational Configurations}.
\newblock \emph{The Astrophysical Journal}, 962\penalty0 (2):\penalty0 136,
  February 2024.
\newblock \doi{10.3847/1538-4357/ad165b}.

\bibitem[{Xu} et~al.(2023){Xu}, {Wang}, {Liu}, {Goldsmith}, {Zhang}, {Juvela},
  {Liu}, {Qin}, {Li}, {Tej}, {Garay}, {Bronfman}, {Li}, {Wu}, {G{\'o}mez},
  {V{\'a}zquez-Semadeni}, {Tatematsu}, {Ren}, {Zhang}, {Toth}, {Liu}, {Yue},
  {Zhang}, {Baug}, {Issac}, {Stutz}, {Liu}, {Fuller}, {Tang}, {Zhang},
  {Dewangan}, {Lee}, {Zhou}, {Xie}, {Jiao}, {Wang}, {Liu}, {Luo}, {Soam}, and
  {Eswaraiah}]{Xu2023}
Feng-Wei {Xu}, Ke~{Wang}, Tie {Liu}, Paul~F. {Goldsmith}, Qizhou {Zhang}, Mika
  {Juvela}, Hong-Li {Liu}, Sheng-Li {Qin}, Guang-Xing {Li}, Anandmayee {Tej},
  Guido {Garay}, Leonardo {Bronfman}, Shanghuo {Li}, Yue-Fang {Wu}, Gilberto~C.
  {G{\'o}mez}, Enrique {V{\'a}zquez-Semadeni}, Ken'ichi {Tatematsu}, Zhiyuan
  {Ren}, Yong {Zhang}, L.~Viktor {Toth}, Xunchuan {Liu}, Nannan {Yue}, Siju
  {Zhang}, Tapas {Baug}, Namitha {Issac}, Amelia~M. {Stutz}, Meizhu {Liu},
  Gary~A. {Fuller}, Mengyao {Tang}, Chao {Zhang}, Lokesh {Dewangan}, Chang~Won
  {Lee}, Jianwen {Zhou}, Jinjin {Xie}, Wenyu {Jiao}, Chao {Wang}, Rong {Liu},
  Qiuyi {Luo}, Archana {Soam}, and Chakali {Eswaraiah}.
\newblock {ATOMS: ALMA Three-millimeter Observations of Massive Star-forming
  regions - XV. Steady accretion from global collapse to core feeding in
  massive hub-filament system SDC335}.
\newblock \emph{Monthly Notices of the Royal Astronomical Society},
  520\penalty0 (3):\penalty0 3259--3285, April 2023.
\newblock \doi{10.1093/mnras/stad012}.

\bibitem[{Xu} et~al.(2020){Xu}, {Li}, {Dai}, {Fuller}, and
  {Yue}]{2020ApJ...894L..20X}
Xuefang {Xu}, Di~{Li}, Y.~Sophia {Dai}, Gary~A. {Fuller}, and Nannan {Yue}.
\newblock {Independent Core Rotation in Massive Filaments in Orion}.
\newblock \emph{The Astrophysical Journal Letters}, 894\penalty0 (2):\penalty0
  L20, May 2020.
\newblock \doi{10.3847/2041-8213/ab8ad7}.

\bibitem[{Zhang} et~al.(2025){Zhang}, {Liu}, {Jiao}, {Zhu}, {Ren}, {Liu},
  {Wang}, {Wu}, {Li}, {Garc{\'\i}a}, {Garay}, {Bronfman}, {Juvela}, {das},
  {Lee}, {Xu}, {T{\'o}th}, {Gorai}, and {Sanhueza}]{Zhang2025}
C.~{Zhang}, Tie {Liu}, Sihan {Jiao}, Feng-Yao {Zhu}, Z.~Y. {Ren}, H.~L. {Liu},
  Ke~{Wang}, J.~W. {Wu}, D.~{Li}, P.~{Garc{\'\i}a}, Guido {Garay}, Leonardo
  {Bronfman}, Mika {Juvela}, Swagat {das}, Chang~Won {Lee}, Feng-Wei {Xu},
  L.~V. {T{\'o}th}, Prasanta {Gorai}, and Patricio {Sanhueza}.
\newblock {ATOMS: ALMA Three-millimeter Observations of massive Star-forming
  regions - XX. Probability distribution function of integrated intensity for
  dense molecular gas tracers}.
\newblock \emph{Monthly Notices of the Royal Astronomical Society},
  538\penalty0 (1):\penalty0 1--10, March 2025.
\newblock \doi{10.1093/mnras/staf176}.

\bibitem[{Zhang} et~al.(2014){Zhang}, {Qiu}, {Girart}, {Liu}, {Tang}, {Koch},
  {Li}, {Keto}, {Ho}, {Rao}, {Lai}, {Ching}, {Frau}, {Chen}, {Li}, {Padovani},
  {Bontemps}, {Csengeri}, and {Ju{\'a}rez}]{2014ApJ...792..116Z}
Qizhou {Zhang}, Keping {Qiu}, Josep~M. {Girart}, Hauyu~Baobab {Liu}, Ya-Wen
  {Tang}, Patrick~M. {Koch}, Zhi-Yun {Li}, Eric {Keto}, Paul T.~P. {Ho},
  Ramprasad {Rao}, Shih-Ping {Lai}, Tao-Chung {Ching}, Pau {Frau}, How-Huan
  {Chen}, Hua-Bai {Li}, Marco {Padovani}, Sylvain {Bontemps}, Timea {Csengeri},
  and Carmen {Ju{\'a}rez}.
\newblock {Magnetic Fields and Massive Star Formation}.
\newblock \emph{The Astrophysical Journal}, 792\penalty0 (2):\penalty0 116,
  September 2014.
\newblock \doi{10.1088/0004-637X/792/2/116}.

\bibitem[{Zhou} et~al.(2022){Zhou}, {Liu}, {Evans}, {Garay}, {Goldsmith},
  {G{\'o}mez}, {V{\'a}zquez-Semadeni}, {Liu}, {Stutz}, {Wang}, {Juvela}, {He},
  {Li}, {Bronfman}, {Liu}, {Xu}, {Tej}, {Dewangan}, {Li}, {Zhang}, {Zhang},
  {Ren}, {Tatematsu}, {Shing Li}, {Won Lee}, {Baug}, {Qin}, {Wu}, {Peng},
  {Zhang}, {Liu}, {Luo}, {Ge}, {Saha}, {Chakali}, {Zhang}, {Kim},
  {Ristorcelli}, {Shen}, and {Li}]{Zhou2022}
Jian-Wen {Zhou}, Tie {Liu}, Neal~J. {Evans}, Guido {Garay}, Paul~F.
  {Goldsmith}, Gilberto~C. {G{\'o}mez}, Enrique {V{\'a}zquez-Semadeni}, Hong-Li
  {Liu}, Amelia~M. {Stutz}, Ke~{Wang}, Mika {Juvela}, Jinhua {He}, Di~{Li},
  Leonardo {Bronfman}, Xunchuan {Liu}, Feng-Wei {Xu}, Anandmayee {Tej}, L.~K.
  {Dewangan}, Shanghuo {Li}, Siju {Zhang}, Chao {Zhang}, Zhiyuan {Ren},
  Ken'ichi {Tatematsu}, Pak {Shing Li}, Chang {Won Lee}, Tapas {Baug}, Sheng-Li
  {Qin}, Yuefang {Wu}, Yaping {Peng}, Yong {Zhang}, Rong {Liu}, Qiu-Yi {Luo},
  Jixing {Ge}, Anindya {Saha}, Eswaraiah {Chakali}, Qizhou {Zhang}, Kee-Tae
  {Kim}, Isabelle {Ristorcelli}, Zhi-Qiang {Shen}, and Jin-Zeng {Li}.
\newblock {ATOMS: ALMA Three-millimeter Observations of Massive Star-forming
  regions - XI. From inflow to infall in hub-filament systems}.
\newblock \emph{Monthly Notices of the Royal Astronomical Society},
  514\penalty0 (4):\penalty0 6038--6052, August 2022.
\newblock \doi{10.1093/mnras/stac1735}.

\end{thebibliography}

\end{document}